\documentclass[preprint,aps,tightenlines,showpacs,superscriptaddress]{revtex4}
\usepackage[dvips]{graphicx}
\usepackage{dcolumn}
\usepackage{epsfig}
\usepackage{color}
\usepackage{bm}
\def\bra#1{\mathinner{\langle{#1}|}}
\def\ket#1{\mathinner{|{#1}\rangle}}

\def\ohalf{{\textstyle{1\over 2}}}

\def\thalf{{\textstyle{3\over 2}}}

\def\ohalf{{\textstyle{1\over 2}}}
\def\thalf{{\textstyle{3\over 2}}}
\def\fhalf{{\textstyle{5\over 2}}}
\def\shalf{{\textstyle{7\over 2}}}
\def\nhalf{{\textstyle{9\over 2}}}

\newcommand{\beq}{\begin{equation}}
\newcommand{\eeq}{\end{equation}}

\begin{document}

\begin{center}

The ANL-Osaka Partial-Wave Amplitudes of $\pi N$ and $\gamma N$ Reactions

\vspace{0.5cm} 
H. Kamano$^{1}$, T.-S. H. Lee$^{2}$, S.X. Nakamura$^{3}$, T. Sato$^{1}$ \\
$^{1}$ Research Center for Nuclear Physics, 10-1 Mihogaoka, Ibaraki, Osaka, Japan \\
$^{2}$Physics Division, Argonne National Laboratory, Argonne, Illinois 60249, USA\\ 
$^{3}$ University of Science and Technology of China, Hefei, 230026, People's Republic of China \\
\vspace{0.5 cm}
Contact Person: T.-S. H. Lee (email:tshlee@anl.gov), September 20, 2019
\vspace{0.5cm}
 ({\bf https://www.phy.anl.gov/theory/research/anl-osaka-pwa})
\end{center}

\begin{center}

Abstract

\end{center}

The determination of the Argonne National Laboratory-Osaka University 
(ANL-Osaka) Partial-Wave Amplitudes (PWA)
of $\pi N$ and $\gamma N$ Reactions is reviewed. The predicted PWA are presented on
a web page ({\bf https://www.phy.anl.gov/theory/research/anl-osaka-pwa}).
The formulas are given for using the predicted PWA to calculate the cross sections of
 (1) meson-baryon ($MB$) scattering $MB \rightarrow M'B'$ with $MB, M'B'= \pi N, 
\eta N, K\Lambda, K\Sigma$, (2) two-pion production 
$\pi N \rightarrow \pi\Delta, \rho N, \sigma N \rightarrow \pi\pi N$,
(3) Meson photoproduction 
$\gamma N \rightarrow \pi N, \eta N, K\Lambda, K\Sigma$, (4) Pion electroproduction
$N(e,e'\pi)N$, (5) inclusive $N(e,e')X$.  
We also present sample results from our fits to the data.

\vspace{1cm}
\vspace{0.5cm}
\section{introduction}
The Argonne National Laboratory-Osaka University (ANL-Osaka) collaboration started in 1996 with a
publication~\cite{sl96} of a meson-exchange model for investigating 
the excitation of the $\Delta$ (1232) resonance in
 $\pi N$ and $\gamma N$ reactions. The predictions~\cite{sl96,sl01}
 from this model, called the Sato-Lee (SL) model in the
community, were found to be  consistent with  
the  data from MIT-Bates, Mainz, Bonn, and JLab.
The results from the SL model strongly suggested 
 that a dynamical model based on a Hamiltonian with bare
$N^*$ states and meson-exchange mechanisms can be used to
\begin{enumerate}
\item describe the data of $\pi N$ and $\gamma N$ reactions up to invariant mass $W=2$ GeV,
\item extract the masses and widths of nucleon resonances ($N^*$) from the
predicted Partial-Wave Amplitudes (PWA) 
 for investigating the structure of
the nucleon and its excited states,
\item apply the constructed Hamiltonian with meson and $N^*$ degrees of freedom to predict
the cross sections of
the production of mesons ($\pi, 2\pi, \eta, K$) from nuclei, which are needed for analyzing
the data of nuclear reactions induced by  hadrons, electrons, and neutrinos in the nucleon resonance
region.
\end{enumerate}

The formulation of the SL model was then extended to include the higher mass bare $N^*$ states and
 the meson-baryon channels which have significant contributions
to the $\pi N$ and $\gamma N$ total cross sections below invariant mass $W=2$ GeV.
Exploratory calculations using this Dynamical  Coupled-Channel (DCC)  model were carried out by
the ANL-Osaka collaboration during 2004-2006. The analysis of the 
world data of $\pi N$ and $\gamma N$ reactions
up to $W=2$ GeV was then carried out at the Excited Baryon Analysis Center (EBAC) at JLab 
during 2006-2012, with 12 publications [3-15]. 
The ANL-Osaka collaboration continued  the analysis to extract~\cite{knls13,knls16}
 22 nucleon resonances during 2013-2016. 
The calculations for the analysis involved solving coupled-channel scattering equations with 8 channels:
$\gamma N, \pi N, \eta N, K\Lambda, K\Sigma$, and $\pi\pi N$ which has
$\pi \Delta, \sigma N, \rho N$ resonant components. The parameters of the model were determined by performing 
$\chi^2$-fits to the world data of 
$\gamma N,\pi N\rightarrow \pi N, \eta N, K\Lambda, K\Sigma$ (about 30,000
data points). The ANL-Osaka model was also extended~\cite{nks15}
to investigate  the electron and neutrino-induced meson production reactions on the nucleon.
The computation resources from DOE's NERSC (about 500,000  hours/year) and Argonne's
LCRC (about 300,000 hours/year) had been used to complete this task.

The results presented on {\bf https://www.phy.anl.gov/theory/research/anl-osaka-pwa} are for
the nuclear and hadron physics communities to get access to the predicted
partial-wave amplitudes of $\gamma N,\pi N\rightarrow \pi N, \eta N, K\Lambda, K\Sigma, 
\pi\Delta,\sigma N,\rho N$. In section II, we describe briefly the ANL-Osaka DCC model and the
resulting formulation of $\pi N $ and $\gamma N$ reactions. The procedures for extracting nucleon resonances
are given  in section III. 
In section IV, we present formula for using
the PWA posted on the web page to calculate the cross sections of the considered meson-baryon reactions .
     The data included in the fits are summarized in section V. 
In section VI and Appendix, we present sample results from our fits to the data.

\section{ANL-Osaka DCC Model}
The ANL-Osaka DCC Model is based on an effective Hamiltonian of the following $energy-independent$ form:
\begin{eqnarray} 
H=H_0+v_{22}+\Gamma_V+h_{\pi\pi N}\,,
\label{eq:h}
\end{eqnarray}
where $H_0=\sum_{\alpha}\sqrt{m^2_\alpha+\vec{p}^{\,\,2}_\alpha}$ with $m_\alpha$ and $\vec{p}_\alpha$
denoting the mass and momentum of
particle $\alpha$, respectively. The interactions are defined as
\begin{eqnarray}
v_{22}&=&\sum_{MB,M'B'}v_{MB,M'B'}+v_{\pi\pi,\pi\pi}\ ,\\
\Gamma_V&=&\{\sum_{N^*}(\sum_{MB}\Gamma_{N^*\rightarrow MB})+\sum_{M^*}h_{M^*\rightarrow \pi\pi}\}
+\{c.c\}\ ,\\
h_{\pi \pi N}&=&\{v_{\pi N,\pi\pi N}+v_{\gamma N,\pi\pi N}\}+\{c.c\}\ ,
\label{eq:hpipi}
\end{eqnarray}
where $v_{MB,M'B'}$ is the meson-baryon ($MB$) interactions, $v_{\pi\pi,\pi\pi}$ is the $\pi\pi$
interactions, $\Gamma_{N^*\rightarrow MB}$ describes the decay of a
$bare$ excited nucleon ($N^*$) into a $MB$ state,  $h_{M^*\rightarrow \pi\pi}$ describes the decay of
a $bare$ meson  ($M^*$) into a $\pi\pi$ state, and
 $\{c.c.\}$ denotes the complex conjugate of the terms on its left-side. For describing the $\pi N$ and $\gamma N$
reactions up to invariant mass $W=$ 2 GeV, we include 
$MB=\gamma N,\pi N,\eta N,K\Lambda, K\Sigma, \pi\Delta,\rho N, \sigma N$, about $20$ bare $N^*$ states,
and $M^*=\rho,\sigma$.

\subsection{Hadronic amplitudes}
\label{sec:dcc-had}

Starting with the Hamiltonian defined by Eqs.~(\ref{eq:h})-(\ref{eq:hpipi}), 
we apply~\cite{msl07} the projection operator method to cast 
the partial-wave amplitudes of the scattering  $T$ matrix
of the meson-baryon reaction, 
$M(\vec k) + B(-\vec k) \to M'(\vec k') + B'(-\vec k')$, 
into the following form
\begin{equation}
T_{M'B',MB}(k',k;E) = t_{M'B',MB}(k',k;E) + t^R_{M'B',MB}(k',k;E),
\label{eq:tmbmb}
\end{equation}
where $E$ is the total energy, $k$ and $k'$ are the 
meson-baryon relative momenta in the center of mass frame, and 
$MB, M'B' = \pi N, \eta N, \pi\Delta, \rho N, \sigma N, K\Lambda, K\Sigma$
are the reaction channels included
 in the analysis. The notation $MB$ also represents the partial-wave quantum numbers:
$[L(s_Ms_B)S]JT]$, where $J$ is the total angular momentum, $T$ the total isospin, $L$ the orbital angular momentum,
$S$ the total spin which is from the coupling 
of the meson spin  $s_M$ and the baryon spin $s_B$.

The direct reaction amplitude $t_{M'B',MB}(k',k;E)$ in 
Eq.~(\ref{eq:tmbmb})
is defined by a set of 
coupled-channel equations
\begin{eqnarray}
t_{M'B',MB}(k',k;E) &=& 
V_{M'B',MB}(k',k;E) 
\nonumber \\
&& 
+ 
\sum_{M^{''} B^{''}} \int_{C_{M''B''}} k''^{2}dk'' V_{M'B',M''B''}(k',k'';E)
\nonumber\\
&&
\qquad\qquad\qquad
\times
G_{M''B''}(k'';E)
t_{M''B'',MB}(k'',k;E).
\label{eq:cc-eq}
\end{eqnarray}
Here $C_{M''B''}$ is the integration path,
which is taken from $0$ to $\infty$ for the physical $E$;
the summation $\sum_{MB}$ runs over the orbital angular momentum and total spin indices
for all $MB$ channels allowed in a given partial wave;
$G_{M''B''}(k;E)$ are the meson-baryon Green functions.
Defining  $E_\alpha(k)=[m^2_\alpha + k^2]^{1/2}$ with $m_\alpha$ being
the mass of a particle $\alpha$,
the meson-baryon Green functions in the above equations are:
\begin{eqnarray}
G_{MB}(k;E)=\frac{1}{E-E_M(k)-E_B(k) + i\epsilon}\,,
\label{eq:prop-stab}
\end{eqnarray}
for the stable $\pi N$, $\eta N$, $K\Lambda$, and $K\Sigma$  channels, and
\begin{equation}
G_{MB}(k;E)=\frac{1}{E-E_M(k)-E_B(k) -\Sigma_{MB}(k;E)}\,,
\label{eq:prop-unstab}
\end{equation}
for the unstable $\pi\Delta$, $\rho N$, and $\sigma N$ channels.
The self energy $\Sigma_{MB}(k;E)$ in Eq.~(\ref{eq:prop-unstab})
is calculated from a vertex
function defining the decay of the considered unstable particle
in the presence of a spectator $\pi$ or $N$ with momentum $k$.
For the $\pi\Delta$ and $\rho N$ channels, the self-energies are explicitly given by
\begin{eqnarray}
\Sigma_{\pi \Delta}(k;E) &=& \frac{m_\Delta}{E_\Delta(k)}
\int q^2 dq \frac{ M_{\pi N}(q)}{[M^2_{\pi N}(q) + k^2]^{1/2}}
\frac{\left|f_{\Delta \to \pi N}(q)\right|^2}
{E-E_\pi(k) -[M^2_{\pi N}(q) + k^2]^{1/2} + i\epsilon},
\label{eq:self-pidelta}
\\
\Sigma_{\rho N}(k;E) &=& \frac{m_\rho}{E_\rho(k)}
\int q^2 dq \frac{ M_{\pi\pi}(q)}{[M^2_{\pi \pi}(q) + k^2]^{1/2}}
\frac{\left|f_{\rho \to \pi \pi}(q)\right|^2}
{E-E_N(k) -[M^2_{\pi \pi}(q) + k^2]^{1/2} + i\epsilon},
\label{eq:self-rhon}
\end{eqnarray}
where $m_\Delta=1280$ MeV, $m_\rho = 812$ MeV, $M_{\pi N}(q) = E_\pi (q) + E_N(q)$, 
and $M_{\pi \pi}(q) = E_\pi (q) + E_\pi(q)$.
The form factors $f_{\Delta\to\pi N}(q)$ and $f_{\rho \to \pi \pi}(q)$ are for describing
the $\Delta\to\pi N$ and $\rho \to \pi \pi$ decays in the $\Delta$ and $\rho$ rest frames,
respectively.
They are parametrized as:
\begin{eqnarray}
f_{\Delta \to \pi N}(q) &=& -i\frac{(0.98)}{[2(m_N+m_\pi)]^{1/2}}
\left(\frac{q}{m_\pi}\right)
\left(\frac{1}{1+[q/(358~\text{MeV})]^2}\right)^2 ,
\\
f_{\rho \to \pi \pi}(q) &=& \frac{(0.6684)}{\sqrt{m_\pi}}
\left(\frac{q}{(461~\text{MeV})}\right)
\left(\frac{1}{1+[q/(461~\text{MeV})]^2}\right)^2 .
\end{eqnarray}
The $\sigma$ self-energy $\Sigma_{\sigma N}(k;E)$  is calculated from a $\pi\pi$ s-wave scattering model with
a vertex function $g(q)$ for the $\sigma \rightarrow \pi\pi$ decay and a
separable interaction $v(q',q)=h_0 h(q')h(q)$.
The resulting form is
\begin{eqnarray}
\Sigma_{\sigma N}(k;E) &=& \langle gG_{\pi\pi}g\rangle(k;E) 
+\tau(k;E)[\langle gG_{\pi\pi}h\rangle(k;E)]^2,
\end{eqnarray}
with
\begin{eqnarray}
\tau(k;E) &=& \frac{h_0}{1-h_0\langle h G_{\pi\pi} h\rangle (k;E)},
\\
\langle hG_{\pi\pi}h\rangle(k;E) &=&
\int dq q^2 \frac{M_{\pi\pi}(q)}{ [M_{\pi\pi}^2(q) + k^2]^{1/2}} 
\nonumber\\
&& \qquad\qquad\qquad \times
\frac{h(q)^2}{E - E_N(k) - [M_{\pi\pi}^2(q) + k^2]^{1/2}+i\varepsilon},
\\
\langle gG_{\pi\pi}g\rangle(k;E) &=& 
\frac{m_\sigma}{E_\sigma(k)}
\int dq q^2 \frac{M_{\pi\pi}(q)}{ [M_{\pi\pi}^2(q) + k^2]^{1/2}} 
\nonumber\\
&& \qquad\qquad\qquad \times
\frac{g(q)^2}{E - E_N(k) - [M_{\pi\pi}^2(q) + k^2]^{1/2}+i\varepsilon},
\\
\langle gG_{\pi\pi}h\rangle(k;E) &=&
\sqrt{\frac{m_\sigma}{E_\sigma(k)}}
\int dq q^2 \frac{M_{\pi\pi}(q)}{ [M_{\pi\pi}^2(q) + k^2]^{1/2}} 
\nonumber\\
&& \qquad\qquad\qquad \times
\frac{g(q)h(q)}{E - E_N(k) - [M_{\pi\pi}^2(q) + k^2]^{1/2}+i\varepsilon}.
\end{eqnarray}
In the above equations, $m_\sigma = 700.0$ MeV and the form factors are 
\begin{eqnarray}
g(p)&=&\frac{g_0}{\sqrt{m_\pi}}\frac{1}{1+(cp)^2},
\\
h(p)&=&\frac{1}{m_\pi}\frac{1}{1+(dp)^2}.
\label{eq:pipi-pot2}
\end{eqnarray}
where
$g_0 = 1.638$, $h_0 = 0.556$, $c=1.02$ fm, and $d=0.514$ fm.

\begin{figure}[b]
\includegraphics[clip,width=0.8\textwidth]{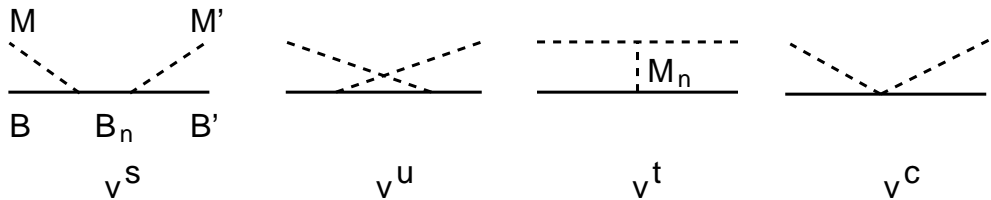}
\caption{\label{fig:mex} Meson-exchange mechanisms.}
\end{figure}

\begin{figure}[t]
\includegraphics[clip,width=0.8\textwidth]{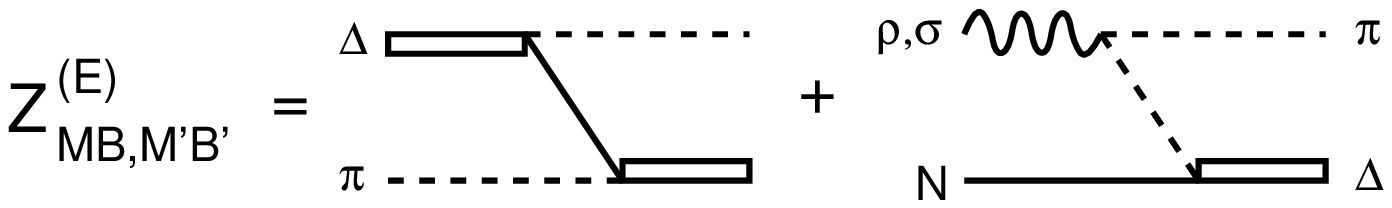}
\caption{\label{fig:z-diag} Z-diagram  mechanisms.}
\end{figure}

The driving terms of Eq.~(\ref{eq:cc-eq}) are
\begin{equation}
V_{M'B',MB}(k',k;E) = v_{M'B',MB}(k',k) + Z^{(E)}_{M'B',MB}(k',k;E)\,, 
\label{eq:veff-mbmb}
\end{equation}
where $v_{M'B',MB}(k',k)$ are the  meson-exchange potentials
derived from the tree-diagrams, as illustrated in Fig.~\ref{fig:mex},   
of phenomenological Lagrangians. 
Within the unitary transformation method
used in the derivation, 
those potentials are energy independent and free of singularities.
The Lagrangians used in our derivations and the
partial-wave expansions of  $v_{M'B',MB}(k',k)$ 
are given in Appendices B and C of Ref.~\cite{knls13}. 

The energy-dependent $Z^{(E)}_{M'B',MB}(k',k;E)$ terms in Eq.~(\ref{eq:veff-mbmb}),
as illustrated in Fig.~\ref{fig:z-diag},
contain the
{\it moving} singularities due to the $\pi\pi N$ cuts. 
The procedures for evaluating the partial-wave
matrix elements of $Z^{(E)}_{M'B',MB}(k',k;E)$
are explained in detail in the Appendix E of Ref.~\cite{msl07}.

The second term in the right-hand-side of Eq.~(\ref{eq:tmbmb}) 
is the $N^*$-excitation amplitude defined by
\begin{eqnarray} 
t^R_{M'B',MB}(k',k;E)= \sum_{N^*_n, N^*_m}
\bar{\Gamma}_{M'B',N^*_n}(k';E) [D(E)]_{n,m}
\bar{\Gamma}_{N^*_m, MB}(k;E) .
\label{eq:tmbmb-r} 
\end{eqnarray}
Here the dressed $N^\ast \to MB$ and $MB \to N^\ast$ decay vertices are, respectively, defined by
\begin{equation}
\bar\Gamma_{MB,N^\ast}(k;E) =
\Gamma_{MB,N^\ast}(k) +
\sum_{M'B'}\int q^2 dq t_{MB,M'B'}(k,q;E) G_{M'B'}(q,E)\Gamma_{M'B',N^\ast}(q),
\label{eq:dress-1}
\end{equation}
\begin{equation}
\bar\Gamma_{N^\ast,MB}(k;E) =
\Gamma_{MB,N^\ast}^\dag(k) +
\sum_{M'B'}\int q^2 dq \Gamma_{M'B',N^\ast}^\dag(q) G_{M'B'}(q,E) t_{M'B',MB}(q,k;E),
\label{eq:dress-2}
\end{equation}
with $\Gamma_{MB,N^\ast}(k)$ being the bare $N^\ast\to MB$ decay vertex
[note that $\Gamma_{N^\ast,MB}(k)= \Gamma_{MB,N^\ast}^\dag(k)$];
the inverse of the dressed $N^\ast$ propagator is defined by
\begin{equation}
[D^{-1}(E)]_{n,m} = (E - M^0_{N^*_n})\delta_{n,m} - [\Sigma_{N^\ast}(E)]_{n,m},
\label{eq:nstar-g}
\end{equation}
where $M_{N^*}^0$ is the mass of the bare $N^*$ and
the $N^\ast$ self-energies $\Sigma_{N^\ast}(E)$ are given by 
\begin{eqnarray}
[\Sigma_{N^\ast}(E)]_{n,m}
&=&\sum_{MB}\int_{C_{MB}}k^2 dk\Gamma_{N^*_n,MB}(k) G_{MB}(k;E)
\bar{\Gamma}_{MB,N^*_m}(k;E).
\label{eq:nstar-sigma}
\end{eqnarray}
It is emphasized that the $N^\ast$ propagator $D(E)$ can have off-diagonal
terms.
The bare vertex functions in Eqs.~(\ref{eq:dress-1})-(\ref{eq:dress-2})
are parametrized as
\begin{eqnarray}
{\Gamma}_{MB(LS),N^\ast}(k)
&=& \frac{1}{(2\pi)^{3/2}}\frac{1}{\sqrt{m_N}}C_{MB(LS),N^\ast}
\left(\frac{\Lambda_{MB(LS),N^\ast}^2}{\Lambda_{MB(LS),N^\ast}^2 + k^2}\right)^{(2+L/2)}
\left(\frac{k}{m_\pi}\right)^{L} ,
\label{eq:gmb}
\end{eqnarray}
where $L$ and $S$ denote the orbital angular momentum and spin of the $MB$ state, respectively.
This vertex function behaves as $k^L$ at $k\sim 0$ and $k^{-4}$ for $k\to\infty$.
The coupling constants $C_{MB(LS),N^*}$, the cutoffs $\Lambda_{MB(LS),N^*}$
and the bare masses $M^0_{N^*}$ are the parameters of the model.

\begin{figure}[t]
\centering
\includegraphics[width=12cm,angle=-0]{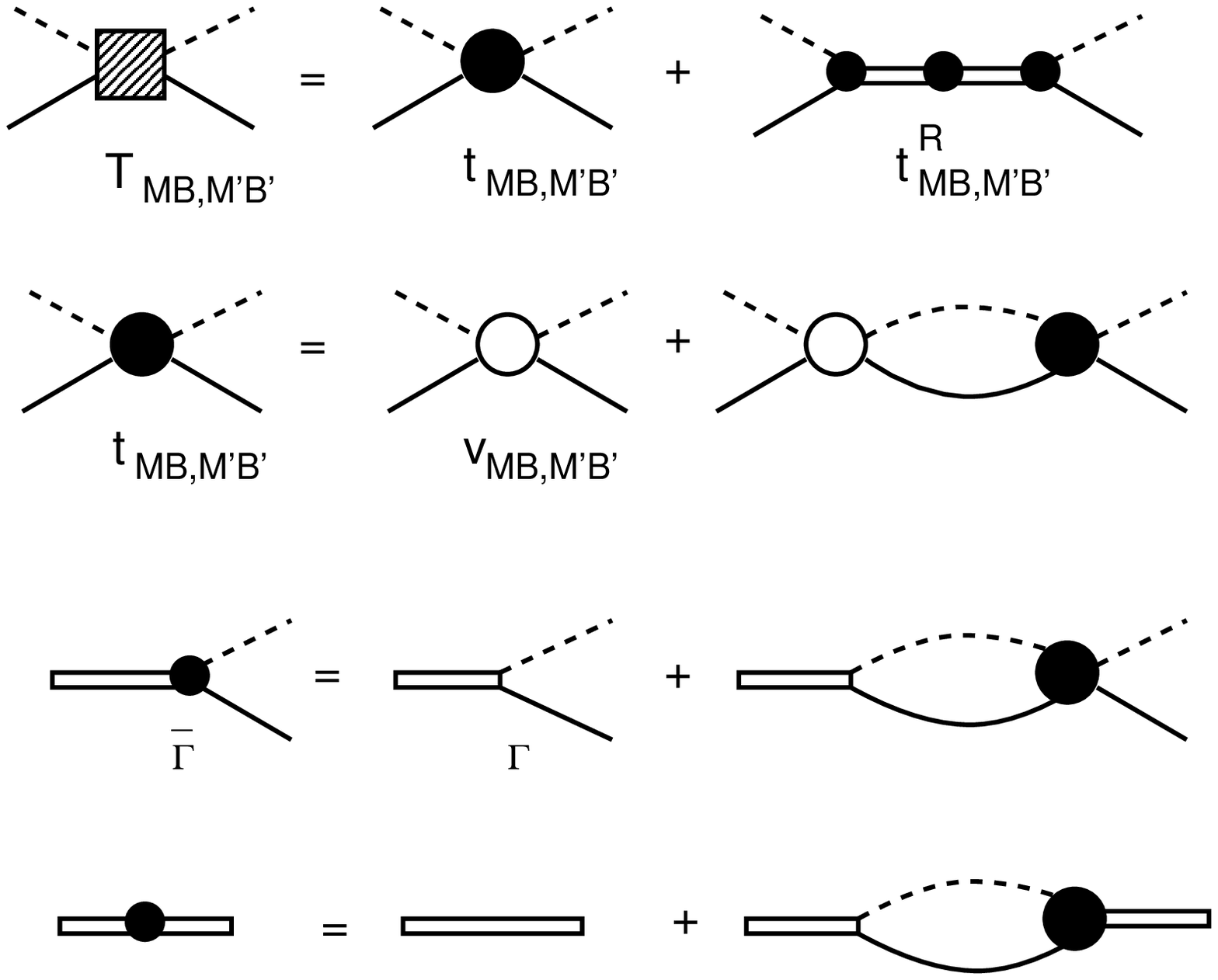}
\caption{Graphical representation of Eqs.~(\ref{eq:tmbmb})-(\ref{eq:nstar-sigma}).}
\label{fig:tmatmbmb}
\end{figure}

Equations~(\ref{eq:tmbmb})-(\ref{eq:nstar-sigma}) define the DCC 
model used in the ANL-Osaka analysis. They are illustrated in Fig.~\ref{fig:tmatmbmb}.
In the absence of theoretical inputs, the DCC model, as well as all hadron reaction models,
has parameters that can only be determined phenomenologically
from fitting the data. 
The meson-exchange interactions $v_{M'B',MB}$ depend on the coupling constants 
and the cutoffs of form factors that regularize their matrix elements. 
While the values of some of the coupling constants
can be estimated from SU(3) and the previous analysis,
we allow most of them to vary in the fits.

\subsection{Electromagnetic amplitudes}
\label{sec:dcc-ele}

\begin{figure}[t]
\includegraphics[clip,width=0.67\textwidth]{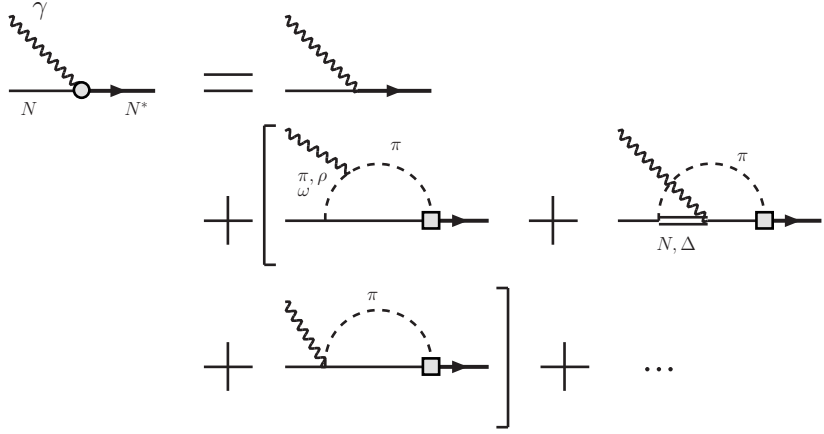}
\caption{\label{fig:gn-nstar}
Dressed $\gamma N \rightarrow N^*$ vertex defined by
Eq.~(\ref{eq:nstar-mb}).
}
\end{figure}

With the hadronic amplitudes $t_{M'B',MB}(k',k;E)$ defined in Eq.~(\ref{eq:cc-eq}),
the partial wave amplitudes for the $\gamma(\vec q) + N(-\vec q) \to M' (\vec k') + B'(-\vec k)$ 
reactions are expressed as~\cite{msl07},
\begin{eqnarray}
T_{M'B',\gamma N} (k',q;E) &=& t_{M'B',\gamma N} (k',q;E) + t^R_{M'B',\gamma N} (k',q;E),
\end{eqnarray}
with
\begin{eqnarray}
t_{M'B',\gamma N} (k',q;E) &=& v_{M'B', \gamma N} (k',q)
\nonumber\\
&&
+\sum_{M''B''} \int p^2 dp t_{M'B',M''B''}(k',p;E)G_{M''B''}(p;E)v_{M''B'', \gamma N} (p,q),
\label{eq:tmbgn}
\\
t^R_{M' B',\gamma N}(k',q; E) &=&
\sum_{n,m} \bar\Gamma_{M'B',N^\ast_n}(k';E) [D(E)]_{n,m} \bar\Gamma_{N^\ast_m,\gamma N}(q;E),
\end{eqnarray}
where the dressed $N^*\rightarrow \gamma N$ vertex is 
\begin{eqnarray}
\bar\Gamma_{N^\ast,\gamma N}(q;E) &=& 
\Gamma_{N^\ast,\gamma N}(q) 
+ \sum_{M'B'}\int p^2 dp \Gamma_{M'B',N^\ast}^\dag(p) G_{M'B'}(p,E) t_{M'B',\gamma N}(p,q;E).
\label{eq:nstargn}
\end{eqnarray}
By using Eq.~(\ref{eq:tmbgn}), the above equation can be written in terms of
dressed $N^*\rightarrow MB$ vertex $\bar{\Gamma}_{N^\ast,M'B'}(p,E)$
defined by Eq.~(\ref{eq:dress-2}) and
\begin{eqnarray}
\bar\Gamma_{N^\ast,\gamma N}(q;E) &=&
\Gamma_{N^\ast,\gamma N}(q)
+ \sum_{M'B'}\int p^2 dp \bar{\Gamma}_{N^\ast,M'B'}(p,E) G_{M'B'}(p,E) v_{M'B',\gamma N}(p,q;E).
\label{eq:nstargn-1}
\end{eqnarray}
Here the transition interaction $v_{MB,\gamma N}$ has the tree-diagram mechanisms shown in Fig.~\ref{fig:mex}.
Thus the pion-loop contributions ($M'B'=\pi N$) to the dressed vertex $\bar\Gamma_{N^\ast,\gamma N}(q;E)$
defined by Eq.~(\ref{eq:nstargn-1}) can be
illustrated in Fig.~\ref{fig:gn-nstar}.
 The procedures for calculating
 $v_{MB,\gamma N}$ are detailed in Ref.~\cite{knls13}. 

 For the bare
 $\gamma N \to N^*$ vertex, we write
 in the helicity representation as 
\begin{eqnarray}
\Gamma_{N^\ast,\gamma N}(q) = \frac{1}{(2\pi)^{3/2}}
\sqrt{\frac{m_N}{E_N(q)}}\sqrt{\frac{q_R}{q_0}} G^{N^\ast}_\lambda(Q^2,q_0)\delta_{\lambda,(\lambda_\gamma-\lambda_N)},
\label{eq:nstar-mb}
\end{eqnarray}
where $q_R$ and $q_0$ are defined by 
$M_{N^\ast}=q_R+E_N(q_R)$ and $W=q_0+E_N(q_0)$, respectively, and
\begin{eqnarray}
G^{N^\ast}_\lambda (Q^2,q_0) 
&=& A^{N^\ast}_\lambda (Q^2,q_0),\qquad \text{for transverse photons},
\\
&=& S^{N^\ast}_\lambda (Q^2,q_0),\qquad \text{for longitudinal photons}.
\end{eqnarray}
The helicity amplitudes $A^{N^*}_\lambda$ and $S^{N^*}_\lambda$ in the above equations
are related to
the multipole amplitudes $E^{N^\ast}_{l\pm},M^{N^\ast}_{l\pm}, S^{N^\ast}_{l\pm}$
of $\gamma N \rightarrow N^*$ processes. For the $N^*$ of spin $J=l+1/2$ with $l$ being the orbital angular momentum
of the $\gamma N$ system, the helicity amplitudes are  
\begin{eqnarray}
A^{N^*}_{3/2}(Q^2,q_0)&=& \frac{\sqrt{l(l+2)}}{2}[-M^{N^\ast}_{l+}(Q^2,q_0) + E^{N^\ast}_{l+}(Q^2,q_0)],
\\
A^{N^*}_{1/2}(Q^2,q_0)&=&-\frac{1}{2}[lM^{N^\ast}_{l+}(Q^2,q_0) + (l+2) E^{N^\ast}_{l+}(Q^2,q_0)],
\\
S^{N^*}_{1/2}(Q^2,q_0)&=&S^{N^\ast}_{l+}(Q^2,q_0),
\end{eqnarray}
For $N^*$ with $J=l-1/2$, we have
\begin{eqnarray}
A^{N^*}_{3/2}(Q^2,q_0)&=&- \frac{\sqrt{(l-1)(l+1)}}{2}[M^{N^\ast}_{l-}(Q^2,q_0) + E^{N^\ast}_{l-}(Q^2,q_0)],
\\
A^{N^*}_{1/2}(Q^2,q_0)&=&\frac{1}{2}[(l+1)M^{N^\ast}_{l-}(Q^2,q_0) - (l-1) E^{N^\ast}_{l-}(Q^2,q_0)],
\\
S^{N^*}_{1/2}(Q^2,q_0)&=&S^{N^\ast}_{l-}(Q^2,q_0).
\end{eqnarray}
The multipole amplitudes are parametrized as
\begin{eqnarray}
M^{N^\ast}_{l\pm}(Q^2,q_0)&=& \left(\frac{q_0}{m_\pi}\right)^l
\left(\frac{\Lambda_{N^\ast,\gamma N}^2 + m_\pi^2}{\Lambda_{N^\ast,\gamma N}^2 + q_0^2}\right)^{(2+l/2)} 
\tilde M^{N^\ast}_{l\pm}(Q^2),
\\
E^{N^\ast}_{l\pm}(k)(Q^2,q_0)&=& \left(\frac{q_0}{m_\pi}\right)^{(l\pm 1)}
\left(\frac{\Lambda_{N^\ast,\gamma N}^2 + m_\pi^2}{\Lambda_{N^\ast,\gamma N}^2 + q_0^2}\right)^{[2+(l\pm 1)/2]} 
\tilde E^{N^\ast}_{l\pm}(Q^2),
\\
S^{N^\ast}_{l\pm}(k)(Q^2,q_0)&=& \left(\frac{q_0}{m_\pi}\right)^{(l\pm 1)}
\left(\frac{\Lambda_{N^\ast,\gamma N}^2 + m_\pi^2}{\Lambda_{N^\ast,\gamma N}^2 + q_0^2}\right)^{[2+(l\pm 1)/2]} 
\tilde S^{N^\ast}_{l\pm}(Q^2),
\end{eqnarray}
where the cutoff $\Lambda_{N^\ast,\gamma N}$ and the coupling constants
$\tilde M_{l\pm}(Q^2)$, $\tilde E_{l\pm}(Q^2)$, $\tilde S_{l\pm}(Q^2)$ are determined in fitting the data.
One significant difference between the above parametrization  and the form used in our previous analysis~\cite{jlmss08}
is that the multipole amplitudes, 
or equivalently the helicity amplitudes, for
the $\gamma N\to N^\ast$ processes now have the dependence on the $\gamma N$ relative momentum $q_0$.

\section{Extractions of Nucleon Resonances}

We follow the earlier works to define that
a nucleon resonance with a complex mass $M_R$ is
an ``eigenstate`` of a Hamiltonian: $H\ket{\psi^R_{N^*}} = M_R \ket{\psi^R_{N^*}}$.
Then from the spectral expansion of the Low Equation for reaction amplitude
$T(E)= H' + H'(E-H)^{-1}H'$, where we have defined $H'=H-H_0$ with
$H_0$ being the non-interacting free Hamiltonian, we have
\begin{eqnarray}
T_{MB, M'B'}(k^0_{MB}, k^0_{M'B'} ; E\to M_R)
&=&\frac{\bra{k^0_{MB}}H'\ket{\psi^R_{N^*}}\bra{\psi^R_{N^*}}H'\ket{k^0_{M'B'}}}{E-M_R} + \cdot\cdot,
\label{eq:em-exp-2}
\end{eqnarray}
 where $k^0_{MB}$ and $k^0_{M'B'}$
are the on-shell momenta defined by
\begin{eqnarray}
E&=&E_{M}(k^0_{MB})+E_{B}(k^0_{MB}) \nonumber \\
&=&E_{M'}(k^0_{M'B'})+
E_{B'}(k^0_{M'B'})\,. \label{eq:on-k}
\end{eqnarray}
Therefore
the resonance positions can also be defined as the poles $M_R$ of
the meson-baryon amplitude $T_{MB,M'B'}(k^0_{MB}, k^0_{M'B'}; E)$
on the complex Riemann $E$-surface.
Because of the quadratic relation between the energy and momentum
variables, each $MB$ channel for a given  $E$ can have a physical ($p$)
sheet
characterized by $Im(k^0_{MB}) >0$ and an unphysical ($u$) sheet by
 $Im(k^0_{MB}) < 0$.
Like all previous works, we only look for poles
close to the physical region 
and have large effects on $\pi N$
scattering  observables. All of these poles are on the unphysical sheet
of the $\pi N$ channel, but could be on either  $(u)$ or
 $(p)$ sheets of other channels.
To find the resonance poles, we
analytically continue Eqs.~(\ref{eq:tmbmb})-(\ref{eq:nstar-sigma})
and Eqs.~(\ref{eq:tmbgn})-(\ref{eq:nstargn})
 to complex $E$-plane by using the method
detailed in Refs.~\cite{ssl09,ssl10}. The main step is to choose
appropriate integration paths $C_{MB}$ of Eq.~(\ref{eq:cc-eq})  in solving
Eqs.~(\ref{eq:tmbmb})-(\ref{eq:nstar-sigma}).

Explicitly, 
as energy approaches a resonance position $M_R$ in the complex $E$-plane, the 
total meson-baryon amplitudes can be written as
\begin{eqnarray}
T_{MB,M'B'}(k^R_{MB},k^R_{M'B'}; E\rightarrow M_R) \rightarrow
 \frac{\tilde{R}_{MB,M'B'}(M_R)}{E-M_R}\,,
\label{eq:res-pole}
\end{eqnarray}
with
\begin{eqnarray}
\tilde{R}_{MB,M'B'}(M_R)=\bar{\Gamma}^R_{M'B'}(k^R_{M'B'},M_R)\bar{\Gamma}^R_{MB}(k^R_{MB},M_R)
\label{eq:res-residue}
\end{eqnarray}
where the on-shell momenta $k^R_{MB},k^R_{M'B'}$ are defined by Eqs.~(\ref{eq:on-k})
with $E=M_R$. In the actual calculation, the residues $\tilde{R}_{MB,M'B'}(M_R)$ are extracted numerically
by using the well known method based on the Cauchy theorem. It is instructive to mention here that
for the partial-wave with only one bare $N^*$ state and the resonance can be extracted
from $t^R_{M'B',MB}(k',k;E)$ of the full amplitude Eq.~(\ref{eq:tmbmb}),
each factor of  the residue defined by Eq.~(\ref{eq:res-residue}) can be related to the dressed
vertex functions Eqs.~(\ref{eq:dress-1})-(\ref{eq:dress-2}) and the self energy Eq.~(\ref{eq:nstar-sigma}): 
\begin{eqnarray}
\bar{\Gamma}^R_{M'B'}(k^R_{M'B'},M_R)&=&\frac{\bar{\Gamma}_{M'B',N^*}(k^R_{M'B'},M_R)}{\sqrt{1-\Sigma'(M_R)}}\\
\bar{\Gamma}^R_{MB}(k^R_{MB},M_R)&=& \frac{\bar{\Gamma}_{N^*,MB}(k^R_{MB},M_R)}{\sqrt{1-\Sigma'(M_R)}}
\label{eq:res-r}
\end{eqnarray}
where $\Sigma'(M_R)=[d\Sigma(E)/dE]_{E=M_R}$.

For the elastic $\pi N \rightarrow \pi N$ case,
it is customary to define
\begin{eqnarray}
R_{\pi N,\pi N}(M_R) & = & \rho_{\pi N}(k^0_{\pi N})\tilde{R}_{\pi N,\pi N}(M_R) \nonumber \\
&=& \rho_{\pi N}(k^0_{\pi N}) \bar{\Gamma}^R_{\pi N}(k^R_{\pi N}, M_R)
\bar{\Gamma}^R_{\pi N}(k^R_{\pi N}, M_R) \,.
\label{eq:pin-resid}
\end{eqnarray}
where $k^0_{\pi N}$ is defined by $M_R=E_\pi(k^0_{\pi N})+E_N(k^0_{\pi N})$ and 
\begin{eqnarray}
\rho_{\pi N}(k^0_{\pi N}) = \pi \frac{k^0_{\pi N}E_\pi(k^0_{\pi N})E_N(k^0_{\pi N})}{E}.
\end{eqnarray}

The helicity amplitudes of $\gamma N \rightarrow N^*$ at the resonance pole $M_R$
are defined as
\begin{eqnarray}
A_{3/2} & = & C\times \bar{\Gamma}^R_{\gamma N}(q_0,M_R,\lambda_\gamma=1,\lambda_N=-1/2),
\label{eq:a32}\\
A_{1/2} & = & C \times
\bar{\Gamma}^R_{\gamma N}(q_0,M_R,\lambda_\gamma=-1,\lambda_N=-1/2) \, \label{eq:a12},
\end{eqnarray}
where  $\lambda_N$ and $\lambda_\gamma$ are the helicities of the initial
nucleon and photon, respectively, and
\begin{eqnarray}
C=\sqrt{\frac{E_N(\vec{q})}{m_N }} \frac{1}{\sqrt{2K}} 
\sqrt{\frac{(2J^R+1)(2\pi)^3(2q_0)}{4\pi}},
\label{eq:coef-c}
\end{eqnarray}
where $J^R$ is the spin of the resonance state,
$q_0 = |\vec{q}|$ and $K  =  (M_R^2 - m_N^2)/(2M_R)$.
In practice, we can use the extracted
residues $\tilde{R}_{\pi N,\gamma N(\lambda)}$ and $\tilde{R}_{\pi N,\pi N}$ to calculate the
determined helicity amplitudes by 
\begin{eqnarray}
A_{3/2} & = & N \times R_{\pi N \gamma N(3/2)},
\\
A_{1/2} & = & N \times R_{\pi N \gamma N(1/2)},
\end{eqnarray}
\begin{equation}
N = a \times \sqrt{\frac{k^{\rm on}_{\pi N}}{K}
\frac{2\pi(2J^R+1)M_R}{ m_N R_{{\pi N, \pi N}}}} ,
\end{equation}
with $a=\sqrt{2/3}$ for the resonance with isospin $I$=3/2,  and $a=-\sqrt{3}$ for $I$=1/2 resonance;
the phase is fixed so that $-\pi/2 \le \arg(N/a) \le \pi/2 $.

\section{Calculation of cross sections}

\subsection{ Meson-baryon scattering}
We follow the convention of Goldberger and Watson to define the meson-baryon scattering amplitudes.
The normalization of states are : $<\vec{k}|\vec{k}^{\,\,'}>=\delta(\vec{k}-\vec{k}^{\,\,'})$ for 
plane wave states, and $<\phi_\alpha|\phi_\beta>=\delta_{\alpha,\beta}$ for bound states.
The $T$-matrix elements are related to $S$-matrix elements by
\begin{eqnarray}
S_{MB,M'B'}=\delta_{MB,M'B'}-2\pi \,i\, \delta(E_{MB}-E_{M'B'})\,T_{MB,M'B'}
\end{eqnarray}
Note that the $"-"$ sign in the right side of the above equation  is opposite to the $"+"$ sign used by
the other partial-wave analysis groups such as SAID. 
 
The formula for calculating the meson-baryon scattering cross sections given in this section are
 in the center of mass system. 
For the process $M(\vec{k})+B(-\vec{k}) \rightarrow M'(\vec{k}^{\,\,'})+B'(-\vec{k}^{\,\,'}) $
the differential cross section can be written as
\begin{eqnarray}
\frac{d\sigma_{MB\rightarrow M'B'}}{d\Omega_{k'}}&=&\frac{(4\pi)^2}{k^2}\rho_{M'B'}(k')\rho_{MB}(k)\frac{1}{(2j_M+1)(2j_B+1)}
 \sum_{m_{j_M}m_{j_B}}\sum_{m'_{j_M}m'_{j_B}}|<M'B'|t(W)|MB>|^2\nonumber \ ,\\
&&
\label{eq:dsc}
\end{eqnarray}
where the incoming and outgoing momentum $k$ and $k'$ are defined by the invariant mass $W$
\begin{eqnarray}
W=E_M(k)+E_B(k)=E_{M'}(k')+E_{B'}(k')\ ,
\end{eqnarray}
where $E_\alpha(k)=\sqrt{m^2_\alpha+\vec{k}^{\,\,2}}$ with $m_\alpha$ being the mass of particle $\alpha$,
and the phase-space factor is
\begin{eqnarray}
\rho_{MB}(k)=\pi\frac{kE_M(k)E_B(k)}{W} 
\label{eq:mb-rho}\ .
\end{eqnarray}
The scattering amplitude $<M'B'|t(W)|MB>$ in Eq.~(\ref{eq:dsc}) can be calculated from the partial-wave amplitudes
$t^{JT}_{L'S'M'B',LSMB}(k',k,W)$ as
\begin{eqnarray}
&&<M'B'|t(W)|MB>\nonumber \\
&&=\sum_{JM_J,T}\,\,\sum_{L'M'_L,S'M'_S}\,\,\sum_{LM_L,SM_S} t^{JT}_{L'S'M'B',`LSMB}(k',k,W)\nonumber \\
&&\times
[<TM_T|i'_M\tau'_Bm'_{i_M}m'_{\tau_B}><JM_J|L'S'm'_Lm'_S><S'm'_S|j'_Mj'_Bm'_{j_M}m'_{j_B}>Y^*_{L'm'_L}(\hat{k}^{\,\,'})]
\nonumber\\
&&\times
[<TM_T|i_M\tau_Bm_{i_M}m_{\tau_B}><JM_J|LSm_Lm_S><Sm_S|j_Mj_Bm_{j_M}m_{j_B}>Y_{Lm_L}(\hat{k})]\ ,
\end{eqnarray}
where $<jm_j|j_1j_2m_{j_1}m_{j_2}>$ is the Clebsch-Gordon coefficient for the $\vec{j}_1+\vec{j}_2=\vec{j}$ coupling,
$[(j_Mm_{j_M}),(i_Mm_{i_M})]$ and $[(j_Bm_{j_B}),(\tau_Bm_{\tau_B})]$ are the spin-isospin quantum numbers of
mesons and baryons, respectively; $(J M_J)$($(TM_T)$) are the total angular momentum (total isospin), 
$(LM_L)$ (($SM_S))$ are the
relative orbital angular momentum (total spin) of the considered two-body systems.

By choosing the incoming momentum $\vec{k}$ in the quantization z-component, 
the total $MB\rightarrow M'B'$ cross sections are
\begin{eqnarray}
\sigma^{tot}_{MB\rightarrow M'B'}(W)=\int
 d\Omega_{k'}\frac{d\sigma_{MB\rightarrow M'B'}}{d\Omega_{k'}}\ .
\end{eqnarray}
By optical theorem and the above partial-wave expansion, one can get the $\pi N\rightarrow X$
 total cross sections averaged over the initial spins:
\begin{eqnarray}
\sigma^{tot}_{\pi N\rightarrow X}(W) =\frac{-4\pi}{(2s_N+1)k^2}\sum_{J,T,L} (2J+1)
\,\rho_{\pi N}(k)\,Im[t^{JT}_{L\frac{1}{2}\pi N,L\frac{1}{2}\pi N}(k,k,W)]
\times [<TM_T|1\frac{1}{2}m_{i_\pi}m_{\tau_N}>]^2 \nonumber \ ,\\
\end{eqnarray}
where $M_T=m_{i_\pi}+m_{\tau_N}$ and $s_N=1/2$ is the nucleon spin .

The ANL-Osaka partial-wave amplitudes $t^{JT}_{L'S'M'B',LSMB}(k',k,W)$ can be obtained from 
the following quantities presented on the  web page:
\begin{eqnarray}
<M'B'|T(W)|MB>=-\rho^{1/2}_{M'B'}(k')t^{JT}_{L'S'M'B',LSMB}(k',k,W)\rho^{1/2}_{MB}(k)\ ,
\end{eqnarray}
  for
$MB,M'B'=\pi N, \eta N, K\Lambda, K\Sigma$ and $W = W_{th} - 2000$ MeV, where 
 $W_{th}$ is the lower one of the two threshold energies $m_M+m_B$ and $m_{M'}+m_{B'}$. 
The phase space factors $\rho_{MB}(k)$ and $\rho_{M'B'}(k')$ are defined by Eq.~(\ref{eq:mb-rho}).

\subsection{Electro-production of pions}

For the pion electroproduction,
( $e(p_e) + N (p_N) \rightarrow e'(p_e') + \pi (k) + N (p_N')$),
the differential cross section is conventionally written as
\begin{eqnarray}
\frac{d\sigma}{dE_e' d\Omega_{e'}d\Omega_{\pi}}
 & = &  \Gamma \frac{d\sigma^v}{d\Omega_{\pi}}\ ,
\label{eq:elec-1}
\end{eqnarray}
where $Q^2  =  - q^2$, $q=p_e-p'_e =(\omega_L, {\bm q}_L)$, $W  =  \sqrt{(p_N + q)^2}$, and
\begin{eqnarray}
\Gamma & = & \frac{\alpha q^\gamma_L}{2\pi^2 Q^2} \frac{E_e'}{E_e}
         \frac{1}{1-\epsilon} .
\end{eqnarray}
Here, we have defined $\alpha=e^2/4\pi=1/137$ and the effective photon
energy in the laboratory system and $\epsilon$ are given as
\begin{eqnarray}
q^\gamma_L & = & \frac{W^2 - m_N^2}{2m_N} \ , \label{eq:ql}\\   
\epsilon & = & [1 + \frac{2 {\bm q}^2_L}{Q^2}\tan^2 \frac{\theta_e}{2}]^{-1}\label{eq:epsi} ,
\end{eqnarray}
where $\theta_e$ is the angle between the outgoing and incoming electrons, and $m_N$ is the nucleon mass and $\bm{q}_L$ is momentum transfer in the laboratory system.

The differential cross section $d\sigma^v/d\Omega_{\pi}$ in Eq.~(\ref{eq:elec-1})
is defined in final $\pi N$ center of mass frame with the following coordinate system:
\begin{eqnarray}
\hat{z}&=&\hat{q}=\frac{\bm{q}}{|{\bm{q}}|} \label{eq:cor-z} \\
\hat{y}&=&\frac{{\bm q}\times {\bm k}}{|{\bm q}\times {\bm k} |} \label{eq:cor-y}\\
\hat{x}&=&\hat{y}\times\hat{z} \label{eq:cor-x}
\end{eqnarray}
We then have the following expression:
\begin{eqnarray}
\frac{d\sigma^v}{d\Omega_{\pi}} & = &
 \frac{d\sigma_T}{d\Omega_{\pi}} + \epsilon \frac{d\sigma_L}{d\Omega_{\pi}}
 + \epsilon \frac{d\sigma_{TT}}{d\Omega_{\pi}} \cos 2 \phi_{\pi}
 + \sqrt{2\epsilon(\epsilon+1)}\frac{d\sigma_{LT}}{d\Omega_{\pi}}
 \cos \phi_{\pi} \nonumber \\
 & &  + h_e \sqrt{2\epsilon(1- \epsilon)}
 \frac{d\sigma_{LT'}}{d\Omega_{\pi}}\sin \phi_{\pi}\,, \label{eq:dsdome}
\end{eqnarray}
where $h_e$ is the helicity of the incoming electron,
$\phi_\pi$ is the pion angle measured from the $e-e'$ scattering plane of electron, and 
\begin{eqnarray}
\frac{d\sigma_T}{d\Omega_{\pi}} & = & \frac{|{\bm k}|}{|{{\bm q}_\gamma}|}\sum_{spin}
             \frac{F^{xx}+F^{yy}}{2}
\label{eq:dsdm-t} \ , \\
\frac{d\sigma_L}{d\Omega_{\pi}} & = & \frac{|{\bm k}|}{|{{\bm q}_\gamma}|}\sum_{spin}
             \frac{Q^2}{{|{\bm q}|^2}} F^{00}\ , \\
\frac{d\sigma_{TT}}{d\Omega_{\pi}} & = & \frac{|{\bm k}|}{|{{\bm q}_\gamma}|}\sum_{spin}
             \frac{F^{xx}-F^{yy}}{2} \ , \\
\frac{d\sigma_{LT}}{d\Omega_{\pi}} & = & \frac{|{\bm k}|}{|{{\bm q}_\gamma}|}\sum_{spin}
             (-1)\sqrt{\frac{Q^2}{{|{\bm q}|^2}}}Re (F^{x0})\ , \\
\frac{d\sigma_{LT'}}{d\Omega_{\pi}} & = & \frac{|{\bm k}|}{|{{\bm q}_\gamma}|}\sum_{spin}
             \sqrt{\frac{Q^2}{{|{\bm q}|^2}}}Im(F^{x0})\ .
\label{eq:dsdm-e}
\end{eqnarray}
Here ${\bm q}$  is the  momentum transfer to the initial nucleon 
and ${\bm k}$  is the
pion momentum in the center of mass system of the final $\pi N$ state:
\begin{eqnarray}
  \omega & = & \frac{W^2 - M_N^2 -Q^2}{2W} \ , \\
 |{\bm q}| & = & \sqrt{Q^2 + \omega^2} \ , \\
  |{\bm k}| & = & \sqrt{
  (\frac{W^2 + m_{\pi}^2 - M_N^2}{2W})^2 - m_{\pi}^2}\ ,
\end{eqnarray}
and
\begin{eqnarray}
  |{\bm q}_\gamma| & = & \frac{W^2 - M_N^2}{2W} \label{eq:qgam} \ .
\end{eqnarray}
Here $\omega$ and $\bm{q}_\gamma$ are the energy transfer and the effective
photon energy in the center of mass system.
Integrating pion angles, Eqs.~(\ref{eq:elec-1}) and (\ref{eq:dsdome}) lead to
\begin{eqnarray}
\frac{d\sigma}{dE_e' d\Omega_{e'}}(Q^2,W)
 & = &  \Gamma [\sigma_T(Q^2,W)+\epsilon\sigma_L(Q^2,W)] .
\label{eq:sigma-tot}
\end{eqnarray}
where
\begin{eqnarray}
  \sigma_{T/L} & = & \sum_{\pi^+,\pi^0}\int d\Omega_{\pi} \frac{d\sigma_{T/L}}{d\Omega_\pi}
\end{eqnarray}

In the coordinate system defined by Eqs.~(\ref{eq:cor-z})-(\ref{eq:cor-x}),
 the pion momentum  ${\bm k}$ is  on the $x-z$ plane.  We thus  can define
\begin{eqnarray}
\hat{k} &=& {\bm k}/|{\bm k}| = \cos\theta \hat{z} + \sin\theta \hat{x}\ ,
\label{eq:kvec}
\end{eqnarray}
where $\theta$ is the angle between the outgoing pion and the virtual photon.
The quantities  $F^{ij}$ with $i,j=x,y,0$ in Eqs.~(\ref{eq:dsdm-t})-(\ref{eq:dsdm-e}) 
are defined as 
\begin{eqnarray}
  \sum_{spin}  F^{ij} & = & \frac{1}{2}\sum_{m_{s_{i}},m_{s_{f}}}
       <m_{s_f}|{\cal F}^i|m_{s_i}><m_{s_f}|{\cal F}^j|m_{s_i}>^* \ ,
\label{eq:fij}
\end{eqnarray}
where $m_s$ is the $z$-component of the nucleon spin, and ${\cal F}^i$ is defined by the 
Chew-Goldberger-Low-Nambu
(CGLN) amplitude ${\cal F}_{CGLN}= {\cal F}^\mu\epsilon_\mu$. 

The CGLN amplitude can be expressed in terms of Pauli operator ${\bm \sigma}$, $\hat{q}$, $\hat{k}$ and
the  photon polarization vector 
$\epsilon^\mu =(\epsilon_0, {\bm \epsilon})$
\begin{eqnarray}
{\cal F}^\mu\epsilon_\mu =  -( i{\bm \sigma}\cdot {\bm \epsilon}_\perp F_1
 + {\bm \sigma}\cdot \hat{k}
   {\bm \sigma}\cdot\hat{q}\times{\bm \epsilon}_\perp F_2
 + i{\bm \sigma}\cdot\hat{q}\hat{k}\cdot{\bm \epsilon}_\perp F_3
 + i{\bm \sigma}\cdot\hat{k}\hat{k}\cdot{\bm\epsilon}_\perp F_4
 \nonumber \\
 + i{\bm\sigma}\cdot\hat{q}\hat{q}\cdot{\bm\epsilon} F_5
 + i{\bm\sigma}\cdot\hat{k}\hat{q}\cdot{\bm \epsilon} F_6)
 + i {\bm \sigma}\cdot\hat{k} \epsilon_0 F_7
 + i {\bm \sigma}\cdot\hat{q}\epsilon_0 F_8
 \,,
  \label{fvec}
\end{eqnarray}
where ${\bm \epsilon}_\perp = \hat{q}\times({\bm \epsilon} \times \hat{q})$. 
By using Eq.~(\ref{eq:kvec}) and 
choosing $\epsilon^\mu =(\epsilon_0, {\bm \epsilon})=(0,1,0,0), (0,0,1,0), (1,0,0,0)$ to evaluate
Eq.~(\ref{fvec}), we then have
\begin{eqnarray}
  {\cal F}^x & = & i\sigma_x (F_1 - \cos\theta F_2 + \sin^2\theta F_4
    + i\sigma_z \sin\theta(F_2+F_3 + \cos\theta F_4)  \ ,\\ 
  {\cal F}^y  & = &  i\sigma_y(F_1 - \cos\theta F_2) - \sin\theta F_2
      \ , \\
 {\cal F}^0 & = & i\sigma_z(\cos\theta F_7 + F_8)
+ i\sigma_y \sin\theta F_7     
 \ .
\end{eqnarray}
The  amplitudes $F_i$ are calculated from 
the  multipole amplitudes $E_{l\pm},M_{l\pm},
S_{l\pm}$ and $L_{l\pm}$ of the $\gamma^*+N \rightarrow \pi N$ process :
\begin{eqnarray}
F_1  & =  & \sum_l[
  P_{l+1}'(x) E_{l+}(Q^2,W)  + P_{l-1}'(x)     E_{l-}(Q^2,W) 
+ lP_{l+1}'(x)M_{l+}(Q^2,W)\nonumber \\
&& + (l+1)P_{l-1}'(x) M_{l-}(Q^2,W)]\,,\label{eq:f1} \\
F_2  & = &  \sum_l[
                    (l+1)P_l'(x) M_{l+}(Q^2,W) + lP_l'(x)  M_{l-}(Q^2,W)]\,,\label{eq:f2} \\
F_3 &  =  & \sum_l[
  P_{l+1}''(x)  E_{l+}(Q^2,W)  + P_{l-1}''(x)    E_{l-}(Q^2,W) 
- P_{l+1}''(x)  M_{l+}(Q^2,W) \nonumber \\
&& + P_{l-1}''(x)    M_{l-}(Q^2,W)]\,,\label{eq:f3}\\
F_4 &  =&   \sum_l[
- P_{l}''(x)  E_{l+}(Q^2,W)  - P_{l}''(x)    E_{l-}(Q^2,W) 
+ P_{l}''(x)  M_{l+}(Q^2,W)  - P_{l}''(x)    M_{l-}(Q^2,W)]\,,\label{eq:f4} \\
F_5 &  =  & \sum_l[
  (l+1) P_{l+1}'(x)  L_{l+}(Q^2,W)  -  l  P_{l-1}'(x)  L_{l-}(Q^2,W)]\,,\label{eq:f5} \\
F_6  & = &  \sum_l[
 -(l+1) P_{l}'(x)  L_{l+}(Q^2,W)  +  l  P_{l}'(x)  L_{l-}(Q^2,W)]\,,\label{eq:f6} \\
F_7  & = &  \sum_l[
 -(l+1) P_{l}'(x)  S_{l+}(Q^2,W)  +  l  P_{l}'(x)  S_{l-}(Q^2,W)]\,, \label{eq:f7} \\
F_8  & = &  \sum_l[
  (l+1) P_{l+1}'(x)  S_{l+}(Q^2,W)  -  l  P_{l-1}'(x)  S_{l-}(Q^2,W)]\,,\label{eq:f8}
\end{eqnarray}
where $x=\hat{q}\cdot\hat{k}$,
 $P_L(x)$ is the Legendre function, $P'_L(x)=dP_L(x)/dx$ and $P''_L(x)=d^2P_L(x)/d^2x$.
For the photo-production process
 $\gamma N \rightarrow \pi N$, 
the differential cross section is $d\sigma_T/d\Omega_\pi$ with $Q^2=0$.

The ANL-Osaka multipole amplitudes $E_{l\pm}(Q^2,W)$, $M_{l\pm}(Q^2,W)$, and $L_{l\pm}(Q^2,W)$ for $W=1080-2000$ MeV and
$Q^2= 0-3$ (GeV/$c$)$^2$ are presented on the web page.

`
\subsection{Photo-production of mesons}

In the center of mass system, the formula for calculating  the un-polarized differential cross section
of the photo-production of pseudo-scalar mesons
 ( $M= \pi, \eta$, $K$) on the nucleon $(N)$,
$\gamma (\bm{q})+ N(-\bm{q})\rightarrow M(\bm{k})+ N(-\bm{k})$, is
\begin{eqnarray}
\frac{d\sigma_0}{d\Omega} = \frac{1}{4}\sum_{m_{s_N}=\pm 1/2}\,\,\sum_{m'_{s_N}=\pm 1/2}\,\,
\sum_{\lambda=\pm 1}
\frac{k}{q}|<m'_{s_N}|{\cal F}_{CGLN}|m_{s_N}>|^2
\end{eqnarray}
where the Chew-Goldberger-Low-Nambu (CGLN) amplitude is defined as
\begin{eqnarray}
{\cal F}_{CGLN}=\sum_{i=1,4}O_iF_i(\theta,W)\,.
\end{eqnarray}
Here we have defined $W=q+E_N(q)=E_M(k)+E_N(k)$, where $E_{\alpha}(p)=\sqrt{m_\alpha^2+p^2}$ and $m_\alpha$ is the
mass of particle $\alpha$, and $\theta$ is the angle between $\bm{k}$ and $\bm{q}$.
The operators in the above equation are:
\begin{eqnarray}
O_1&=&-i\bm{\sigma}\cdot\bm{\epsilon}_\lambda \nonumber \\
O_2&=&-[\bm{\sigma}\cdot\hat{k}][\bm{\sigma}\cdot\hat{q}\times\bm{\epsilon}_\lambda]
\nonumber \\
O_3&=&-i[\bm{\sigma}\cdot\hat{q}][\hat{k}\cdot\bm{\epsilon}_\lambda]
\nonumber \\
O_4&=&-i[\bm{\sigma}\cdot\hat{k}][\hat{k}\cdot\bm{\epsilon}_\lambda]
\nonumber
\end{eqnarray}
where $\hat{k}=\bm{k}/|\bm{k}|$ and $\hat{q}=\bm{q}/|\bm{q}|=\hat{z}$, and 
$\bm{\sigma}$ is the standard Pauli operator.
 The quantization direction is chosen
to be in the $\hat{z}$-direction,
and the photon polarization  vectors are $\bm{\epsilon}_{\pm}=
\frac{\mp}{\sqrt{2}}[\hat{x}\pm i \hat{y}]$.
The CGLN  amplitudes $F_i$  can be calculated from the multipole amplitudes with $Q^2=0$, $E_{l\pm}(W)$
and $M_{l\pm}(W)$, by using Eqs.~(\ref{eq:f1})-(\ref{eq:f4}).

The ANL-Osaka multipole amplitudes $E_{l\pm}(W)$ and $M_{l\pm}(W)$ at $Q^2=0$ for 
$\gamma N \rightarrow \pi N, \eta N, \Lambda N ,\Sigma N$ and $W=1080-2000$ MeV 
are presented on the web page.

The formula for using the multipole amplitudes to calculate the polarization observables can be found
in Ref.\cite{shkl11}.

\subsection{Inclusive $N(e,e')$ cross sections}

For the inclusive process,
 $e(p_e) + N (p_N) \rightarrow e'(p_e') + X$,
the differential cross section is  written by using structure functions
$W_i$ as
\begin{eqnarray}
\frac{d\sigma}{dE_e' d\Omega_{e'}}(Q^2,W)=\frac{\alpha^2}{4E_e^2\sin^4\frac{\theta_e}{2}}
[W_2(Q^2,W)\cos^2\frac{\theta_e}{2}+2W_1(Q^2,W)\sin^2\frac{\theta_e}{2}]
\end{eqnarray}
The structure functions are defined from the hadron tensor $W^{\mu\nu}$ as
\begin{eqnarray}
  W^{\mu\nu} & = &
  \sum_{\bar{i}}\sum_f (2\pi)^3 \delta^4(P + q - P')\frac{E_N(P)}{m_N}
 <f(P')|J_{em}^\mu|N(P)><f(P')|J_{em}^{\nu}|N(P)>^* \nonumber \\
 & = &
  (- g^{\mu\nu} + \frac{q^\mu q^\nu}{q^2})W_1(Q^2,W)
  + \frac{W_2(Q^2,W)}{m_N^2}\hat{P}^\mu \hat{P}^\nu,
\label{eq:hadron-tensor}
\end{eqnarray}
where $\hat{P}^\mu = P^\mu - q^\mu (P\cdot q)/q^2$.

The structure functions are also expressed by virtual photon
cross sections $\sigma^X_T$ and $\sigma^X_L$ as
\begin{eqnarray}
W_1(Q^2,W)&=&\frac{q^\gamma_L}{4\pi^2\alpha}\sigma^X_T(Q^2,W)\\
W_2(Q^2,W)&=&\frac{q^\gamma_L}{4\pi^2\alpha}\frac{Q^2}{\bm{q}_L^2}
[\sigma^X_T(Q^2,W)+\sigma^X_L(Q^2,W)].
\end{eqnarray}
The inclusive cross section can then be written as
\begin{eqnarray}
\frac{d\sigma}{dE_e' d\Omega_{e'}}(Q^2,W)
 & = &  \Gamma [\sigma^X_T(Q^2,W)+\epsilon\sigma^X_L(Q^2,W)] .
\label{eq:elec-2}
\end{eqnarray}
where $\epsilon$ is defined in Eq.~(\ref{eq:epsi}).
If only the single pion contribution to $W_1$ and $W_2$ is kept in evaluating Eq.~(\ref{eq:hadron-tensor}), 
$\sigma^X_{T/L}$ will be identical to $\sigma_{T/L}$ defined in Eq.~(\ref{eq:sigma-tot}).

The ANL-Osaka structure functions $W_1(Q^2,W)$ and $W_2(Q^2,W)$ for
$Q^2= 0-3$ (GeV/$c$)$^2$ and $W=1080-2000$ MeV
are presented on the web page.

\subsection{$\pi N \rightarrow \pi\pi N$ cross sections}

In the center of mass frame, the momentum variables of
the $\pi N \rightarrow \pi\pi N $ reaction with invariant mass  $W$
can be specified as
\begin{eqnarray}
a(\vec{p}_a)+ b(\vec{p}_b)\rightarrow 
c(\vec{p}_c) + d(\vec{p}_d) + e(\vec{p}_e)\ ,
\label{eq:process}
\end{eqnarray}
where
$\vec{p}_a =-\vec{p}_b=\vec{k}$ with $k$ defined by $W=E_a(k)+E_b(k)$,
 $\vec{p}_c + \vec{p}_d =-\vec{p}_e=\vec{k'}$, and
 $(c+d+e)$ can be any possible charged states formed from two pions and
one nucleon. The total cross section of the process Eq.~(\ref{eq:process}) can
 be  written as
\begin{eqnarray}
\sigma^{rec}_{ab \rightarrow cde}& = & \int_{m_c+m_d}^{W-m_e}
\frac{d\sigma^{rec}}{dM_{cd}}dM_{cd} \ ,
\label{eq:sigma-t}
\end{eqnarray}
with
\begin{eqnarray}
\frac{d\sigma^{rec}}{dM_{cd}}
=
\frac{\rho_i}{k^2}16\pi^3
\int d\Omega_{k_{cd}} d\Omega_{k'}
\frac{k_{cd}k'}{W}
\frac{1}{(2s_a +1)(2s_b+1)}
\sum_{{i},{f}}|\sqrt{E_cE_dE_e}
\langle \vec{p}_c\vec{p}_d\vec{p}_e,f|T|\vec{k},i\rangle |^2 \ ,
\nonumber\\
\label{eq:sigma-ta}
\end{eqnarray}
where $\rho_i =\pi\frac{kE_a(k)E_b(k)}{W}$,$i,f$ denote all spin $(s_a, s_{az})$ and isospin $(t_a, t_{az})$
quantum numbers, and $\sum_{{i},{f}}$ means summing over only
spin quantum numbers.
For a given invariant mass $M_{cd}$,  
$\vec{k}_{cd}$ is the relative momentum between $c$ and $d$ in the center of mass of
the sub-system $(cd)$ .
It follows that $k'$ and $k_{cd}$ are defined by $W$ and $M_{cd}$:
\begin{eqnarray}
M_{cd} &=& E_c(k_{cd}) + E_d(k_{cd}) \, , \nonumber\\
W      &=& E_e(k') + E_{cd}(k') \,, \nonumber \\
E_{cd}(k')&=& \sqrt{M_{cd}^2 + (k')^2}\, . 
\end{eqnarray}

The $T$-matrix elements in the Eq.~(\ref{eq:sigma-ta})  are 
of the following form
\begin{eqnarray}
\langle \vec{p}_c\vec{p}_d\vec{p}_e,f|T|\vec{k},i\rangle  & = & \sum_{s_{R_z},t_{R_z}}
\frac{
\langle \vec{p}_c,s_{cz},t_{cz};\vec{p}_d,s_{dz},t_{dz}|H_I|
\vec{k}',s_{Rz},t_{R_z} \rangle 
}{W - E_e(k') - E_R(k') - \Sigma_{eR}(k',E)} \nonumber \\
& &\times 
\langle \vec{k}',s_{Rz},t_{Rz};-\vec{k}',s_{ez},t_{ez}|
T|\vec{k},s_{az},t_{az};-\vec{k},s_{bz},t_{bz}\rangle \ ,
\label{eq:pin-pipin-1}
\end{eqnarray}
where $R$ is a bare state which has $R\rightarrow cd$ decay channel.
For the $eR=\pi\Delta$ and $eR=\rho N$ channels, the self-energies are explicitly given by
\begin{eqnarray}
\Sigma_{\pi \Delta}(k;W) &=& \frac{m_\Delta}{E_\Delta(k)}
\int q^2 dq \frac{ M_{\pi N}(q)}{[M^2_{\pi N}(q) + k^2]^{1/2}}
\frac{\left|f_{\Delta \to \pi N}(q)\right|^2}
{W-E_\pi(k) -[M^2_{\pi N}(q) + k^2]^{1/2} + i\epsilon},
\label{eq:self-pidelta-1}
\\
\Sigma_{\rho N}(k;W) &=& \frac{m_\rho}{E_\rho(k)}
\int q^2 dq \frac{ M_{\pi\pi}(q)}{[M^2_{\pi \pi}(q) + k^2]^{1/2}}
\frac{\left|f_{\rho \to \pi \pi}(q)\right|^2}
{W-E_N(k) -[M^2_{\pi \pi}(q) + k^2]^{1/2} + i\epsilon},
\label{eq:self-rhon-1}
\end{eqnarray}
where $m_\Delta=1280$ MeV, $m_\rho = 812$ MeV, $M_{\pi N}(q) = E_\pi (q) + E_N(q)$, 
and $M_{\pi \pi}(q) = E_\pi (q) + E_\pi(q)$.
The form factors $f_{\Delta\to\pi N}(q)$ and $f_{\rho \to \pi \pi}(q)$ are for describing
the $\Delta\to\pi N$ and $\rho \to \pi \pi$ decays in the $\Delta$ and $\rho$ rest frames,
respectively.
They are parametrized as:
\begin{eqnarray}
f_{\Delta \to \pi N}(q) &=& -i\frac{(0.98)}{[2(m_N+m_\pi)]^{1/2}}
\left(\frac{q}{m_\pi}\right)
\left(\frac{1}{1+[q/(358~\text{MeV})]^2}\right)^2 ,
\label{eq:fdpin}
\\
f_{\rho \to \pi \pi}(q) &=& \frac{(0.6684)}{\sqrt{m_\pi}}
\left(\frac{q}{(461~\text{MeV})}\right)
\left(\frac{1}{1+[q/(461~\text{MeV})]^2}\right)^2 .
\label{eq:frhopp}
\end{eqnarray}
The $\sigma$ self-energy $\Sigma_{\sigma N}(k;E)$  is calculated from a $\pi\pi$ s-wave scattering model with
a vertex function $g(q)$ for the $\sigma \rightarrow \pi\pi$ decay and a
separable interaction $v(q',q)=h_0 h(q')h(q)$.
The resulting form is
\begin{eqnarray}
\Sigma_{\sigma N}(k;W) &=& \langle gG_{\pi\pi}g\rangle(k;W) 
+\tau(k;E)[\langle gG_{\pi\pi}h\rangle(k;W)]^2,
\end{eqnarray}
with
\begin{eqnarray}
\tau(k;W) &=& \frac{h_0}{1-h_0\langle h G_{\pi\pi} h\rangle (k;W)},
\\
\langle hG_{\pi\pi}h\rangle(k;W) &=&
\int dq q^2 \frac{M_{\pi\pi}(q)}{ [M_{\pi\pi}^2(q) + k^2]^{1/2}} 
\nonumber\\
&& \qquad\qquad\qquad \times
\frac{h(q)^2}{W - E_N(k) - [M_{\pi\pi}^2(q) + k^2]^{1/2}+i\varepsilon},
\\
\langle gG_{\pi\pi}g\rangle(k;W) &=& 
\frac{m_\sigma}{E_\sigma(k)}
\int dq q^2 \frac{M_{\pi\pi}(q)}{ [M_{\pi\pi}^2(q) + k^2]^{1/2}} 
\nonumber\\
&& \qquad\qquad\qquad \times
\frac{g(q)^2}{W - E_N(k) - [M_{\pi\pi}^2(q) + k^2]^{1/2}+i\varepsilon},
\\
\langle gG_{\pi\pi}h\rangle(k;W) &=&
\sqrt{\frac{m_\sigma}{E_\sigma(k)}}
\int dq q^2 \frac{M_{\pi\pi}(q)}{ [M_{\pi\pi}^2(q) + k^2]^{1/2}} 
\nonumber\\
&& \qquad\qquad\qquad \times
\frac{g(q)h(q)}{W - E_N(k) - [M_{\pi\pi}^2(q) + k^2]^{1/2}+i\varepsilon}.
\end{eqnarray}
In the above equations, $m_\sigma = 700.0$ MeV and the form factors are 
\begin{eqnarray}
g(p)&=&\frac{g_0}{\sqrt{m_\pi}}\frac{1}{1+(cp)^2},
\label{eq:fsigmapp}
\\
h(p)&=&\frac{1}{m_\pi}\frac{1}{1+(dp)^2}.
\label{eq:pipi-pot2p}
\end{eqnarray}
where
$g_0 = 1.638$, $h_0 = 0.556$, $c=1.02$ fm, and $d=0.514$ fm.

For any spins and isospins and c.m. momenta $\vec{p}$ and $\vec{p}^{\,'}$, 
the $MB\rightarrow M'B'$ $T$-matrix elements in Eq.~(\ref{eq:pin-pipin-1}) are in general defined
by
\begin{eqnarray}
&&
\langle \vec{p}',s_{M'z},t_{M'z}; -\vec{p}',s_{B'z},t_{B'z}|
T|\vec{p},s_{Mz},t_{Mz};-\vec{p},s_{Bz},t_{Bz}\rangle  \nonumber 
\\
&&= \sum_{JM,TT_z}\,\,\sum_{L'M'_L,S'S'_z}\,\,\sum_{LM_L,SS_z} Y_{L',M_L'}(\hat{p}')Y_{L,M_L}^*(\hat{p})
\nonumber\\
&& \times
\langle s_{M'},s_{B'},s_{M'z},s_{B'z}|S',S_z'\rangle \langle L',S',M_L',S_z'|J,M\rangle  
\langle t_{M'},t_{B'},t_{M'z},t_{B'z}|T,T_z\rangle \nonumber \\
& & \times
\langle s_M,s_B,s_{Mz},s_{Bz}|S,S_z\rangle \langle L,S,M_L,S_z|J,M\rangle 
\langle t_M,t_B,t_{Mz},t_{Bz}|T,T_z\rangle 
 \nonumber \\
&& \times
t^{JT}_{L'S'M'B',LSMB}(p',p,W),
\end{eqnarray}
where 
the matrix elements
$t^{JT}_{L'S'M'B',LS\pi N}(p',p,W) $ for $M'B'=\pi\Delta,\sigma N, \rho N$ are the PWA from
the ANL-Osaka model.

The matrix elements of $H_I$ of Eq.~(\ref{eq:pin-pipin-1}) describe the decay of a resonance 
$R=\Delta,\rho,\sigma$ into
a two-particle state $cd$. It is of the following expression 
\begin{eqnarray}
&&
\langle \vec{p}_c,s_{cz},t_{cz};\vec{p}_d,s_{dz},t_{dz}|H_I|\vec{k}',s_{Rz},t_{Rz}\rangle 
\nonumber\\
&=& \delta(\vec{p}_c+\vec{p}_d - \vec{k}')
\sqrt{\frac{E_c(k_{cd})E_d(k_{cd})M_R}
{{E_c(p_c)E_d(p_d)E_R(k')}}}
\langle \vec{k}_{cd},s_{cz},t_{cz};-\vec{k}_{cd},s_{dz},t_{dz}|
H_I|\vec{0},s_{Rz},t_{Rz}\rangle ,
\end{eqnarray}
with
\begin{eqnarray}
&&
\langle \vec{k}_{cd},s_{cz},t_{cz};-\vec{k}_{cd},s_{dz},t_{dz}|
H_I|\vec{0},s_{Rz},t_{Rz}\rangle 
\nonumber\\
&&=
\sum_{L_{cd},S_{cd},m_{cd},S_{cdz}}
[\langle s_c,s_d,s_{cz},s_{dz}|S_{cd},S_{cdz}\rangle 
\langle L_{cd},S_{cd},m_{cd},S_{cdz}|s_R,s_{Rz}\rangle  \nonumber \\
& & \,\,\,\,\,\, \times\langle t_{c},t_{d},t_{cz},t_{dz}|t_R,t_{Rz}\rangle 
Y_{L_{cd},m_{cd}}(\hat{k}_{cd})
F_{L^R_{cd},S^R_{cd}}(k_{cd})]
\delta_{L_{cd},L^R_{cd}}\delta_{S_{cd},S^R_{cd}}
\ .
\end{eqnarray}
The vertex functions are
\begin{eqnarray}
F_{L^\Delta_{\pi N}, S^\Delta_{\pi N}}(q) &=& if_{\Delta\rightarrow \pi N}(q)
\label{eq:ff-pid} \ , \\
F_{L^\sigma_{\pi \pi}, S^\sigma_{\pi \pi}}(q) 
&=& \sqrt{2}g(q) \label{eq:ff-sn}
\ , \\
F_{L^\rho_{\pi \pi}, S^\rho_{\pi \pi}}(q) &=& 
(-1)\sqrt{2}f_{\rho\rightarrow \pi\pi}(q)\ , \label{eq:ff-rn}
\end{eqnarray}
where $L^\Delta_{\pi N}=1, S^\Delta_{\pi N}=3/2$, 
$L^\sigma_{\pi \pi}=0, S^\sigma_{\pi \pi}=0$, 
$ L^\rho_{\pi \pi}=1, S^\rho_{\pi \pi}=1$.
Here it is noted that the factor $\sqrt{2}$ in 
Eqs.~(\ref{eq:ff-sn})-(\ref{eq:ff-rn}) comes from the Bose symmetry
of pions, and the phase factor $i$ and $(-1)$ are chosen to be
consistent with the non-resonant interactions involving
$\pi N\Delta$, $\sigma \pi\pi$ and $\rho\pi\pi$ vertex
interactions.
 The form factors $f_{\Delta\rightarrow \pi N}(q)$ and $f_{\rho\rightarrow \pi\pi}(q)$
have been in Eqs.~(\ref{eq:fdpin})-(\ref{eq:frhopp}) and $g(q)$ in Eq.~(\ref{eq:fsigmapp}).

With the above equations, the contribution from $\pi N\rightarrow \pi\Delta\rightarrow \pi\pi N$ to
the total cross section $\sigma^{rec}_{\pi N \rightarrow \pi\pi N}$,
as defined by Eq.~(\ref{eq:sigma-t})-(\ref{eq:sigma-ta}), can be written as
\begin{eqnarray}
 \sigma^{rec}_{\pi \Delta}(W)&=&\int_{m_N+m_\pi}^{W-m_\pi}dM_{\pi N}
{{M_{\pi N}\over E_\Delta(k)}}
\frac{\Gamma_{\Delta}/(2\pi)}{|W-E_\pi(k)-E_\Delta(k)-\Sigma_{\pi\Delta}(k,W)|^2}
\times \sigma_{\pi N\rightarrow \pi \Delta}\ ,
\label{eq:tot-pid}
\end{eqnarray}
where $k$ and $E_\Delta(k)$ are  defined by $W$ and $M_{\pi N}$
\begin{eqnarray}
k&=&\frac{1}{2W}[(W^2-M^2_{\pi N}-m^2_\pi)^2-4M^2_{\pi N}m^2_\pi]^{1/2}\,, \\
E_\Delta(k)&=&[{m^2_{\Delta}}+k^2]^{1/2}\,,
\end{eqnarray}
$\Sigma_{\pi\Delta}(k,W)$is defined in Eq.~(\ref{eq:self-pidelta-1}), 
{$\Gamma_{\Delta}=-2Im[\Sigma_{\pi\Delta}(k=0,W)]$}, and
\begin{eqnarray}
\sigma_{\pi N\rightarrow \pi \Delta}&=&{4\pi\over {k_0^2}}
\sum_{J{T},L'S',LS}\frac{2J+1}{(2S_N+1)(2S_\pi+1)}
|\rho^{1/2}_{\pi\Delta}(k)t^{J{T}}_{L'S'\pi\Delta, LS\pi N}(k,k_0;W)
\rho^{1/2}_{\pi N}(k_0)|^2
\nonumber \\
 &&
  {\times \langle t_\pi, t_N, t_\pi^z, t_N^z | T, T^z  \rangle^2}
\ ,
\label{eq:tot-pid-a}
\end{eqnarray}
where $k_0$ is defined by $W=E_\pi(k_0)+E_N(k_0)$ and $ \rho_{ab}(k)=\pi k E_a(k)E_b(k)/W$.
Similarly, 
the contributions of $\pi N\rightarrow \rho N\rightarrow \pi\pi N$ 
and $\pi N\rightarrow \sigma N\rightarrow \pi\pi N$ to
the total cross section $\sigma^{rec}_{\pi N \rightarrow \pi\pi N}$ are
\begin{eqnarray}
\sigma^{rec}_{aN}(W)&=&\int_{2m_\pi}^{W-m_N}dM_{\pi \pi}
{{M_{\pi\pi}\over E_a(k)}}
\frac{\Gamma_{a}/(2\pi)}{|W-E_N(k)-E_a(k)-\Sigma_{aN}(k,W)|^2}
\times \sigma_{\pi N\rightarrow aN}\ ,
\label{eq:tot-an}
\end{eqnarray}
where $a=\rho, \sigma$, $k$ is defined by $M_{\pi \pi}$ and $W$
\begin{eqnarray}
k&=&\frac{1}{2W}[(W^2-M^2_{\pi \pi}-m^2_N)^2-4M^2_{\pi \pi}m^2_N]^{1/2}\,, \\
E_a(k)&=&[{m^2_{a}}+k^2]^{1/2}\,,
\end{eqnarray}
$\Sigma_{aN}(k,W)$ for $aN=\rho N, \sigma N$ are is defined in Eqs.(104) and (107),
{$\Gamma_{a}=-2Im[\Sigma_{aN}(k=0,W)]$}, and
\begin{eqnarray}
\sigma_{\pi N\rightarrow aN}&=&{4\pi \over {k^2_0}}
\sum_{J{T},L'S',LS}\frac{2J+1}{(2S_N+1)(2S_\pi+1)}
|\rho^{1/2}_{aN}(k)t^{J{T}}_{L'S'aN, LS\pi N}(k,k_0;W)\rho^{1/2}_{\pi N}(k_0)|^2
\nonumber \\
 &&  {\times \langle t_\pi, t_N, t_\pi^z, t_N^z | T, T^z  \rangle^2}
\ .
\label{eq:tot-an-a}
\end{eqnarray}

To perform calculations, we need to have
the partial-wave amplitudes $t^{JT}_{L'S'M'B',LS\pi N}(p,k,W)$ for $M'B'=\pi\Delta,\rho N,\sigma N$.
These PWA from ANL-Osaka model  can be obtained from the web page which present the following  :
\begin{eqnarray}
<\pi\Delta|T(W)|\pi N> =-\rho^{1/2}_{\pi\Delta}(p_\Delta)t^{JT}_{L'S'\pi\Delta,LS\pi N}(p_\Delta,k,W)\rho^{1/2}_{\pi N}(k)\nonumber \ , \\
<\rho N|T(W)|\pi N> =-\rho^{1/2}_{\rho N}(p_\rho)t^{JT}_{L'S'\rho N,LS\pi N}(p_\rho,k,W)\rho^{1/2}_{\pi N}(k)\nonumber \ , \\
<\sigma N|T(W)|\pi N> =-\rho^{1/2}_{\sigma N}(p_\sigma)t^{JT}_{L'S'\sigma N,LS\pi N}(p_\sigma,k,W)\rho^{1/2}_{\pi N}(k)\nonumber \ , \\
\end{eqnarray}
 Here the phase space factors account for the effects due to $\Delta\rightarrow \pi N$,  
$\sigma \rightarrow \pi\pi$ and
$\rho \rightarrow \pi\pi$ decays.  Explicitly, we have
\begin{eqnarray}
\rho_{\pi\Delta}(p_\Delta)&=&\pi\frac{p_\Delta E_{\Delta}(p_\Delta)E_\pi(p_\Delta)}{W}\ ,
\label{eq:rho-pid}
\end{eqnarray}
where $p_\Delta$ and $E_{\Delta}(p_\Delta)$ are 
 defined by $W$ and the invariant mass $M_{\pi N}$ in the integrations of Eqs.~(\ref{eq:sigma-t}) and
(\ref{eq:tot-pid})  
\begin{eqnarray}
p_\Delta&=&\frac{1}{2W}[(W^2-M^2_{\pi N}-m^2_\pi)^2-4M^2_{\pi N}m^2_\pi]^{1/2}\,, \\
E_\Delta(p_\Delta)&=&[M^2_{\pi N}+p^2_\Delta]^{1/2}\,,
\end{eqnarray}
For the calculations of Eqs.~(\ref{eq:sigma-t}) and
(\ref{eq:tot-pid}), we thus present $<\pi\Delta|T(W)|\pi N>$ in the range $0\leq p_\Delta \leq p_{\Delta,max}$ with
\begin{eqnarray}
p_{\Delta,max}=\frac{1}{2W}[(W^2-(m_\pi+m_N)^2-m^2_\pi)^2-4(m_\pi+m_N)^2m^2_\pi ]^{1/2}\ .
\end{eqnarray}

For the $\rho  N$ and $\sigma N$ channels, we have
\begin{eqnarray}
\rho^{1/2}_{\sigma N}(p_\sigma)&=& \pi\frac{p_\sigma E_{\sigma}(p_\sigma)E_N(p_\sigma)}{W} \ ,\\
\rho^{1/2}_{\rho N}(p_\rho)&=&\pi\frac{p_\rho E_{\rho}(p_\rho)E_N(p_\rho)}{W} \ ,
\end{eqnarray}
For $a=\sigma$ and $\rho$, we have 
\begin{eqnarray}
p_a&=&\frac{1}{2W}[(W^2-M^2_{\pi \pi}-m^2_N)^2-4M^2_{\pi \pi}m^2_N]^{1/2}\,, \\
E_a(p_a)&=&[M^2_{\pi \pi}+p^2_a]^{1/2}\,.
\end{eqnarray}
For the calculation of Eqs.~(\ref{eq:sigma-t}) and
(\ref{eq:tot-an}), we thus present $<\rho N|T(W)|\pi N>$
and $<\sigma N|T(W)|\pi N>$ in the range $0\leq p_a\leq p_{a,max}$ with
\begin{eqnarray}
p_{a,max}=\frac{1}{2W}[(W^2-(2m_\pi)^2-m^2_N)^2-4(2m_\pi)^2m^2_N ]^{1/2}\ .
\label{eq:pmax-rn}
\end{eqnarray}

\begin{figure}[h]
\centering
\includegraphics[clip,width=16cm]{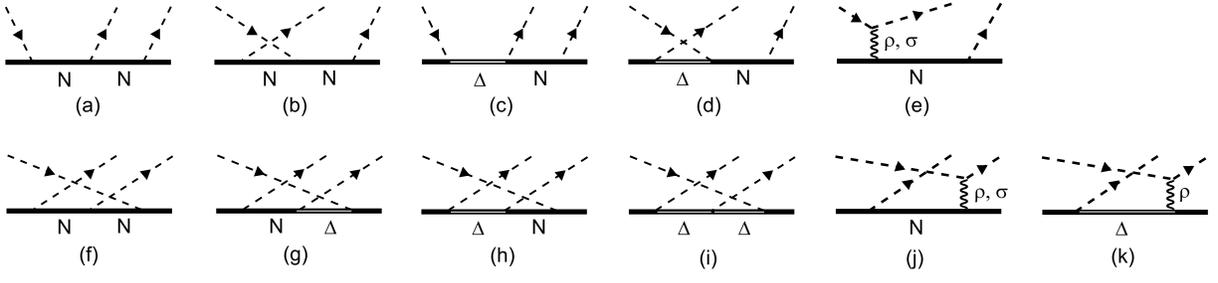}
\caption{The considered $v_{\pi N,\pi\pi N}$. }
\label{fig:mbpipin}
\end{figure}

The above equations are for the calculations of the $\pi N \rightarrow \pi\pi N$ through the
resonant $\pi\Delta$, $\sigma N$ and $\rho N$ channels.  There are also weaker contributions from
the direct production mechanisms, as illustrated in Fig.~\ref{fig:mbpipin},  which can be calculated by using
the procedures explained in Ref.~\cite{kjlms09}.

\clearpage
\section{data}

The parameters of the ANL-Osaka DCC model were determined by performing  $\chi^2$-fits to the data
of $\pi N, \gamma N\rightarrow \pi N, \eta N, K\Lambda, K\Sigma$. 
The calculations involve solving the coupled-channel  Eqs.~(\ref{eq:tmbmb})-(\ref{eq:nstar-sigma}).
 The meson-baryon partial-waves included in the calculations 
are listed in Table~\ref{tab:pw}.
The total number of the data included in the fits
 are about 30,000 data points, as listed in table~\ref{tab:pwa-data}-\ref{tab:data-chi2}.

Note that the 1940 data points for  $\pi N\rightarrow \pi N$ 
listed in Table~\ref{tab:pwa-data} are in fact contain information of about 30,000 data  
in the  SAID analysis: 
14196  of $\pi^+ p \rightarrow \pi^+ p$, 13895 of
$\pi^- p \rightarrow \pi^- p$, and 2877 of
$\pi^- p \rightarrow \pi^0 n$.  In addition, the partial-wave amplitudes of $\pi N\rightarrow \pi \pi N$
determined
 from 
the very extensive bubble-chamber data of $\pi N\rightarrow \pi \pi N$ were also used in the
earlier analysis of SAID. Thus the determined
$PWA$ for $\pi N\rightarrow \pi N, \pi\Delta, \rho N, \sigma N$ are rather reliable.

On the other hand, the data points for  $\pi N\rightarrow \eta N, K\Lambda, K\Sigma $ listed in
Table~\ref{tab:ha-data}  are much less, 
only 294, 941, and 1262, respectively. Thus the determined PWA for these processes need to be improved by using
more extensive data which could be available from experiments at J-PARC in near future. 
 
In Tables~\ref{tab:em-data} and \ref{tab:data-chi2}, 
 we see that the data points for $\gamma N\rightarrow \pi N$ are much more
than those of  $\gamma N\rightarrow \eta N, K\Lambda, K\Sigma$. Therefore, the $\gamma N \rightarrow N^*$ 
couplings are mainly determined by the predicted multipole amplitudes of $\gamma N\rightarrow \pi N$.
It will be interesting to include more new JLab data of $\gamma N\rightarrow K\Lambda, K\Sigma$
in the analysis to further improve the determination of $\gamma N \rightarrow N^*$
couplings which are crucial to test the predictions from LQCD and various hadron models.

\begin{table}[b]
\caption{
The orbital angular momentum $(L)$ and total spin ($S$) of each $MB$ channel allowed in a given partial wave.
In the first column, partial waves are denoted with the conventional notation $l_{2I2J}$ as well as ($I$,$J^P$).
\label{tab:pw}}
\begin{ruledtabular}
\begin{tabular}{ccccccccccc}
$l_{2I2J}$ $(I,J^P)$ &\multicolumn{10}{c}{$(L,S)$ of the considered partial waves}\\
\cline{2-11}
& $\pi N$     & $\eta N$   &\multicolumn{2}{c}{$\pi \Delta $}& $\sigma N$   &\multicolumn{3}{c}{$\rho N$} & $K\Lambda$   &$K\Sigma$ \\
&             &            &$(\pi \Delta)_1$&$(\pi \Delta)_2$&              &$(\rho N)_1$&$(\rho N)_2$& $(\rho N)_3$&              &          \\
\hline
$S_{11}$ $(1,\ohalf^-)$&($0,\ohalf$)&($0,\ohalf$)&($2,\thalf$)& --         &($1,\ohalf$)&($0,\ohalf$)&($2,\thalf$)& --         &($0,\ohalf$)&($0,\ohalf$)\\
$S_{31}$ $(3,\ohalf^-)$&($0,\ohalf$)&   --       &($2,\thalf$)&--          &    --      &($0,\ohalf$)&($2,\thalf$)& --         &--          &($0,\ohalf$)\\
$P_{11}$ $(1,\ohalf^+)$&($1,\ohalf$)&($1,\ohalf$)&($1,\thalf$)&--          &($0,\ohalf$)&($1,\ohalf$)&($1,\thalf$)& --         &($1,\ohalf$)&($1,\ohalf$)\\
$P_{13}$ $(1,\thalf^+)$&($1,\ohalf$)&($1,\ohalf$)&($1,\thalf$)&($3,\thalf$)&($2,\ohalf$)&($1,\ohalf$)&($1,\thalf$)&($3,\thalf$)&($1,\ohalf$)&($1,\ohalf$)\\
$P_{31}$ $(3,\ohalf^+)$&($1,\ohalf$)&   --       &($1,\thalf$)&--          &    --      &($1,\ohalf$)&($1,\thalf$)& --         &--          &($1,\ohalf$)\\
$P_{33}$ $(3,\thalf^+)$&($1,\ohalf$)&   --       &($1,\thalf$)&($3,\thalf$)&     --     &($1,\ohalf$)&($1,\thalf$)&($3,\thalf$)&--          &($1,\ohalf$)\\
$D_{13}$ $(1,\thalf^-)$&($2,\ohalf$)&($2,\ohalf$)&($0,\thalf$)&($2,\thalf$)&($1,\ohalf$)&($2,\ohalf$)&($0,\thalf$)&($4,\thalf$)&($2,\ohalf$)&($2,\ohalf$)\\
$D_{15}$ $(1,\fhalf^-)$&($2,\ohalf$)&($2,\ohalf$)&($2,\thalf$)&($4,\thalf$)&($3,\ohalf$)&($2,\ohalf$)&($2,\thalf$)&($4,\thalf$)&($2,\ohalf$)&($2,\ohalf$)\\
$D_{33}$ $(3,\thalf^-)$&($2,\ohalf$)&    --      &($0,\thalf$)&($2,\thalf$)& --         &($2,\ohalf$)&($0,\thalf$)&($2,\thalf$)&--          &($2,\ohalf$)\\
$D_{35}$ $(3,\fhalf^-)$&($2,\ohalf$)&    --      &($2,\thalf$)&($4,\thalf$)&    --      &($2,\ohalf$)&($2,\thalf$)&($4,\thalf$)&--          &($2,\ohalf$)\\
$F_{15}$ $(1,\fhalf^+)$&($3,\ohalf$)&($3,\ohalf$)&($1,\thalf$)&($3,\thalf$)&($2,\ohalf$)&($3,\ohalf$)&($1,\thalf$)&($3,\thalf$)&($3,\ohalf$)&($3,\ohalf$)\\
$F_{17}$ $(1,\shalf^+)$&($3,\ohalf$)&($3,\ohalf$)&($3,\thalf$)&($5,\thalf$)&($4,\ohalf$)&($3,\ohalf$)&($3,\thalf$)&($5,\thalf$)&($3,\ohalf$)&($3,\ohalf$)\\
$F_{35}$ $(3,\fhalf^+)$&($3,\ohalf$)&    --      &($1,\thalf$)&($3,\thalf$)& --         &($3,\ohalf$)&($1,\thalf$)&($3,\thalf$)&--          &($3,\ohalf$)\\
$F_{37}$ $(3,\shalf^+)$&($3,\ohalf$)&  --        &($3,\thalf$)&($5,\thalf$)&  --        &($3,\ohalf$)&($3,\thalf$)&($5,\thalf$)&--          &($3,\ohalf$)\\     
$G_{17}$ $(1,\shalf^-)$&($4,\ohalf$)&($4,\ohalf$)&($2,\thalf$)&($4,\thalf$)&($3,\ohalf$)&($4,\ohalf$)&($2,\thalf$)&($4,\thalf$)&($4,\ohalf$)&($4,\ohalf$)\\
$G_{19}$ $(1,\nhalf^-)$&($4,\ohalf$)&($4,\ohalf$)&($4,\thalf$)&($6,\thalf$)&($5,\ohalf$)&($4,\ohalf$)&($4,\thalf$)&($6,\thalf$)&($4,\ohalf$)&($4,\ohalf$)\\
$G_{37}$ $(3,\shalf^-)$&($4,\ohalf$)&   --       &($2,\thalf$)&($4,\thalf$)&--          &($4,\ohalf$)&($2,\thalf$)&($4,\thalf$)&--          &($4,\ohalf$)\\
$G_{39}$ $(3,\nhalf^-)$&($4,\ohalf$)&  --        &($4,\thalf$)&($6,\thalf$)&  --        &($4,\ohalf$)&($4,\thalf$)&($6,\thalf$)&--          &($4,\ohalf$)\\   
$H_{19}$ $(1,\nhalf^+)$&($5,\ohalf$)&($5,\ohalf$)&($3,\thalf$)&($5,\thalf$)&($4,\ohalf$)&($5,\ohalf$)&($3,\thalf$)&($5,\thalf$)&($5,\ohalf$)&($5,\ohalf$)\\
$H_{39}$ $(3,\nhalf^+)$&($5,\ohalf$)&($5,\ohalf$)&($3,\thalf$)&($5,\thalf$)&--          &($5,\ohalf$)&($3,\thalf$)&($5,\thalf$)&--          &($5,\ohalf$)      
\end{tabular}
\end{ruledtabular}
\end{table}

\begin{table}[h]
\caption{\label{tab:pwa-data} 
Number of the data points of $\pi N \rightarrow \pi N$ amplitudes 
included in our fits. The data are from SAID analysis of 
14196 data points of $\pi^+ p \rightarrow \pi^+ p$, 13895 of
$\pi^- p \rightarrow \pi^- p$, and 2877 of
$\pi^- p \rightarrow \pi^0 p$.
}
\begin{ruledtabular}
\begin{tabular}{lrrrrr}
Partial Wave  & & Partial Wave&&\\\hline
$S_{11}$ & 65$\times$2 & $S_{31}$ & 65$\times$2 &\\
$P_{11}$ & 65$\times$2 & $P_{31}$ & 61$\times$2 &\\
$P_{13}$ & 61$\times$2 & $P_{33}$ & 65$\times$2 &\\
$D_{13}$ & 61$\times$2 & $D_{33}$ & 55$\times$2 &\\
$D_{15}$ & 61$\times$2 & $D_{35}$ & 45$\times$2 &\\
$F_{15}$ & 48$\times$2 & $F_{35}$ & 43$\times$2 &\\
$F_{17}$ & 32$\times$2 & $F_{37}$ & 44$\times$2 &\\
$G_{17}$ & 42$\times$2 & $G_{37}$ & 32$\times$2 &\\
$G_{19}$ & 28$\times$2 & $G_{39}$ & 32$\times$2 &\\
$H_{19}$ & 34$\times$2 & $H_{39}$ & 31$\times$2 &\\
& &  & &\\
Sum     & 994 &&946&1940
\end{tabular}
\end{ruledtabular}
\end{table}

\begin{table}[h]
\caption{\label{tab:ha-data}
Number of data points of hadronic processes included in our fits.
Data are from the compilation of Bonn-Gatchina.
}
\begin{ruledtabular}
\begin{tabular}{lrrrrr}
  & $d\sigma/d\Omega$ & $P$ & $R$& $a$ &sum \\\hline
$\pi^- p \rightarrow \eta p$& 294    &  --    &   --    \ &  --& 294\\
                             &       &       &        &     &    \\
$\pi^- p \rightarrow K^0 \Lambda$& 587 &  354 &   --   &  --  & 941\\
$\pi^- p \rightarrow K^0\Sigma^0$& 259 &  90 &      -- &   --& 349\\
$\pi^+p \rightarrow K^+\Sigma^+$& 609 &  304 &  --     &   -- &  913\\
                             &       &       &        &     &    \\
 Sum                         & 1749      & 748      & --  & --    & 2497 \\
\end{tabular}
\end{ruledtabular}
\end{table}
\vspace{0.5cm}

\begin{table}[h]
\caption{\label{tab:em-data}
The number of data points of photoproduction processes included in our fits.
The data are from compilation of Bonn-Gatchina.
 }
\begin{ruledtabular}
\begin{tabular}{lrrrrrrrrrrrrr}
  & $d\sigma/d\Omega$ & $\Sigma$ & $T$& $P$ & $ G$  & $H$ &
$E$& $F$& $O_{x'}$& $O_{z'}$
&$C_{x'}$& $C_{z'}$ &sum \\\hline
$\gamma p\rightarrow \pi^0 p$&4414 &  1866 &   389   &  607 &75 & 71    &140  &
--&     7 &    7& --& --& 7576 \\
$\gamma p\rightarrow \pi^+ n$&2475 &  899 &   661   &  252 &86 & 128    &231  &
--&     -- &    --&--&-- &4732\\
        &              &     &        &       &   &       &   &&       &  && \\
$\gamma p\rightarrow \eta p$&780 &  151 &   50   & -- &-- & --    &--  &
--&     -- &    --&--&-- &981 \\
        &              &     &        &       &   &       &   &&       &   &&\\
$\gamma p\rightarrow K^+\Lambda $&1320 & 118 &   66   &  1336 &-- & --    &--  &
--&    160 &159&66&66  & 3291  \\
$\gamma p\rightarrow K^+\Sigma^0 $&1280 & 87 &   --   &  95 &-- & --    &--  &
--&    -- &--  &94 & 94  &1650   \\
$\gamma p\rightarrow K^0\Sigma^+ $&276 & 15 &   --   &  72 &-- & --    &--  &
--&    -- &--  &-- & --  &363   \\
        &              &     &        &       &   &       &   &&       &   &&\\
Sum     & 10545 & 3136 &   1166   &  2362 & 161  & 199    &371  &
--&   167 &166&160&160  &18593   \\
\end{tabular}
\end{ruledtabular}
\end{table}

\begin{table}[h]
\caption{\label{tab:data-chi2}
Observables and number of the data points considered in this coupled-channels analysis.
The data are taken from the database of the INS DAC Services.
}
\begin{ruledtabular}
\begin{tabular}{lcc}
Reactions &Observables & Number of data points\\
\hline
$\gamma \text{`}n\text{'} \to \pi^- p$ &$d\sigma/d\Omega$& 2305  \\
                       &$\Sigma$& 308  \\
                       &$T$& 94  \\
                       &$P$& 88  \\
$\gamma \text{`}n\text{'} \to \pi^0 n$ &$d\sigma/d\Omega$& 148 \\
                       &$\Sigma$& 216 \\
  & & \\
Sum & & 3159 \\
\end{tabular}
\end{ruledtabular}
\end{table}

\begin{table}[h]
\caption{\label{tab:eep-data}
Number of data points of $p(e,e'\pi)N$ included in our fits.
The data are for about 25 $Q^2$ with $W < 2000$ MeV.
}
\begin{ruledtabular}
\begin{tabular}{lrrrrr}
  & $\sigma_T+\epsilon \sigma_L$ & $\sigma_{TT}$ & $\sigma_{LT}$ & $\sigma'_{LT}$ &sum \\\hline
$p(e,e'\pi^0)p$& 5830    & 5830    &  5830    \ &  240 & 17730\\
                             &       &       &        &     &    \\
$p(e,e'\pi^+)n$& 2614    & 2614    &  2614    \ &  566 & 8408\\
                             &       &       &        &     &    \\
 Sum                         & 8444      & 8444      & 8444  & 806    & 26138 \\
\end{tabular}
\end{ruledtabular}
\end{table}
\vspace{0.5cm}
\clearpage

\vspace{0.5cm}
\section{ Results}

The  nucleon resonance properties extracted from ANL-Osaka amplitudes were
finalized in 2016 and published in  Ref.~\cite{knls16}.
In Table~\ref{tab:pole}, we list the pole positions and the
residues $R_{\pi N,\pi N}$ of the extracted resonances. In Ref.~\cite{knls13}, the residues for
$R_{\pi N, \eta N}$, $R_{\pi N, K\Lambda}$, and $R_{\pi N, K\Sigma}$ are also given. However these results are
not as solid as that of $R_{\pi N, \pi N}$ because the data of $\pi N \rightarrow \eta N, K\Lambda, K\Sigma$
included in the fits 
are not as accurate  as the data of $\pi N\rightarrow \pi N$. The residues associated with the unstable channels
$\pi \Delta, \rho N$ and $\sigma N$ are also not given because the analytic continuation of the PWA associated
with these channels to complex $E$-plane is rather complex and a rigorous way to do this remains to be explored.

The determined helicity amplitudes for the $\gamma p\to N^*$ transition at resonance pole positions, published in 
Ref.~\cite{knls16}, are listed in Table~\ref{tab:helicity_p}. However the $Q^2$-dependence of
$\gamma p\to N^*$ transitions at resonance pole was not extracted because only the data
at few $Q^2$ were included in the fits.

To illustrate the quality of our fits to the data, we present sample results
from our analysis in the Appendix:
(A) Total cross sections of $\pi N$ reactions, (B) Total cross sections of $\gamma N$ reactions,
(C) Differential cross sections of $\pi N\rightarrow \pi N, \eta N,\ K\Lambda, K\Sigma$,
(D) Differential cross sections of $\gamma N\rightarrow \pi N, \eta N,\ K\Lambda, K\Sigma$,
(E) Differential cross sections of $\gamma^* N\rightarrow \pi N$, (F) Inclusive cross sections of
$p(e,e')$.

\begin{table}
\caption{\label{tab:pole}
The extracted  nucleon resonance pole mass ($M_R$) and $\pi N$ elastic residue ($R_{\pi N, \pi N}$).
$M_R$ is listed as $\left(\mathrm{Re}(M_R),-\mathrm{Im}(M_R)\right)$ in units of MeV,
while $R_{\pi N,\pi N}=|R_{\pi N,\pi N}|e^{i\phi}$ is listed as $\left(|R_{\pi N,\pi N}|,\phi\right)$ 
in units of MeV for 
$|R_{\pi N,\pi N}|$ and degree for $\phi$.
The range of $\phi$ is taken to be $-180^\circ \leq \phi < 180^\circ$.
The $N^*$ resonances for which the asterisk (*) is marked locate in the complex energy plane
slightly off the sheet closest to the physical real energy axis, yet are still expected
to visibly affect the physical observables. 
}
\begin{ruledtabular}
\begin{tabular}{lccccc}
&$J^P(L_{2I2J})$  &$M_R$ & &$R_{\pi N,\pi N}$ & \\
\hline
$N^*$ 
&$1/2^-(S_{11})$ &(1490, 102)& &($  70$,  $-$42)&\\ 
&                &(1652,  71)&  &($  45$,  $-$74)&\\ 
&$1/2^+(P_{11})$ &(1376,  75)&  &($  38$,  $-$70)&\\ 
&                &(1741, 139)&  &($  15$,     80)&\\ 
&$3/2^+(P_{13})$ &(1708,  65)&  &($   9$,   $-$4)&\\ 
&                &(1765, 160)&  &($  30$, $-$105)&\\ 
&$3/2^-(D_{13})$ &(1509,  48)&  &($  30$,   $-$10)&\\ 
&                &(1702, 148)*& &($ <1 $, $-$161)&\\ 
&$5/2^-(D_{15})$ &(1651,  68)&&($  26$,  $-$27)&\\ 
&$5/2^+(F_{15})$ &(1665,  52)& &($  36$,  $-$22)&\\ 
\\                                                   
$\Delta^*$                                      
&$1/2^-(S_{31})$ &(1597,  69)& &($  21$, $-$111)&\\ 
&                &(1713, 187)&  &($  20$,     73)&\\ 
&$1/2^+(P_{31})$ &(1857, 145)&  &($  11$, $-$118)&\\ 
&$3/2^+(P_{33})$ &(1212,  52)&  &($  55$,  $-$47)&\\ 
&                &(1733, 162)&  &($  16$, $-$108)&\\ 
&$3/2^-(D_{33})$ &(1577, 113)&  &($  13$,  $-$67)&\\ 
&$5/2^-(D_{35})$ &(1911, 130)&  &($   4$,  $-$30)&\\ 
&$5/2^+(F_{35})$ &(1767,  88)& &($  11$,  $-$61)&\\ 
&$7/2^+(F_{37})$ &(1885, 102)&  &($  49$,  $-$30)&\\ 
\end{tabular}
\end{ruledtabular}
\end{table}

\begin{table}
\caption{\label{tab:helicity_p}
The determined helicity amplitudes for the $\gamma p\to N^*$ transition at resonance pole positions.
The presented values follow the notation: 
$A_{1/2,3/2} = \bar A_{1/2,3/2}\times e^{i\phi}$ with $\phi$ taken to be in the range $-90^\circ \leq \phi < 90^\circ$.
The units of $\bar A_{1/2,3/2}$ and $\phi$ are $10^{-3}\ {\rm GeV}^{-1/2}$ and degree, respectively.
Each resonance is specified by the isospin and spin-parity quantum numbers 
as well as the real part of the resonance pole mass.
}
\begin{ruledtabular}
\begin{tabular}{lrrrrrrrr}
\\
& \multicolumn{2}{c}{}
& \multicolumn{2}{c}{}
& \multicolumn{2}{c}{}
& \multicolumn{2}{c}{}
\\
Particle $J^P(L_{2I2J})$
&$\bar A_{1/2}$&$\phi$ && &$\bar A_{3/2}$&$\phi$ &&
\\
\hline
$N(1490) 1/2^-(S_{11})$    &$  160 $&$   8  $&&&-&    -   &      &      \\ 
$N(1652) 1/2^-(S_{11})$    &$   36 $&$ -28  $&&&-&    -   &      &      \\ 
$N(1376) 1/2^+(P_{11})$    &$ - 40 $&$  -8  $&&&-&    -   &      &      \\ 
$N(1741) 1/2^+(P_{11})$    &$ - 47 $&$ -24  $&&&-&    -   &      &      \\ 
$N(1708) 3/2^+(P_{13})$    &$  131 $&$   7  $&&&$- 33 $&$   12   $&&\\ 
$N(1765) 3/2^+(P_{13})$    &$  123 $&$- 11  $&&&$ - 71 $&$    3 $&&\\ 
$N(1509) 3/2^-(D_{13})$    &$ - 28 $&$<  1  $&&&$  102 $&$    4 $&&\\ 
$N(1703) 3/2^-(D_{13})$    &$   13 $&$  50  $&&&$   31 $&$ - 71 $&&\\ 
$N(1651) 5/2^-(D_{15})$    &$    8 $&$  19  $&&&$   49 $&$ - 12 $&&\\ 
$N(1665) 5/2^+(F_{15})$    &$ - 44 $&$ -11  $&&&$   60 $&$ -  2 $&&\\ 
\\                                                              
$\Delta(1597)1/2^-(S_{31})$&$  105 $&$   1  $&&&  -     &   -    &      &      \\  	 
$\Delta(1713)1/2^-(S_{31})$&$   40 $&$  13  $&&&  -     &   -    &      &      \\  	 
$\Delta(1857)1/2^+(P_{31})$&$  -1 $&$  -78  $&&&  -     &   -    &      &      \\  	 
$\Delta(1212)3/2^+(P_{33})$&$ -134 $&$ -16  $&&&$- 257 $&$  -3  $&&\\  	 
$\Delta(1733)3/2^+(P_{33})$&$ - 48 $&$  63  $&&&$ - 94 $&$   74 $&&\\  	 
$\Delta(1577)3/2^-(D_{33})$&$  128 $&$  19  $&&&$  119 $&$   46 $&&\\  	 
$\Delta(1911)5/2^-(D_{35})$&$   48 $&$ - 22 $&&&$   11 $&$  -36 $&&\\  	 
$\Delta(1767)5/2^+(F_{35})$&$   38 $&$  - 7 $&&&$  -24 $&$  -80 $&&\\  	 
$\Delta(1885)7/2^+(F_{37})$&$ - 69 $&$  -14 $&&&$ - 83 $&$    2 $&&\\  	 
\end{tabular}
\end{ruledtabular}
\end{table}

\begin{table}
\caption{\label{tab:helicity_n-iso}
The isovector and isoscalar 
helicity amplitudes for $\gamma N\to N^*$ are defined as
$A^{T=1}_\lambda=(A^p_\lambda-A^n_\lambda)$, and
$A^{T=0}_\lambda=(A^p_\lambda+A^n_\lambda)$, where
$A^p_\lambda$ and $A^n_\lambda$ are the helicity amplitudes of $\gamma p \rightarrow N^*$ and
$\gamma n \rightarrow N^*$, respectively.
See the caption of Table~\ref{tab:helicity_p} for the notation of the table.
}
\begin{ruledtabular}
\begin{tabular}{lrrrrrrrr}
\\
Particle $J^P(L_{2I2J})$
&$\bar A^{T=1}_{1/2}$&$\phi$
&$\bar A^{T=0}_{1/2}$&$\phi$
&$\bar A^{T=1}_{3/2}$&$\phi$
&$\bar A^{T=0}_{3/2}$&$\phi$
\\
\hline
$N(1490) 1/2^-(S_{11})$ &$ 136 $&$  11 $&$  26 $&$- 10 $&   -   &   -   &    -  &    -  \\ 
$N(1652) 1/2^-(S_{11})$ &$  19 $&$ -29 $&$  18 $&$- 28 $&   -   &   -   &    -  &    -  \\ 
$N(1376) 1/2^+(P_{11})$ &$- 68 $&$ -13 $&$  28 $&$- 21 $&   -   &   -   &    -  &    -  \\ 
$N(1741) 1/2^+(P_{11})$ &$-120 $&$ -11 $&$  75 $&$-  3 $&   -   &   -   &    -  &    -  \\ 
$N(1708) 3/2^+(P_{13})$ &$  95 $&$   7 $&$  36 $&$   8 $&$  -9 $&$  68 $&$ -29 $&$  -2 $\\ 
$N(1765) 3/2^+(P_{13})$ &$  78 $&$ -10 $&$  45 $&$ -14 $&$ -55 $&$  4  $&$ -16 $&$  -2 $\\ 
$N(1509) 3/2^-(D_{13})$ &$   7 $&$  -2 $&$- 35 $&$  -1 $&$ 106 $&$ 4   $&$- 5  $&$   8 $\\ 
$N(1703) 3/2^-(D_{13})$ &$  20 $&$ -28 $&$ -22 $&$ -63 $&$  54 $&$ -61 $&$ -24 $&$ -48 $\\ 
$N(1651) 5/2^-(D_{15})$ &$  42 $&$   4 $&$- 34 $&$   1 $&$  44 $&$ -8  $&$   6 $&$ -37 $\\ 
$N(1665) 5/2^+(F_{15})$ &$ -39 $&$ -11 $&$  -5 $&$  -8 $&$  58 $&$- 3  $&$   2 $&$  18 $\\ 
\end{tabular}
\end{ruledtabular}
\end{table}

\clearpage

\begin{acknowledgments}
This work is  supported
by the U.S. Department of Energy, Office of Science, Office of Nuclear Physics, Contract No. DE-AC02-06CH11357.

\end{acknowledgments}

\appendix

\section{Total Cross sections of $\pi N$ reactions}
\begin{figure}[htbp] \vspace{-0.cm}
\begin{center}
\includegraphics[width=0.6\textwidth]{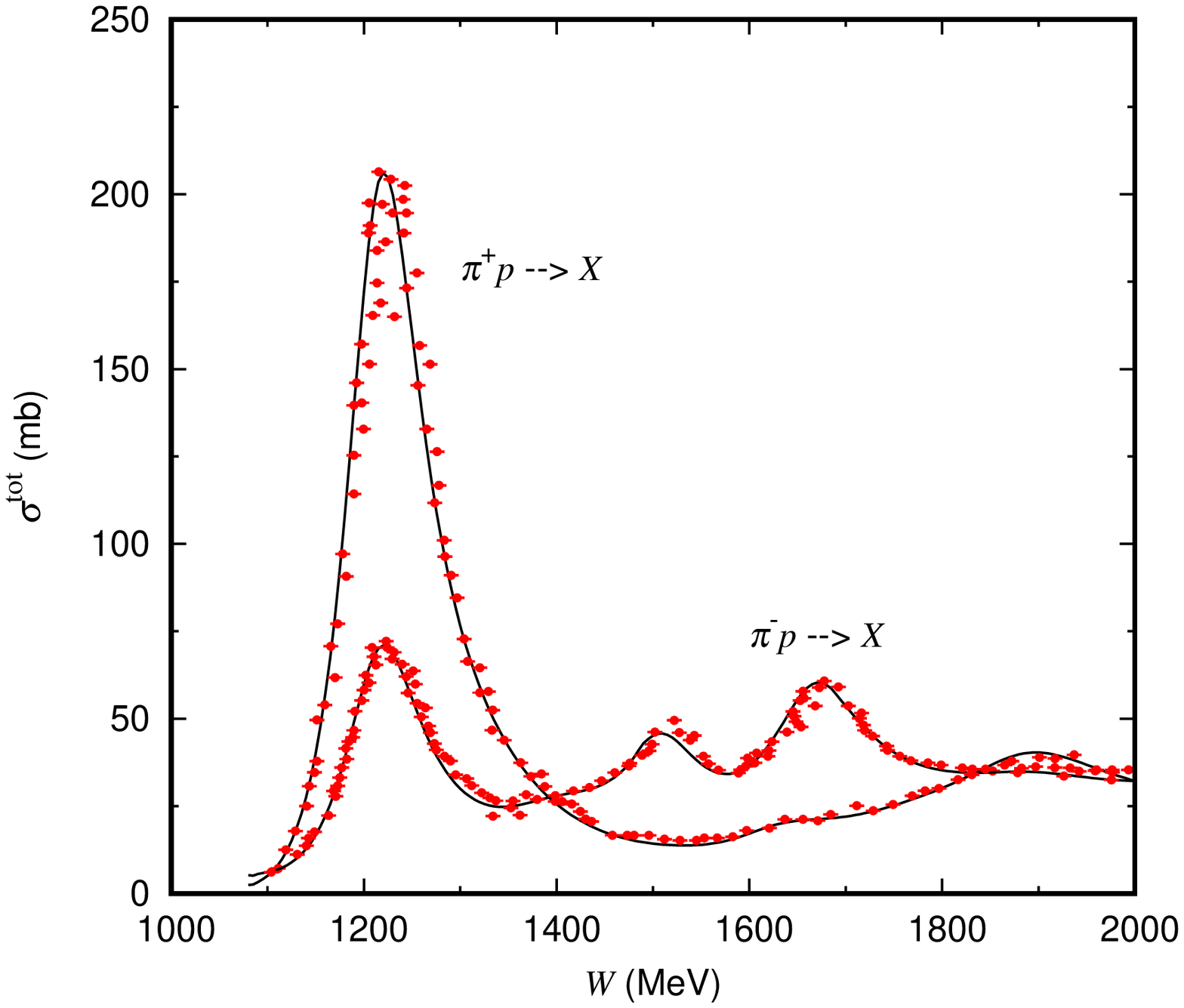}
\includegraphics[width=0.4\textwidth]{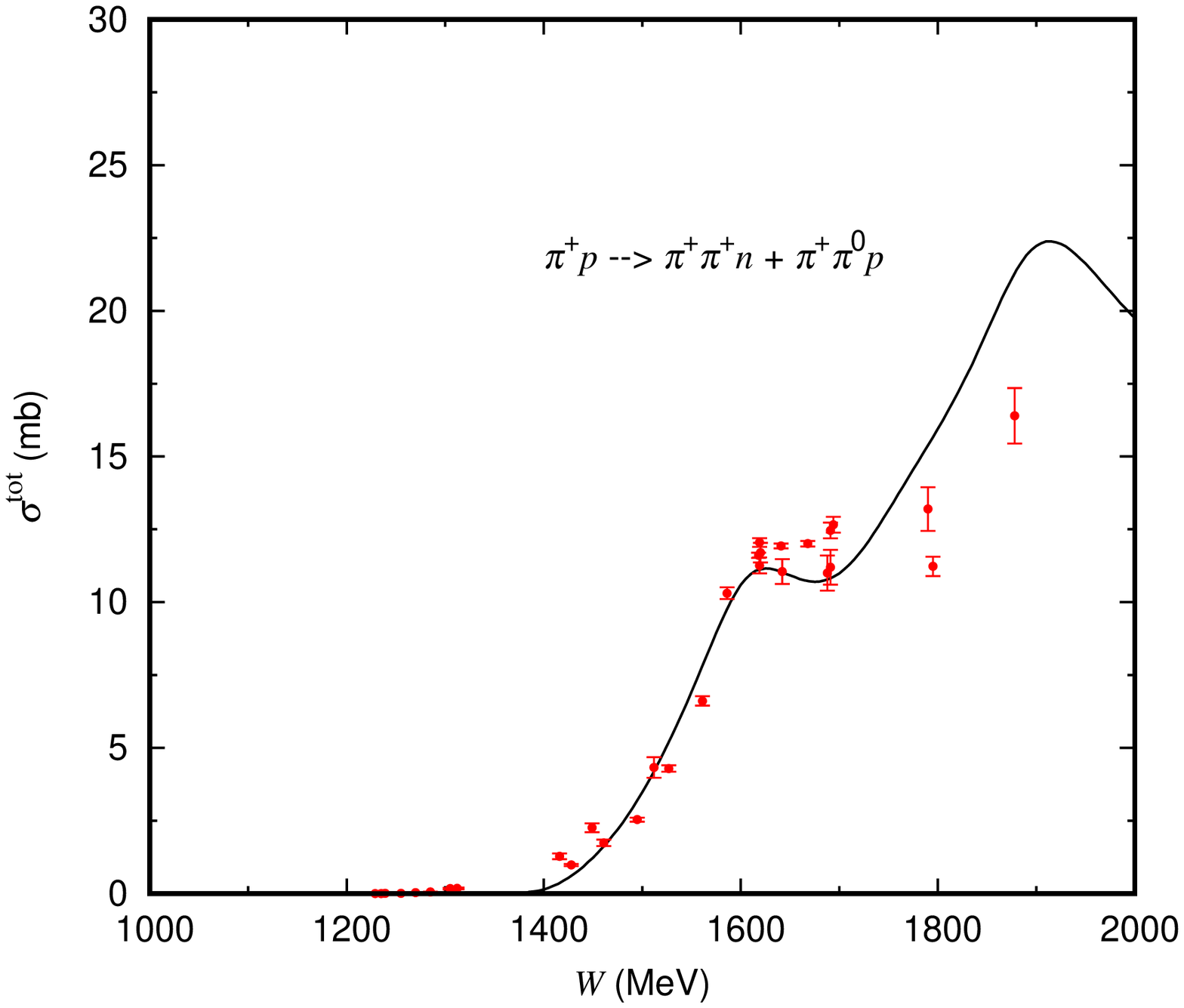}
\includegraphics[width=0.4\textwidth]{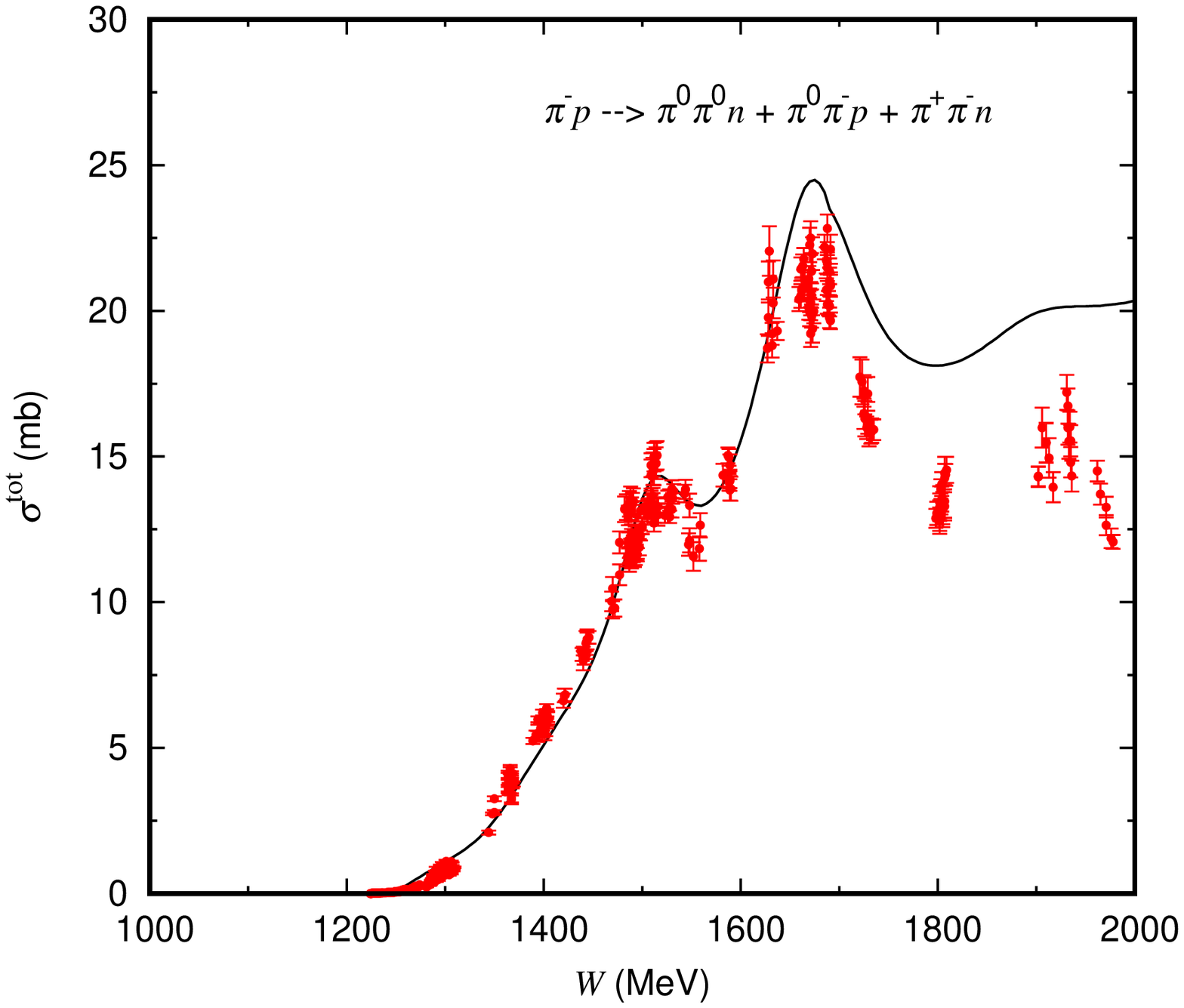}
\caption{$\sigma^{tot}$ of $\pi^{\pm} p\rightarrow X,\pi\pi N$ 
}
 \label{fig:pip-x}
\end{center}
\end{figure}

\clearpage

\begin{figure}[htbp] \vspace{-0.cm}
\begin{center}
\includegraphics[width=0.4\textwidth]{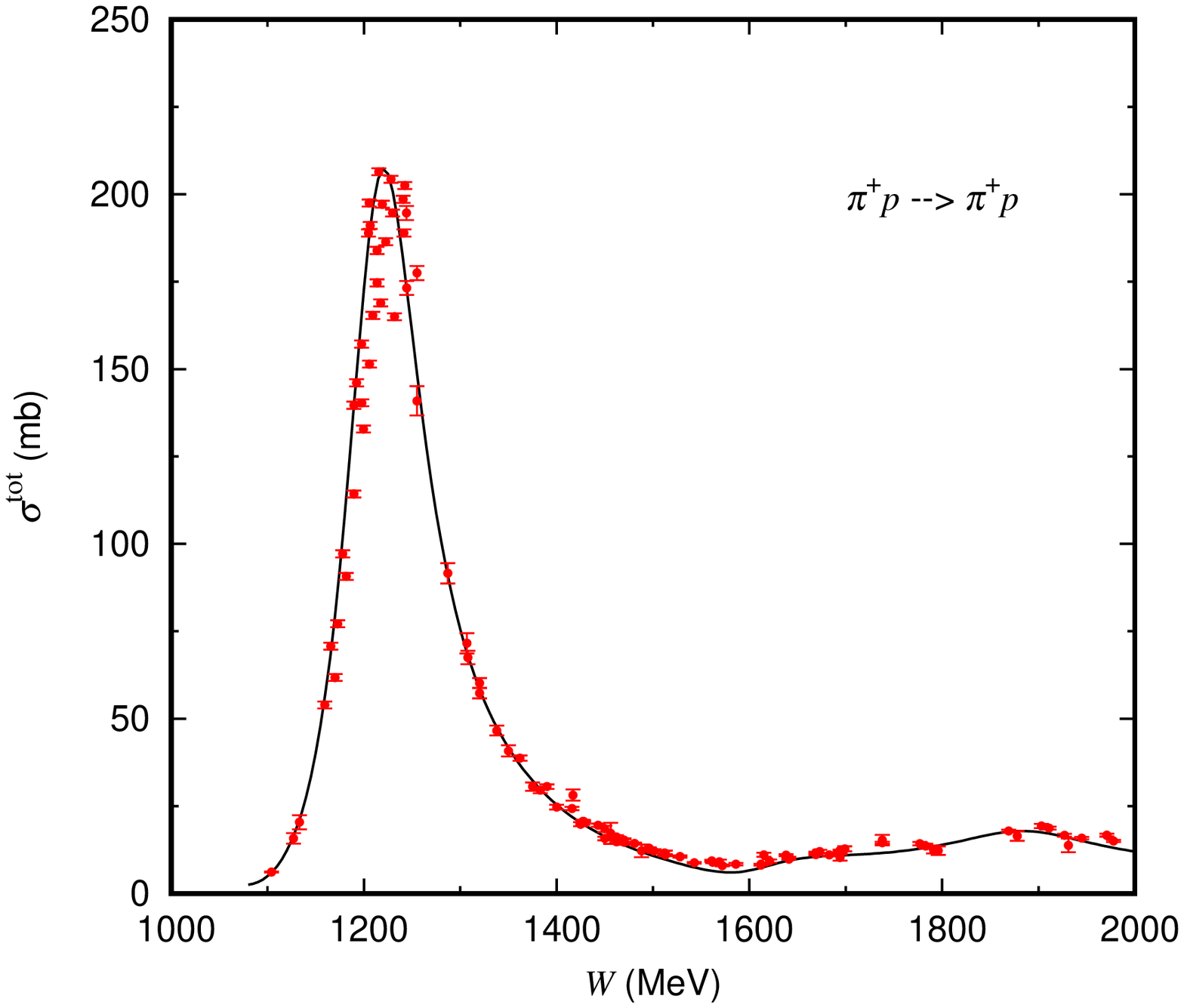}
\includegraphics[width=0.4\textwidth]{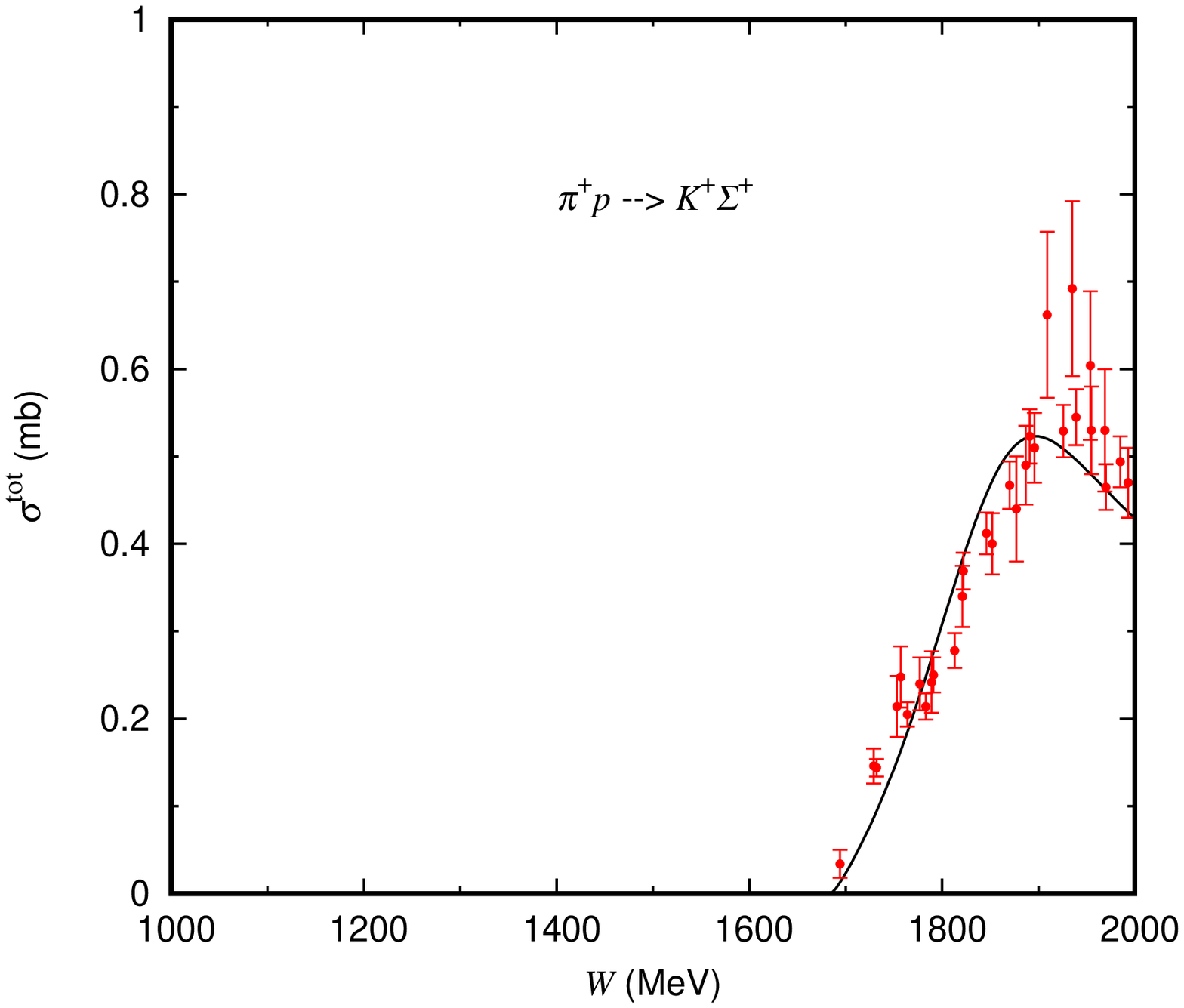}
\includegraphics[width=0.4\textwidth]{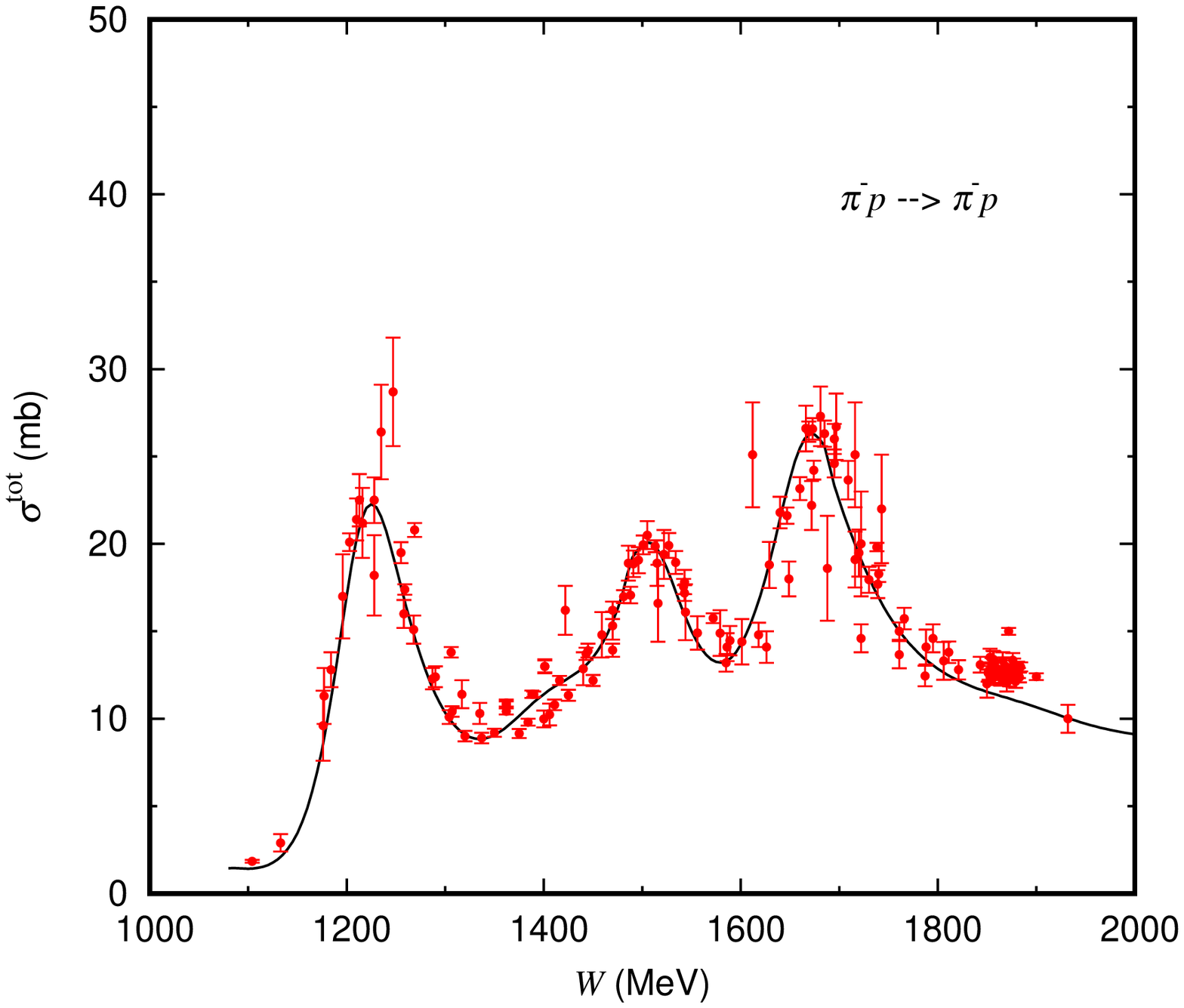}
\includegraphics[width=0.4\textwidth]{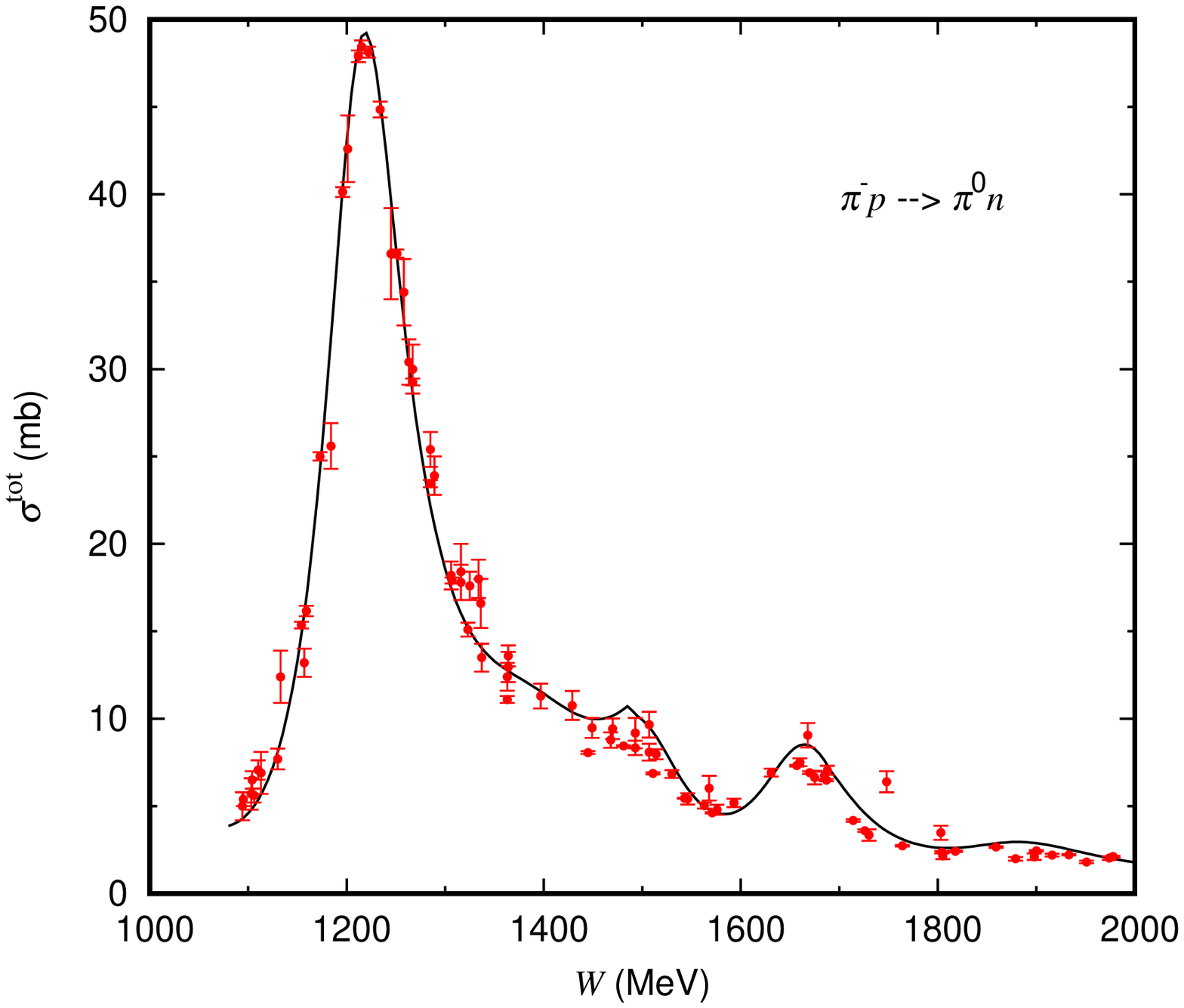}
\includegraphics[width=0.4\textwidth]{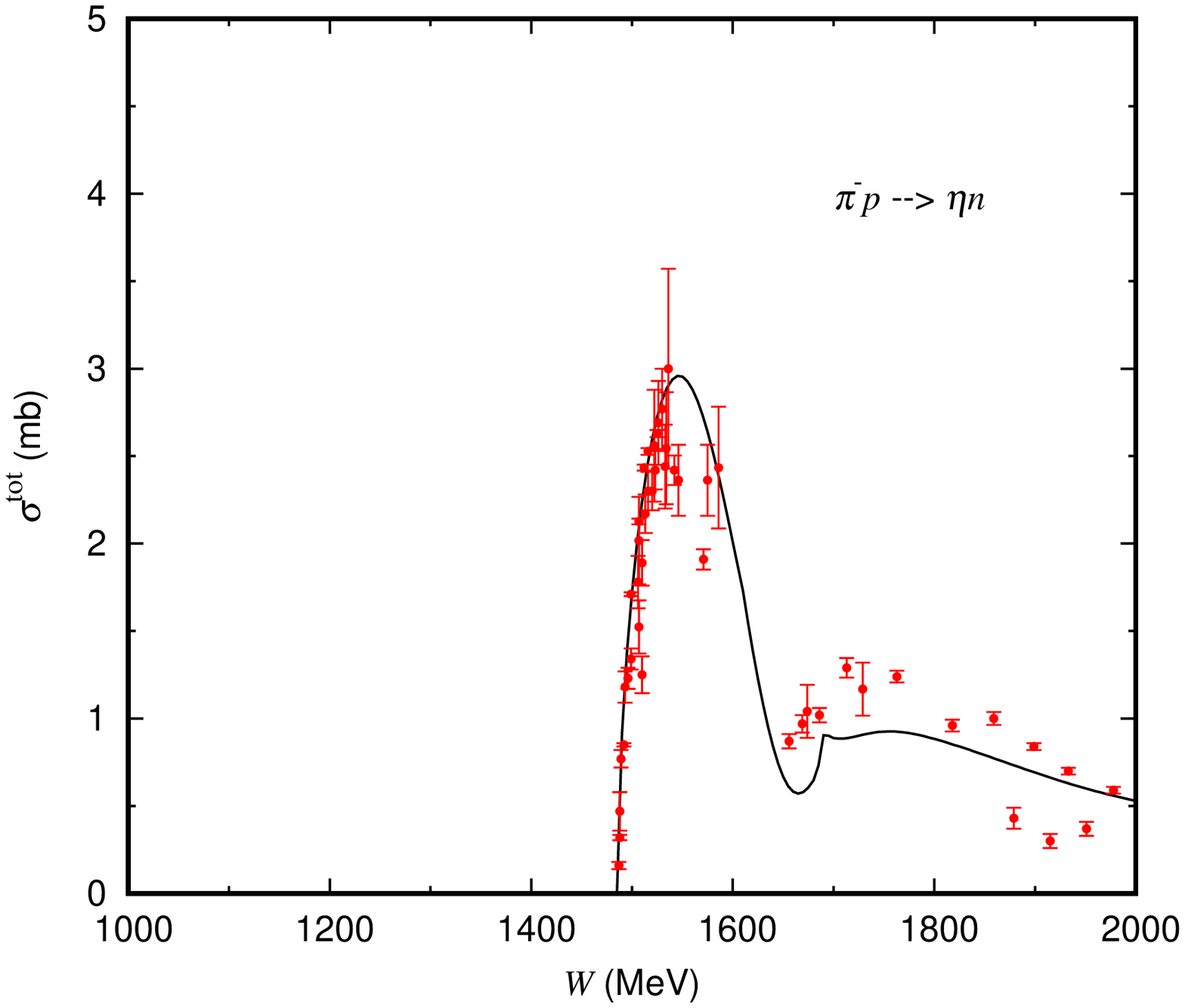}
\includegraphics[width=0.4\textwidth]{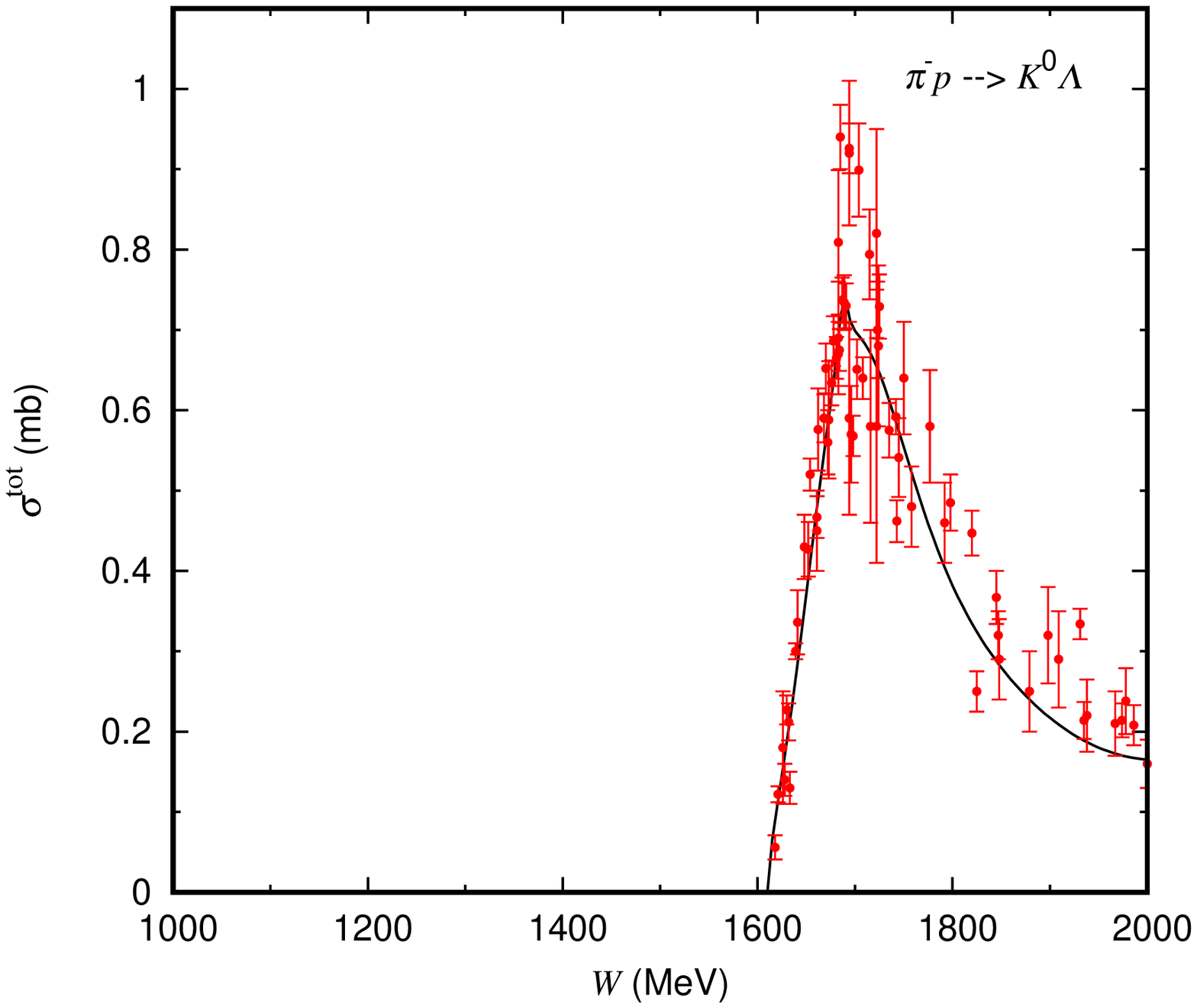}
\includegraphics[width=0.4\textwidth]{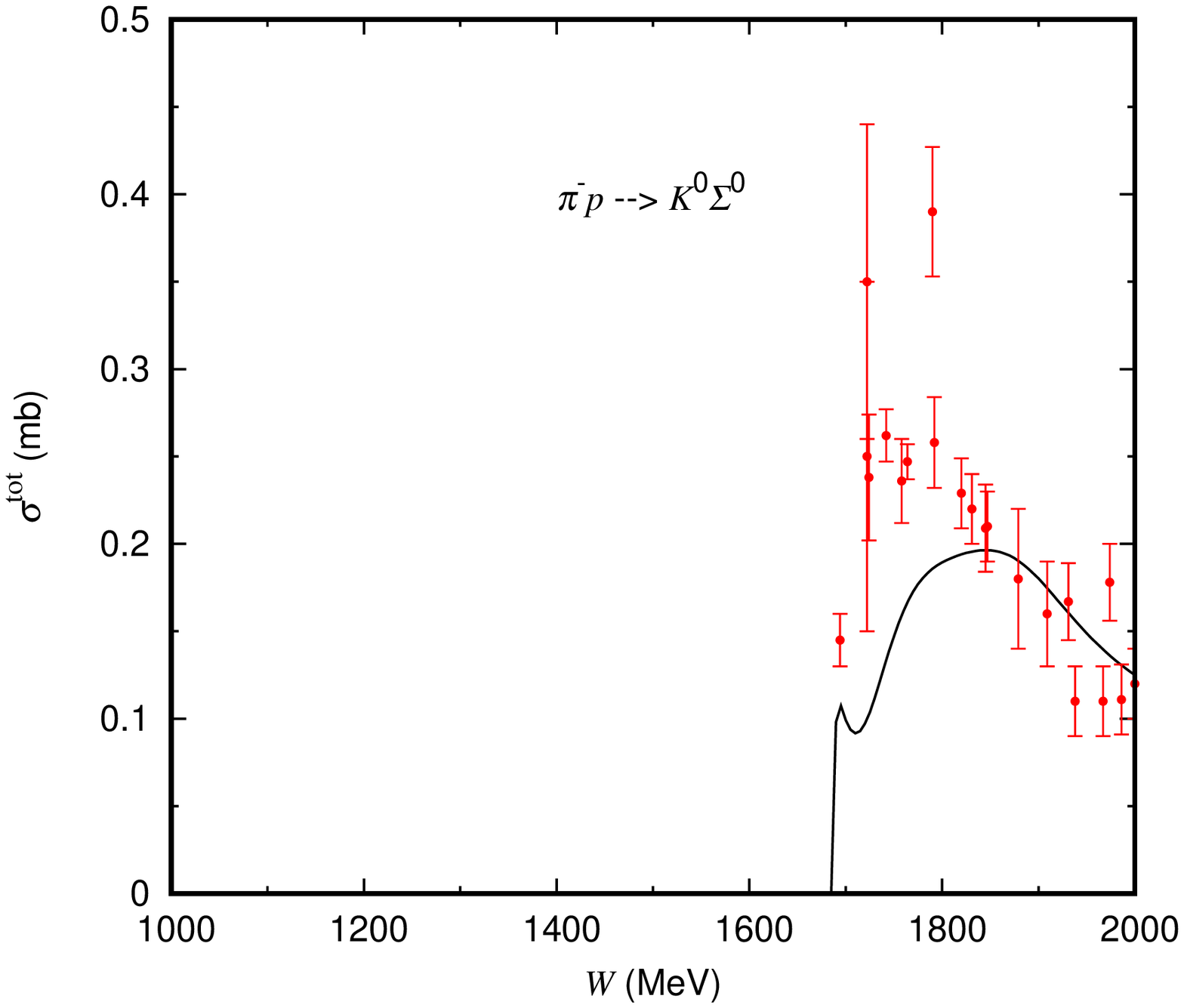}
\includegraphics[width=0.4\textwidth]{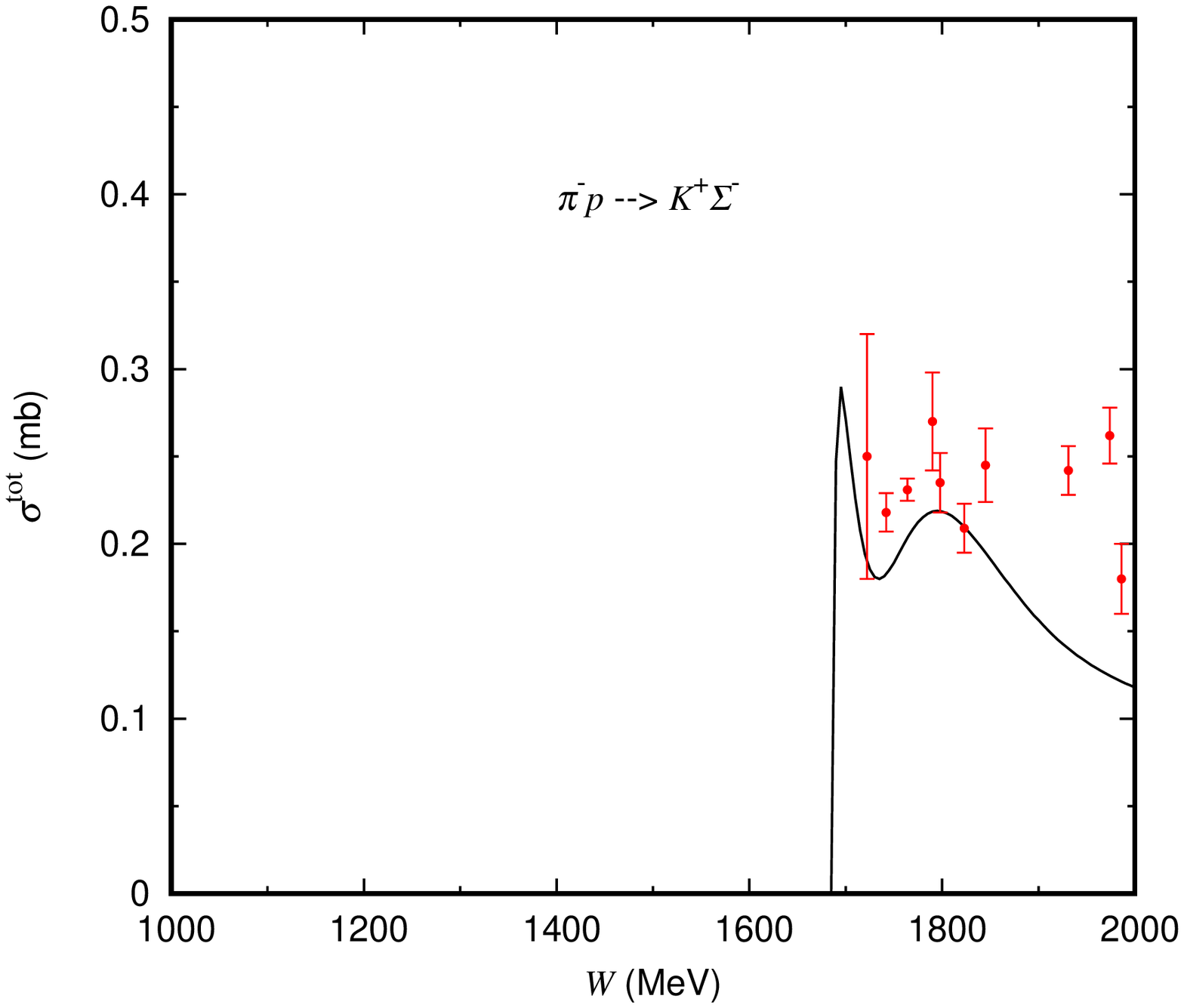}
\caption{$\sigma^{tot}$ of $\pi^\pm p\rightarrow \pi N, \eta N, K\Lambda, K\Sigma$}
 \label{fig:pimp-mb-1}
\end{center}
\end{figure}

\clearpage

\begin{figure}[h]
\includegraphics[clip,width=0.4\textwidth]{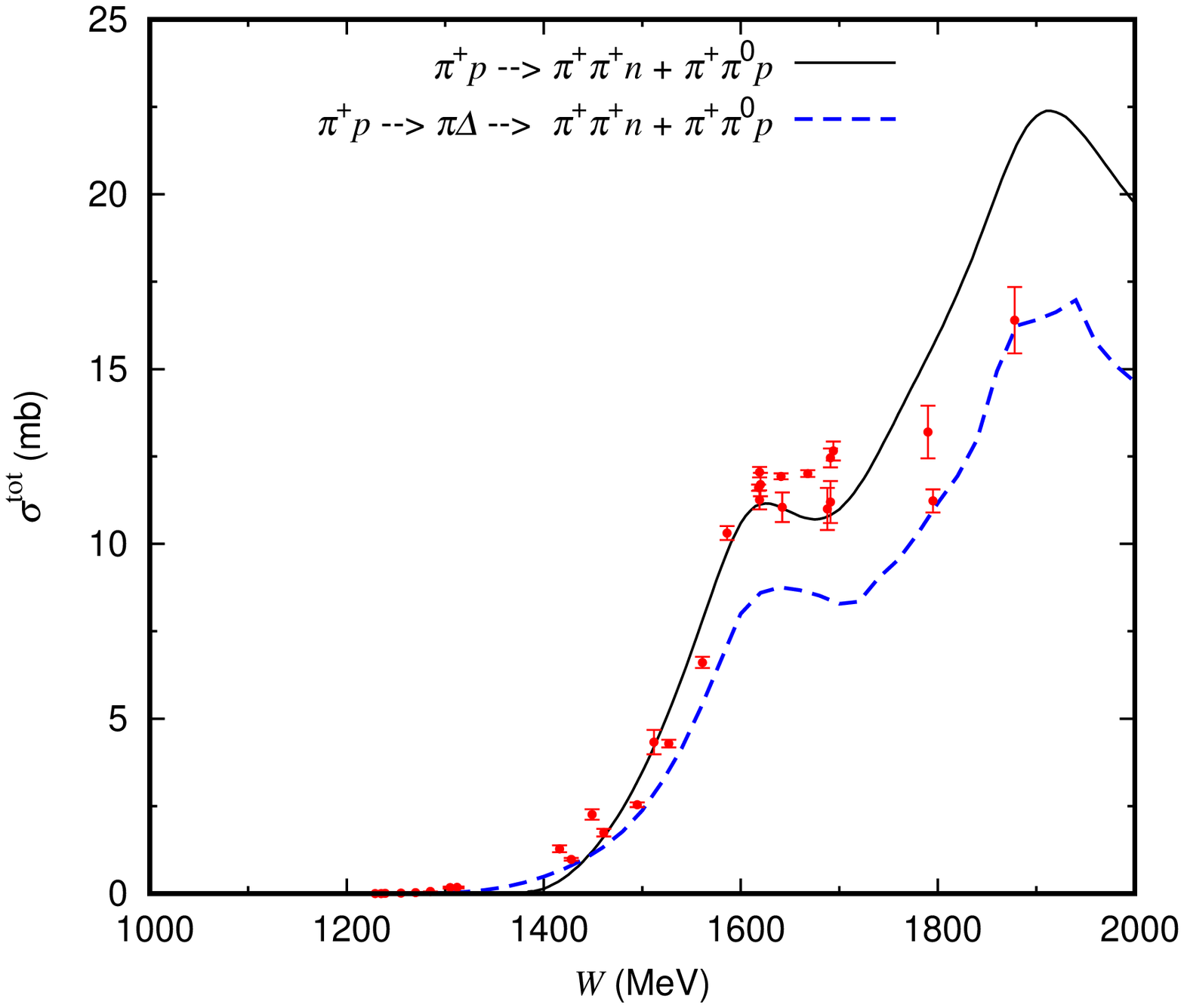}
\includegraphics[clip,width=0.4\textwidth]{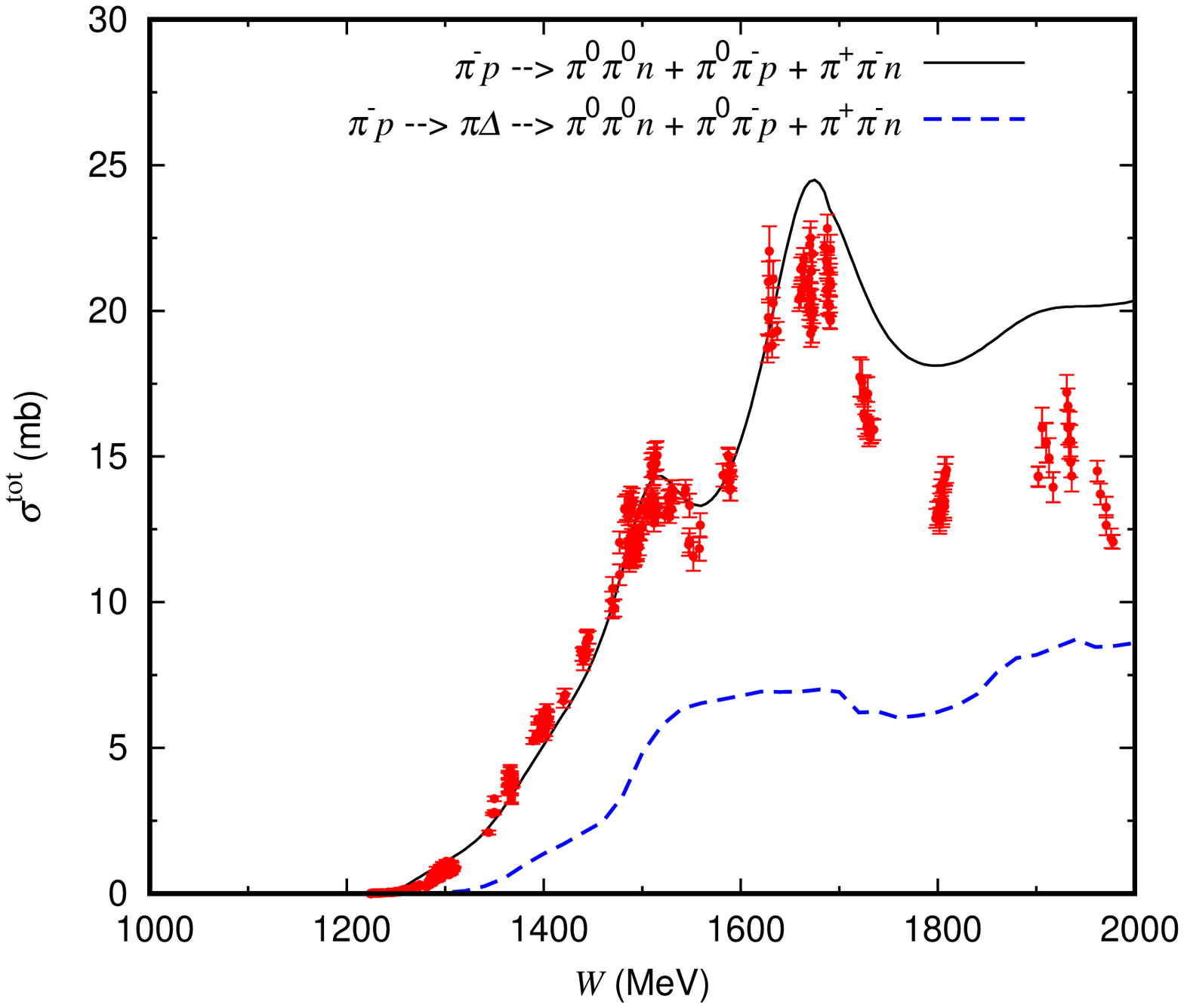}
\includegraphics[clip,width=0.4\textwidth]{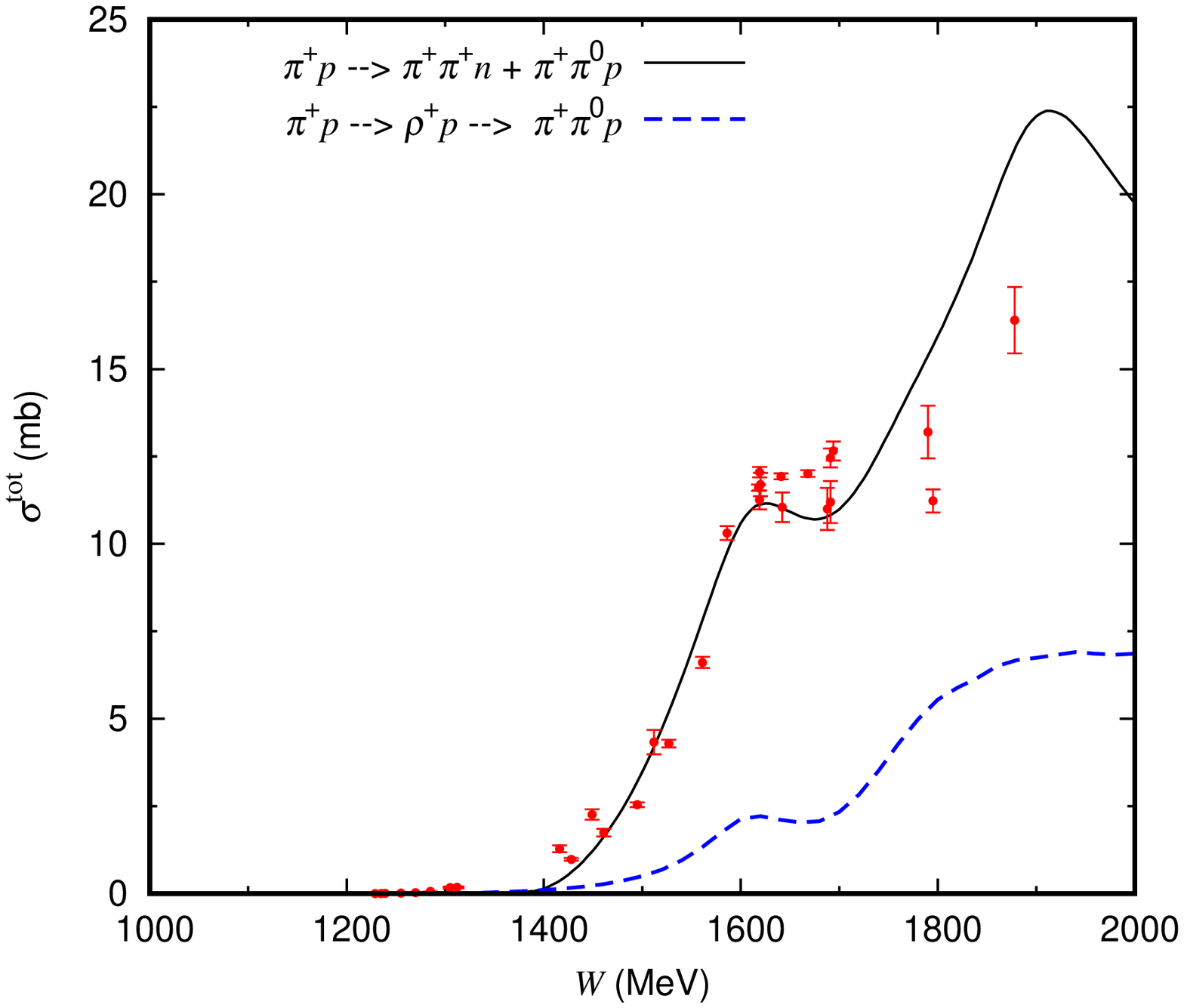}
\includegraphics[clip,width=0.4\textwidth]{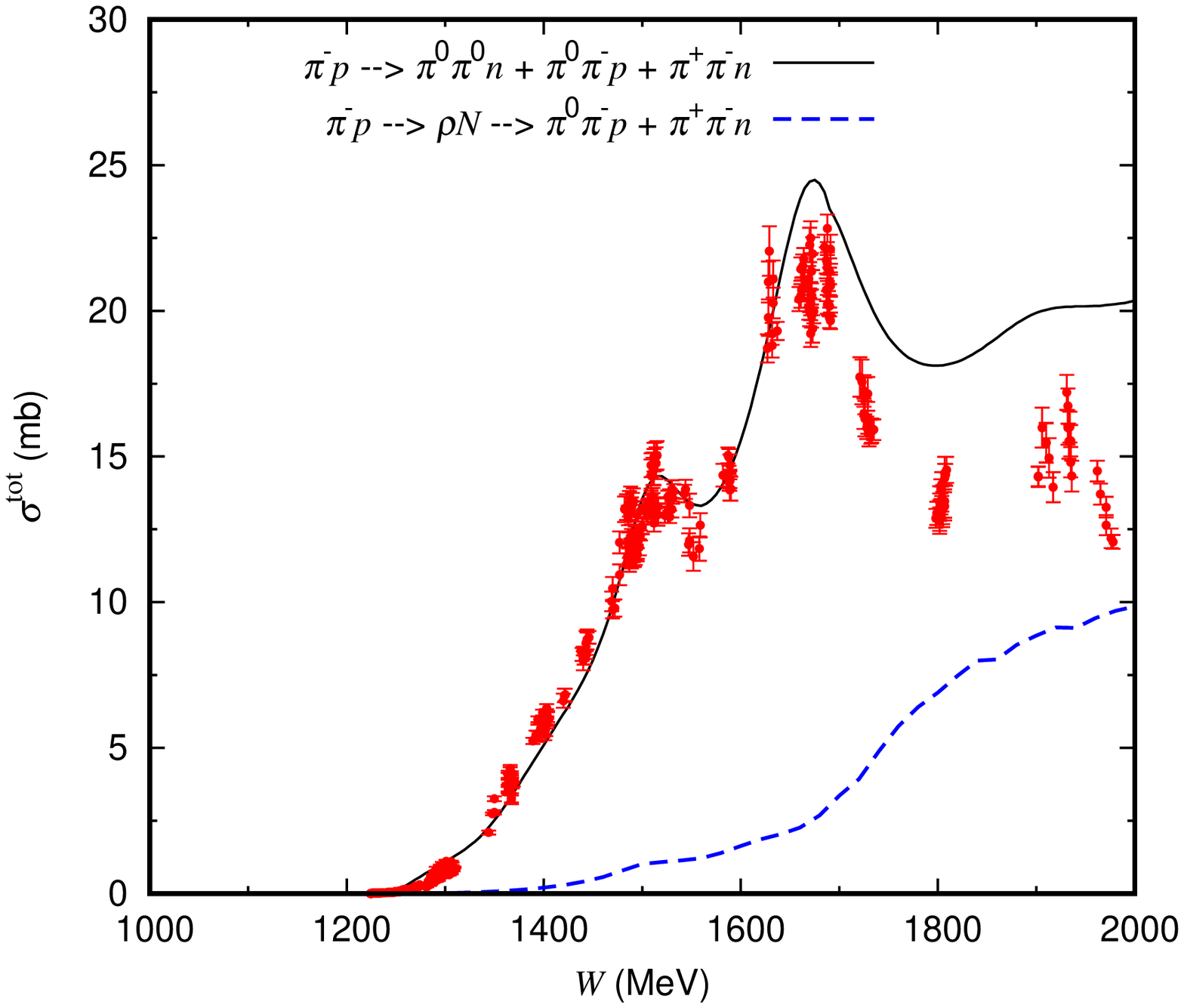}
\includegraphics[clip,width=0.4\textwidth]{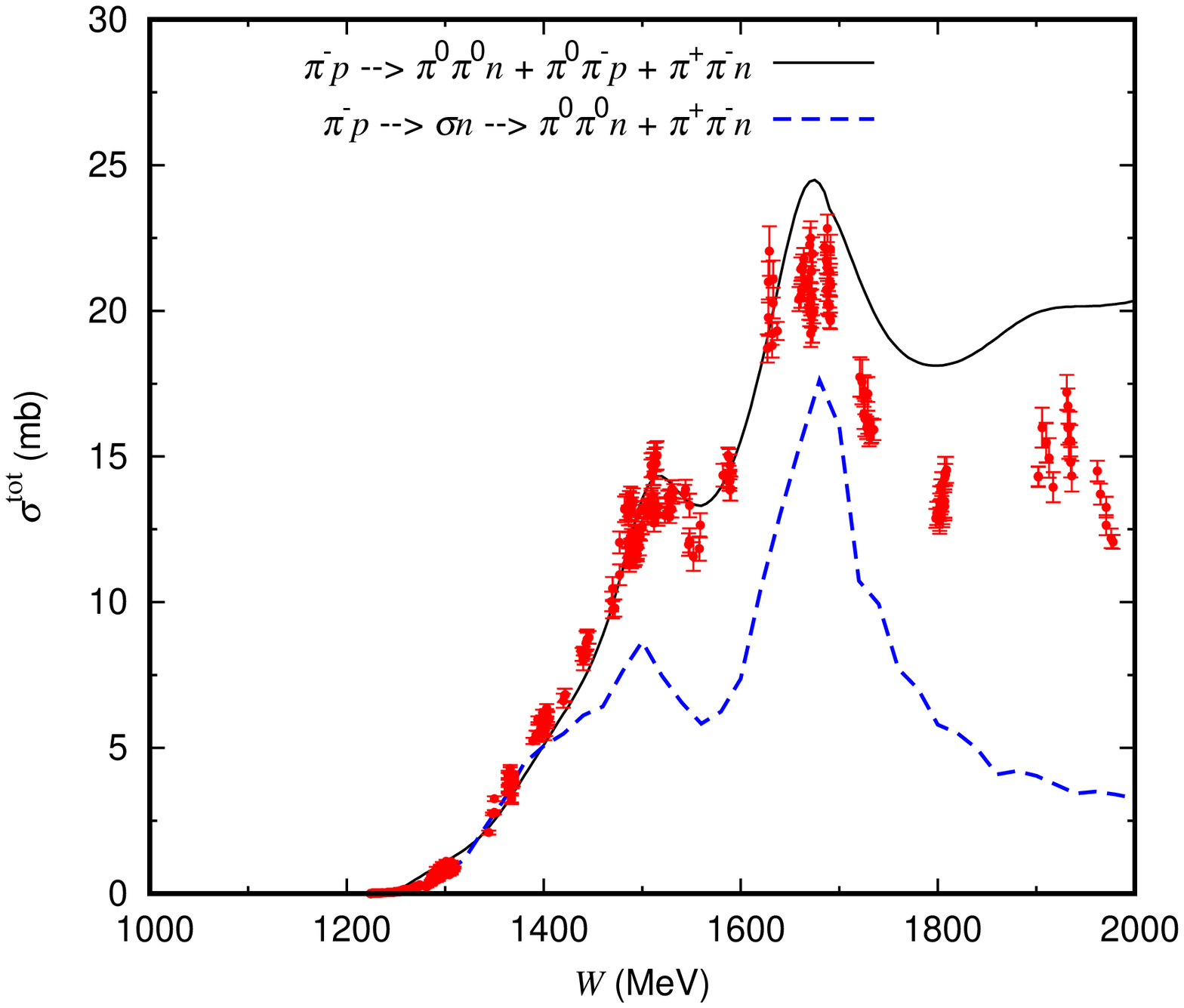}
\caption{ Total cross sections  for  $\pi^\pm p\rightarrow (MB)^R \rightarrow \pi\pi N$ (dashed curves)
are compared with $\pi^\pm p \rightarrow \pi\pi N$(solid curves).
Top row: $(MB)^R=\pi\Delta$; middle row: $(MB)^R=\rho N$; bottom row:$(MB)^R=\sigma N$.
}
\label{fig:pipp-2pi}
\end{figure}

\clearpage
\section{Total Cross sections of $\gamma N$ reactions}

\begin{figure}[h] \vspace{0.1cm}
\begin{center}
\includegraphics[width=0.6\textwidth]{gptot.eps}
\includegraphics[width=0.4\textwidth]{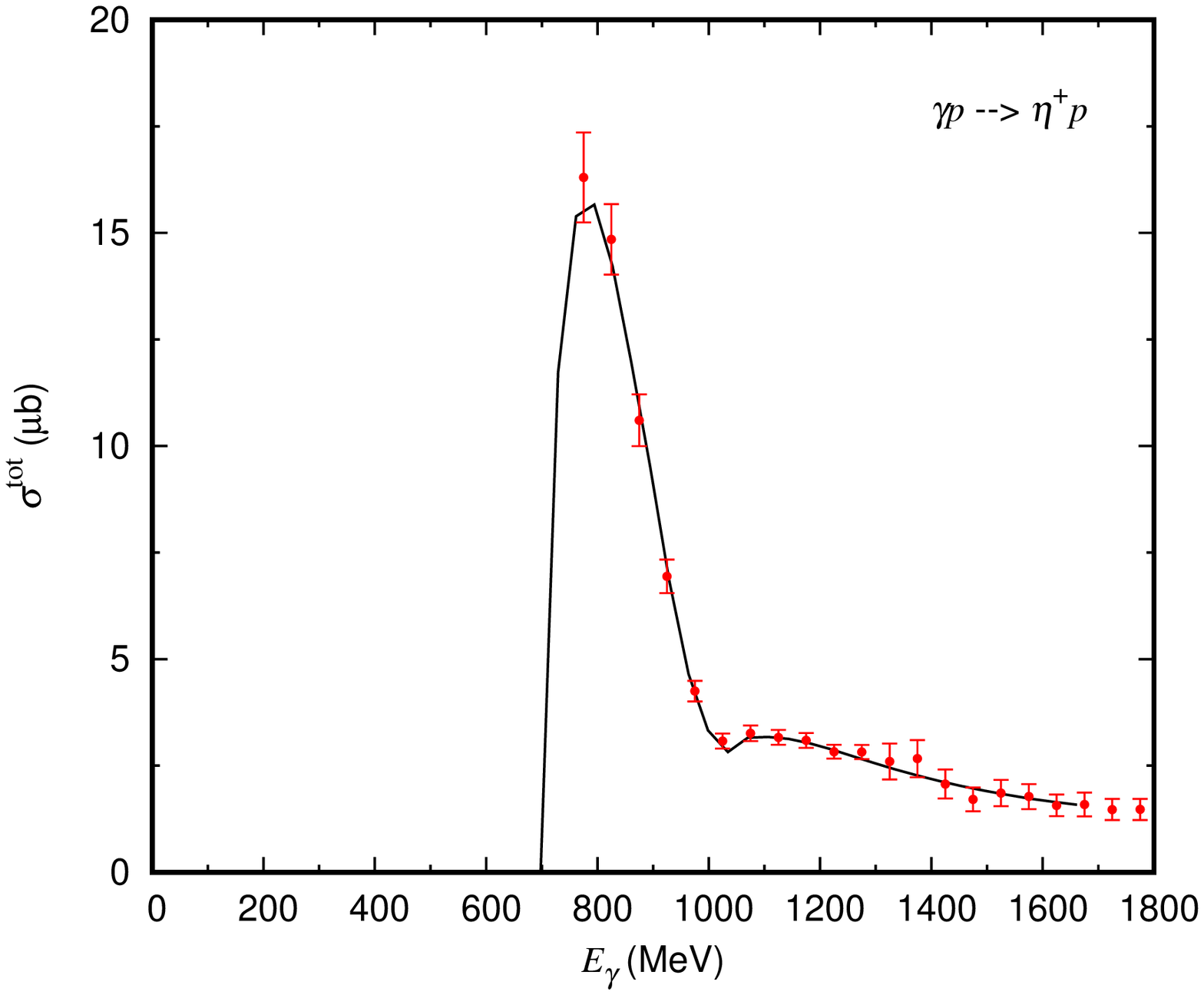}
\includegraphics[width=0.4\textwidth]{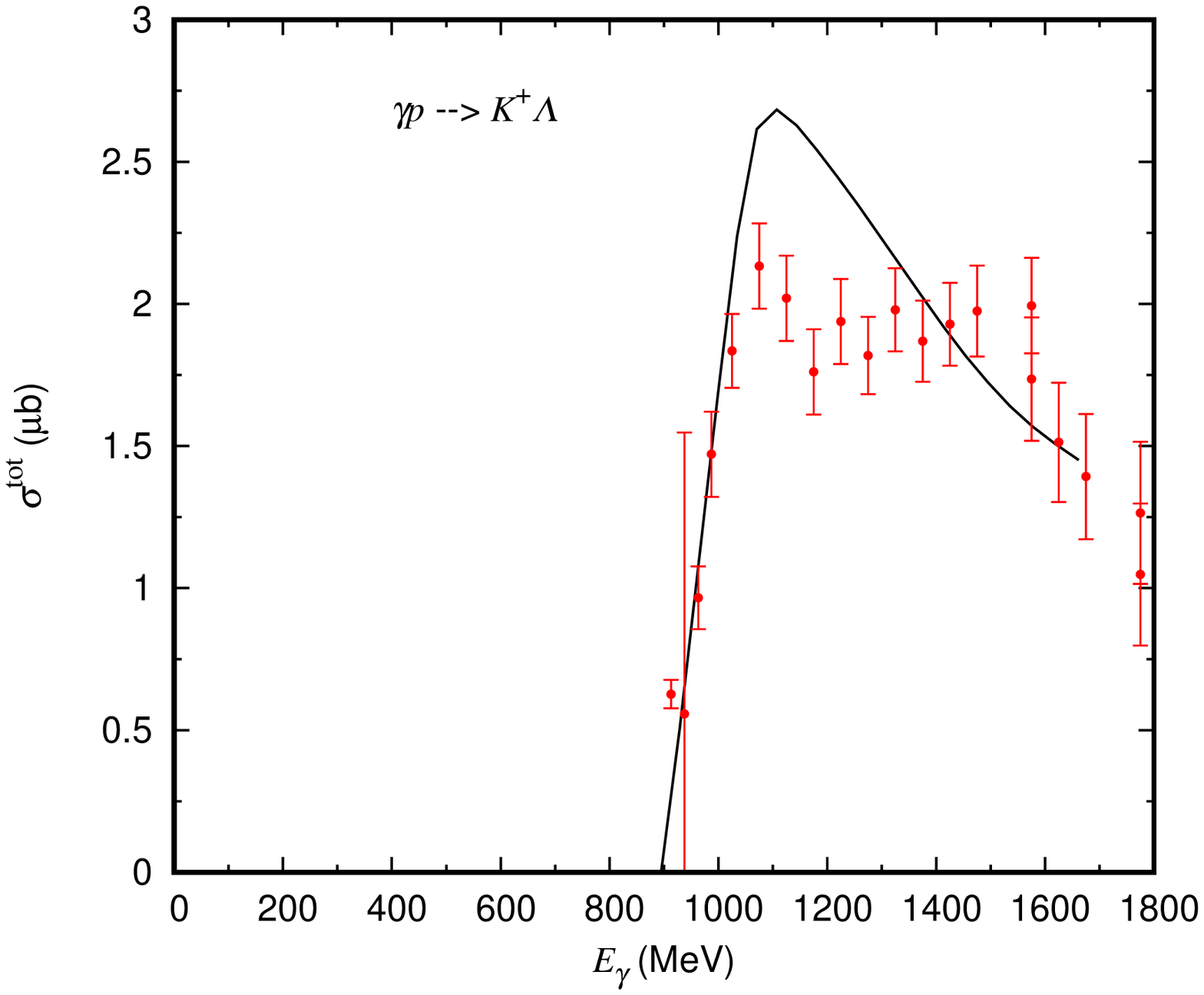}
\includegraphics[width=0.4\textwidth]{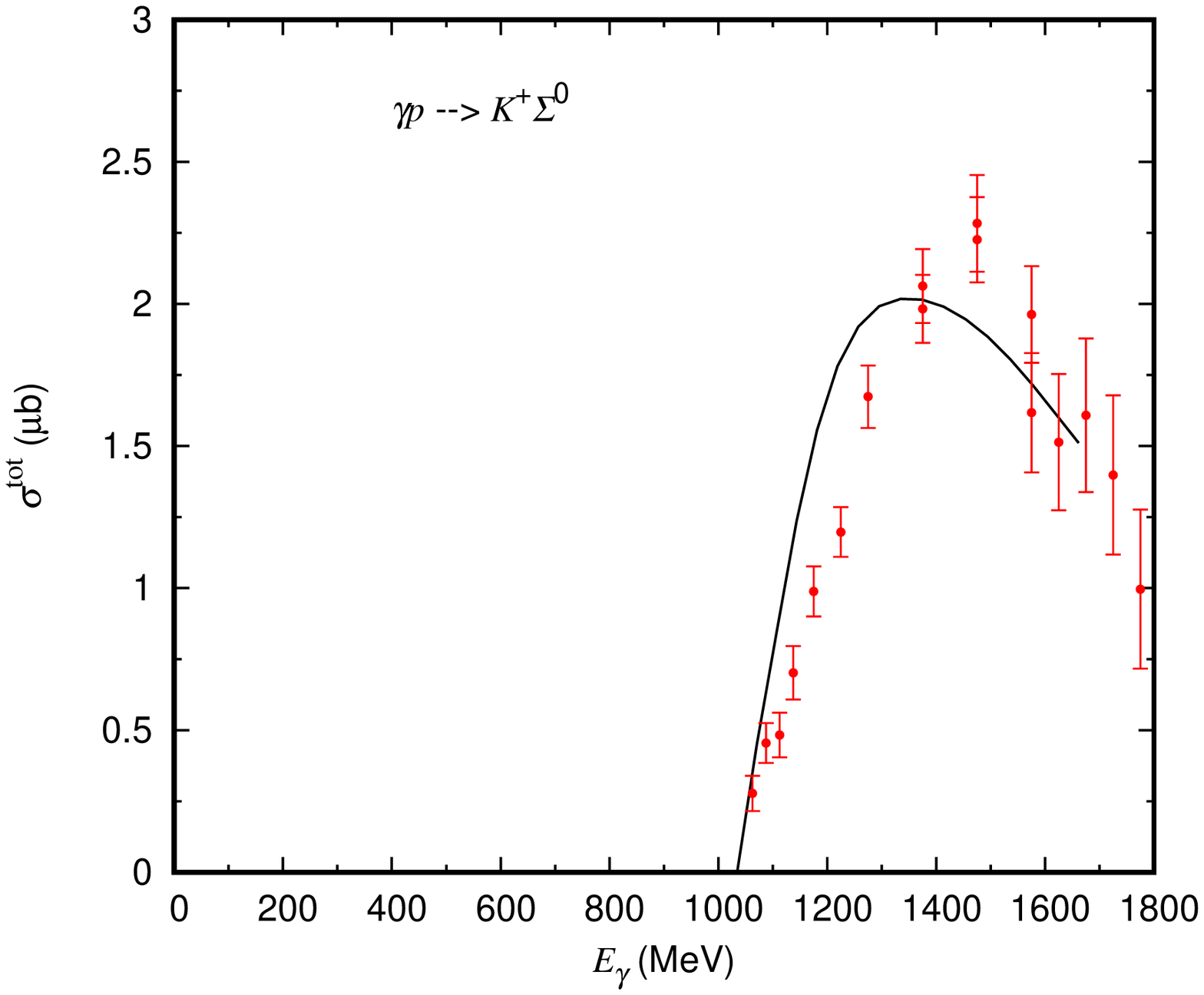}
\includegraphics[width=0.4\textwidth]{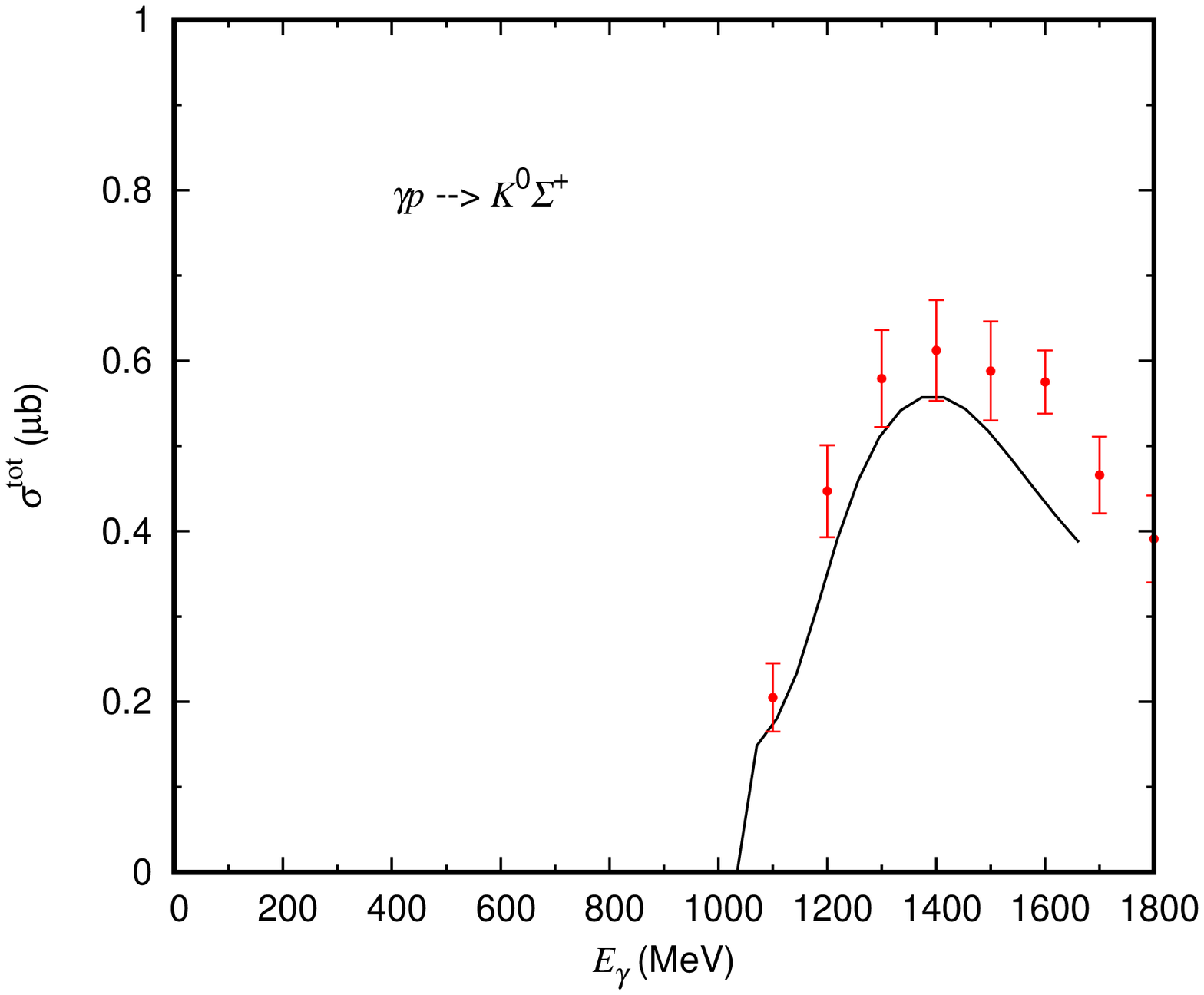}
\caption{
Top :$\gamma p \rightarrow X,\pi^0p,\pi^+n$, dashed curves are our previous analysis;
Lower four figures: $\gamma p \rightarrow \eta p, K\Lambda, K^+\Sigma^0, K^0\Sigma^+$. }
 \label{fig:gp-x}
\end{center}
\end{figure}

\clearpage
\section{ Differential cross sections of $\pi N$ reactions}
\subsection{$\pi N \rightarrow \pi N$}
\begin{figure}[h]
\includegraphics[clip,width=1.0\textwidth]{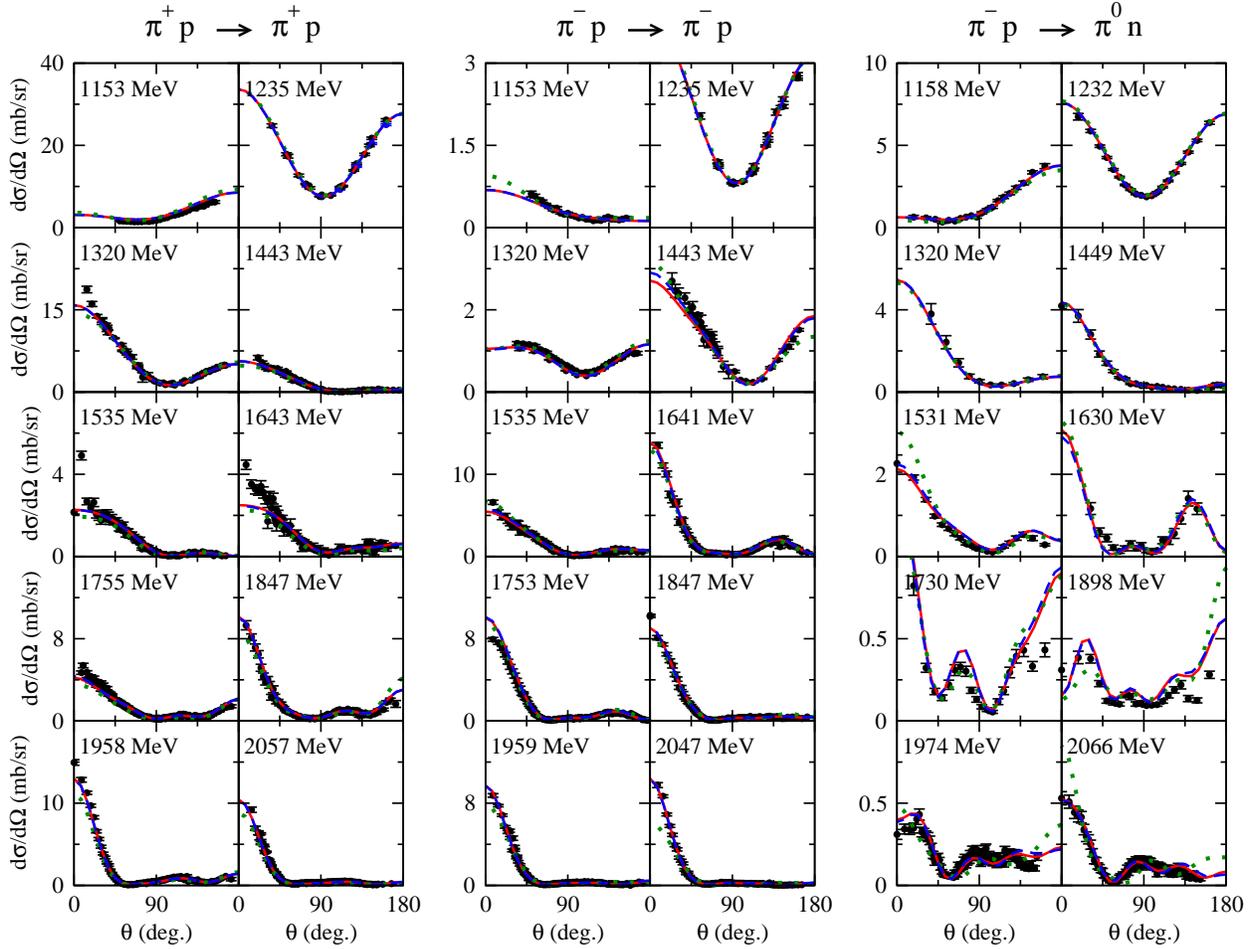}
\caption{Differential cross section for $\pi N \rightarrow \pi N$.
The red solid curves are the current results while the blue dashed curves are
from our previous analysis of 2007.
}
\label{fig:pin-dcs}
\end{figure}
\begin{figure}[h]
\includegraphics[clip,width=1.0\textwidth]{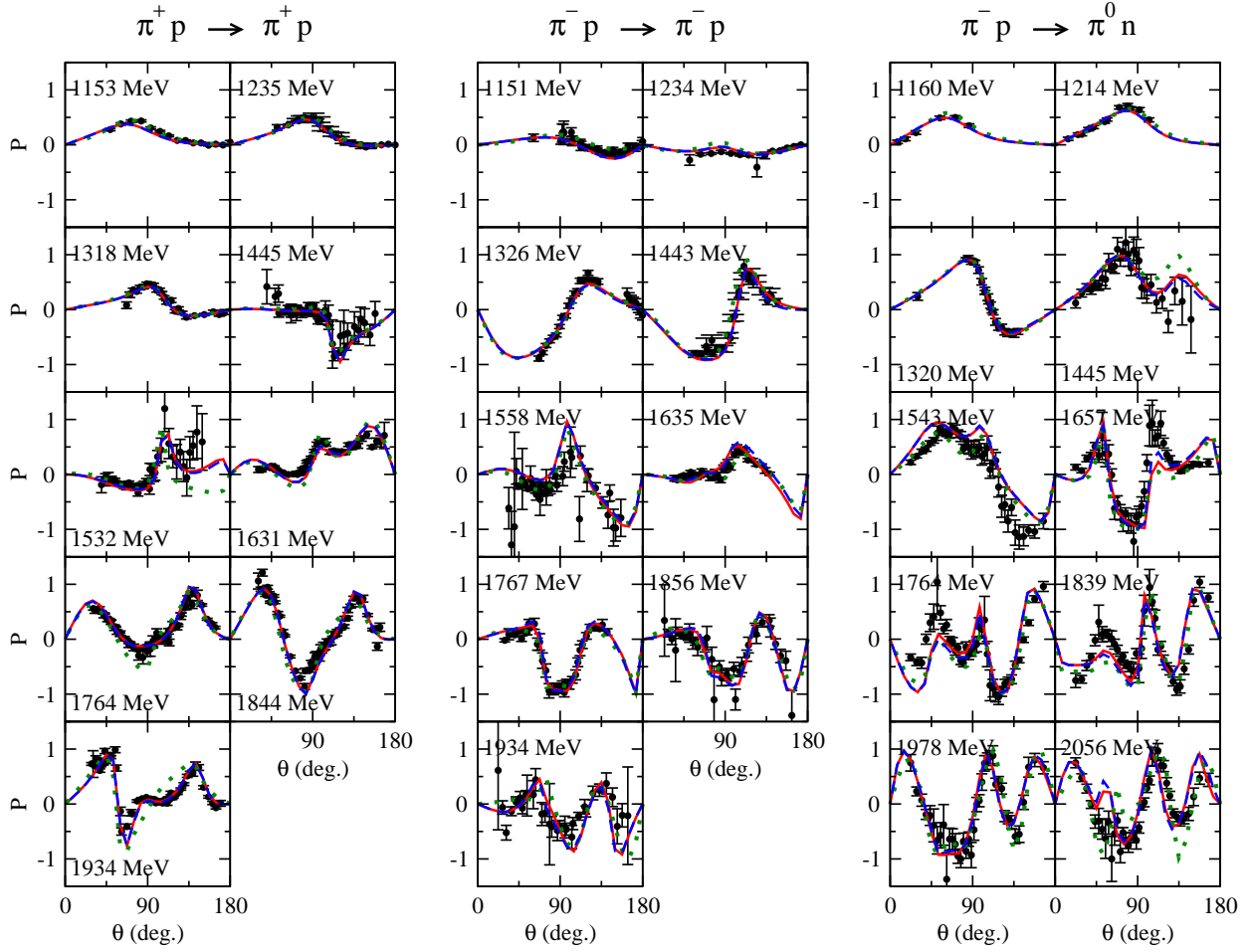}
\caption{\label{fig:pin-p}
The target polarization $P$ of $\pi N \rightarrow \pi N$.
The red solid curves are the current results while the blue dashed curves are
from our previous analysis of 2007.
}
\end{figure}
\clearpage
\subsection{$\pi N \rightarrow \eta N$}
\begin{figure}[h]
\includegraphics[clip,width=0.80\textwidth]{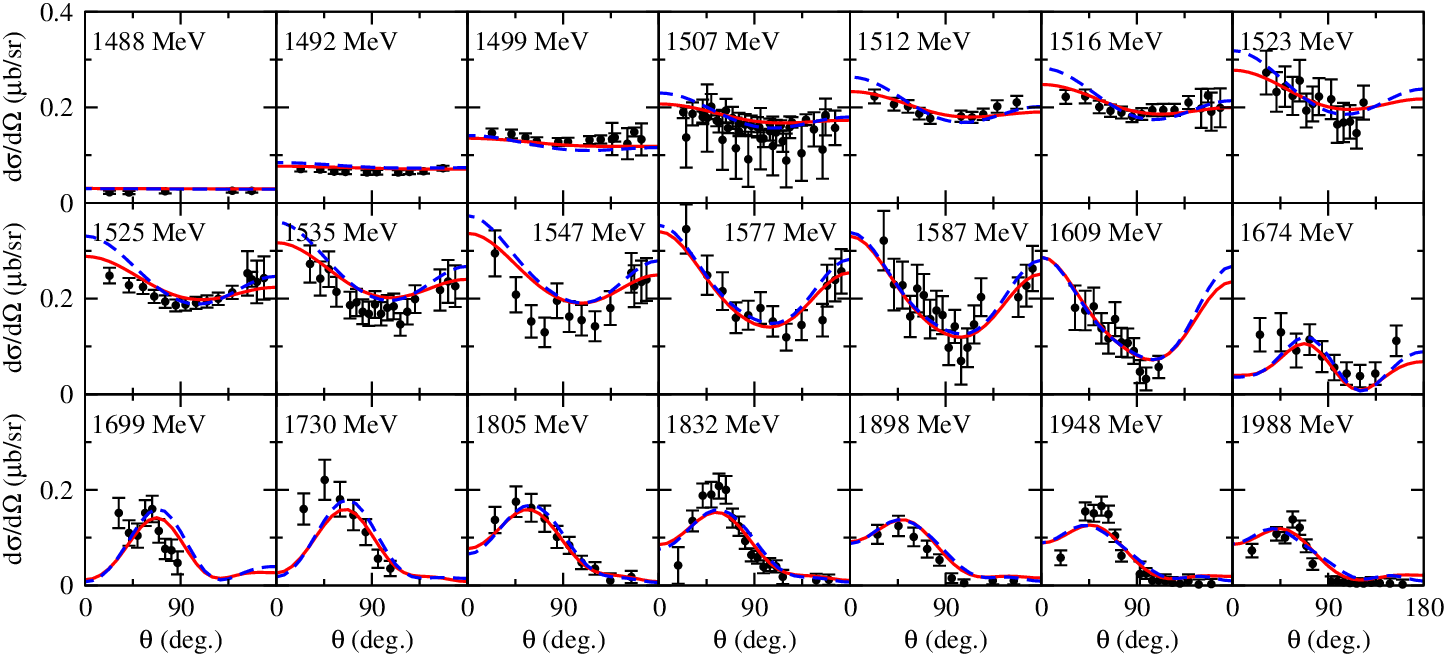}
\caption{Differential cross sections of $\pi^- p\to \eta n$.
The red solid curves are the current results while the blue dashed curves are
from our previous analysis of 2007.}
\label{fig:pin-etan-dcs}
\end{figure}

\subsection{$\pi N \rightarrow K\Lambda$}
\begin{figure}[h]
\includegraphics[clip,width=0.80\textwidth]{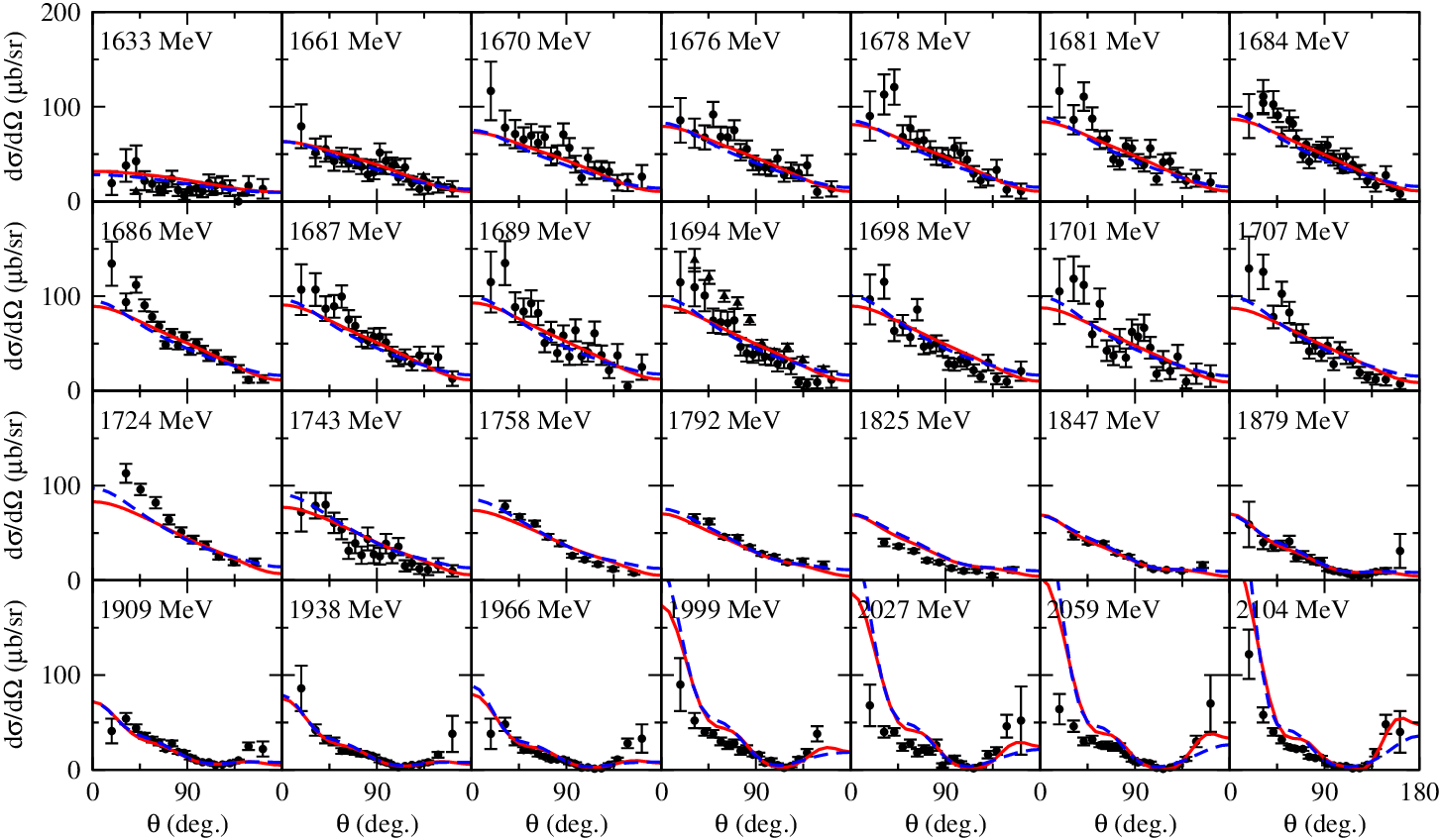}
\caption{Differential cross sections of $\pi^- p \rightarrow K^0\Lambda^0$.
The red solid curves are the current results while the blue dashed curves are
from our previous analysis of 2007.}
\label{fig:pin-k0k0-dcs}
\end{figure}
\begin{figure}[h]
\includegraphics[clip,width=0.80\textwidth]{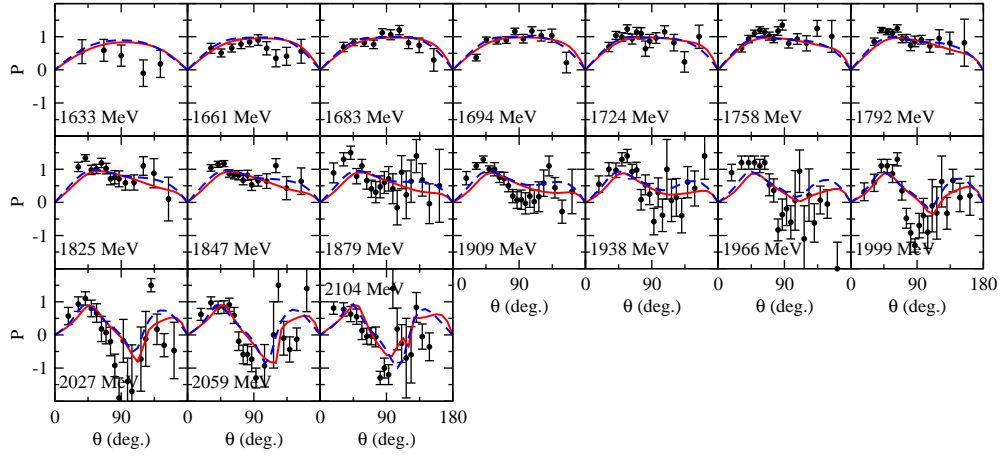}
\caption{Differential cross sections of $\pi^- p \rightarrow K^0\Lambda^0$.
The red solid curves are the current results while the blue dashed curves are
from our previous analysis of 2007.}
\label{fig:pin-k0k0-p}
\end{figure}

\subsection{$\pi N \rightarrow K\Sigma$}
\begin{figure}[h]
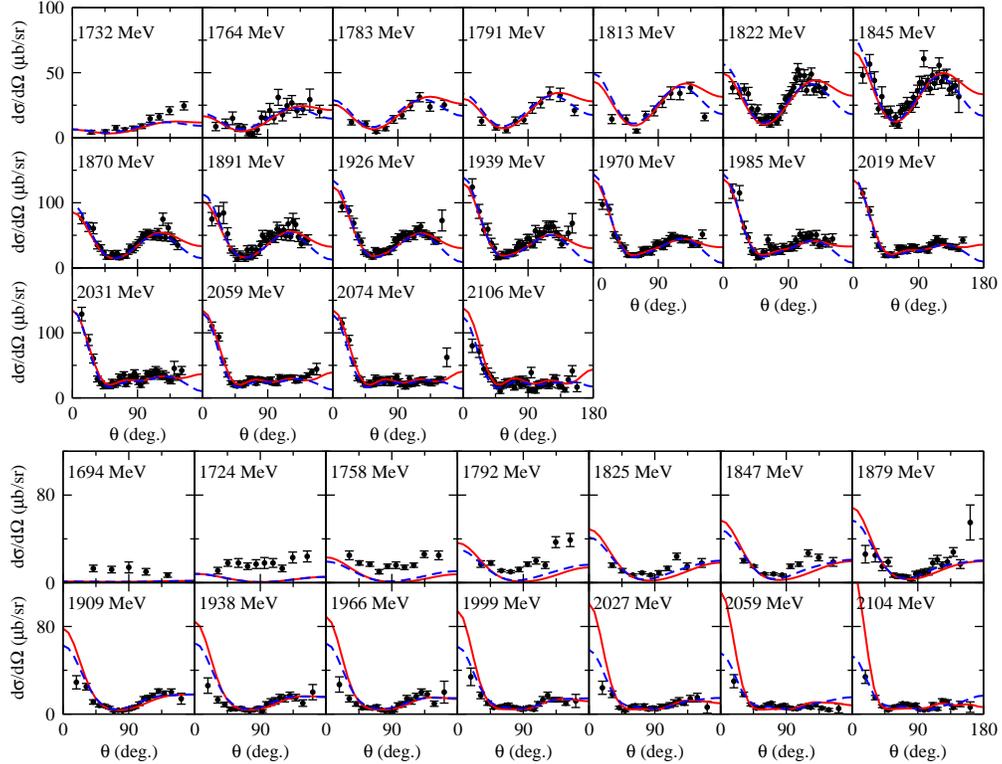

\includegraphics[clip,width=0.80\textwidth]{kpsp-DC}
\includegraphics[clip,width=0.80\textwidth]{k0s0-DC}
\caption{Differential cross sections
of $\pi^+ p \rightarrow K^+\Sigma^+$(upper)
and $\pi^- p \rightarrow K^0\Sigma^0$(lower).
The red solid curves are the current results while the blue dashed curves are
from our previous analysis of 2007.}
\label{fig:pin-pks-dcs}
\end{figure}
\begin{figure}[h]
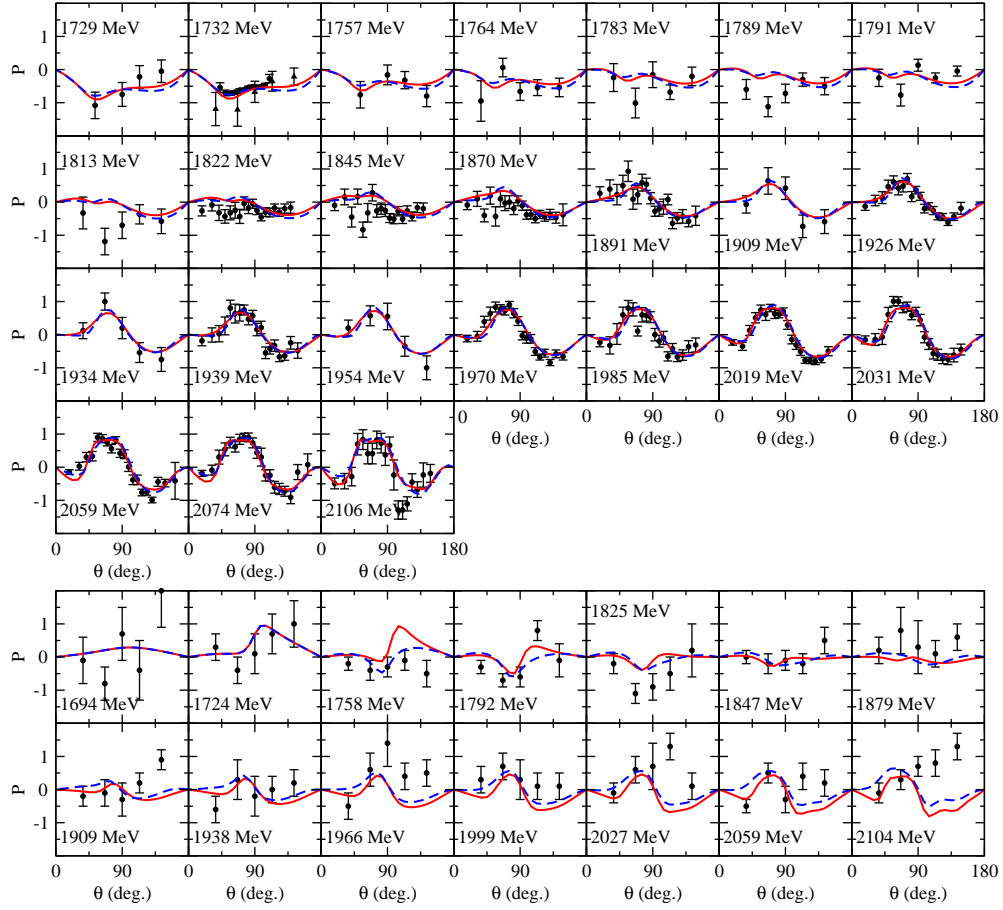

\includegraphics[clip,width=0.80\textwidth]{kpsp-P}
\includegraphics[clip,width=0.80\textwidth]{k0s0-P}
\caption{Polarization $P$
of $\pi^+ p \rightarrow K^+\Sigma^+$(upper)
and $\pi^- p \rightarrow K^0\Sigma^0$(lower).
The red solid curves are the current results while the blue dashed curves are
from our previous analysis of 2007.}
\label{fig:pin-pks-p}
\end{figure}

\clearpage

\section{Differential Cross sections of $\gamma N \rightarrow \pi N, \eta N, K\Lambda, K\Sigma$}
\subsection{$\gamma p \rightarrow \pi^0 p$}

\begin{figure}[h]
\includegraphics[clip,width=0.67\textwidth]{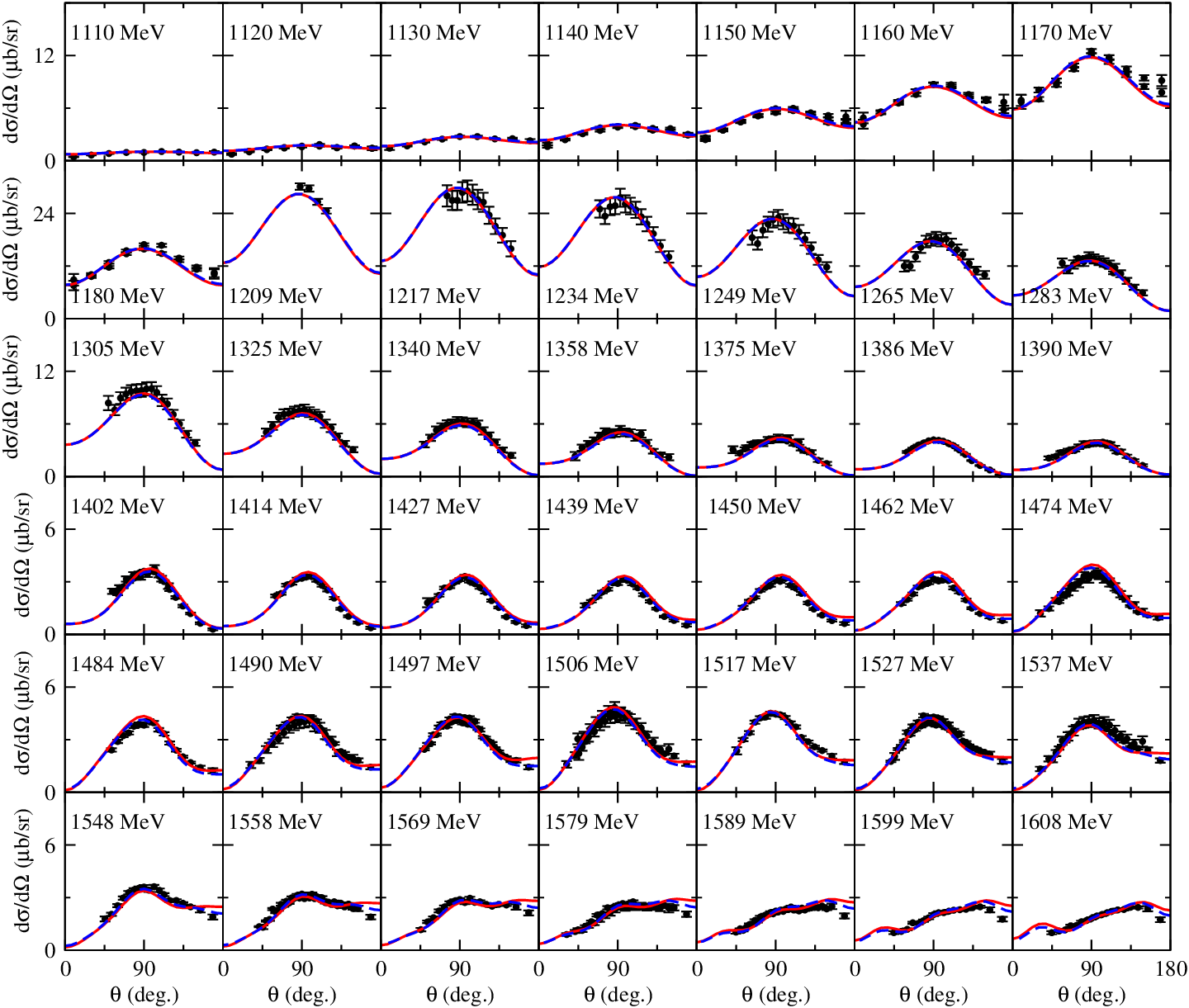}
\includegraphics[clip,width=0.67\textwidth]{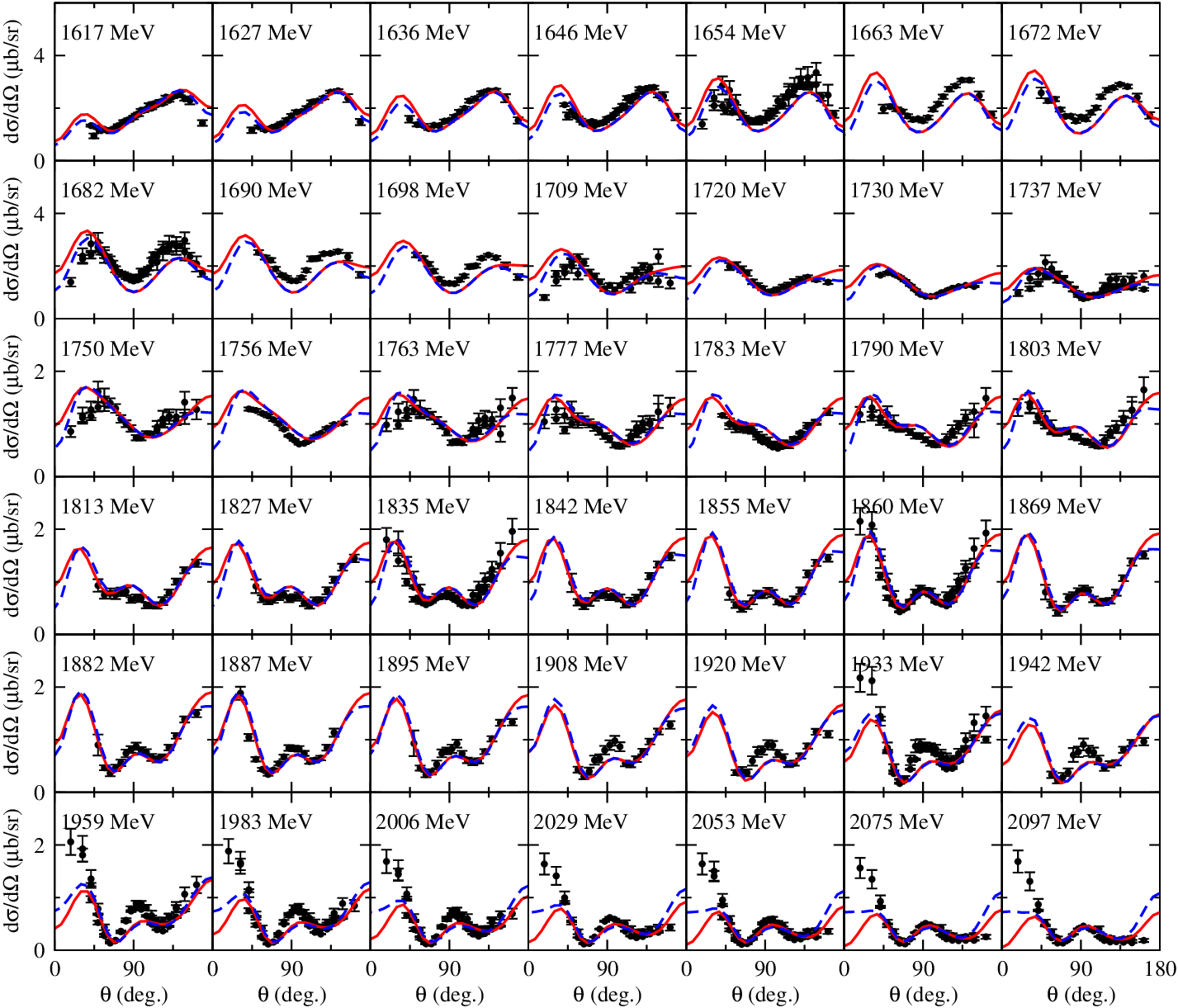}
\caption{\label{fig:gp-pi0p-dcs}
Differential cross sections of $\gamma p\to \pi^0 p$.
The red solid curves are the current results while the blue dashed curves are
from our previous analysis.
}
\end{figure}

\begin{figure}[h]
\includegraphics[clip,width=0.80\textwidth]{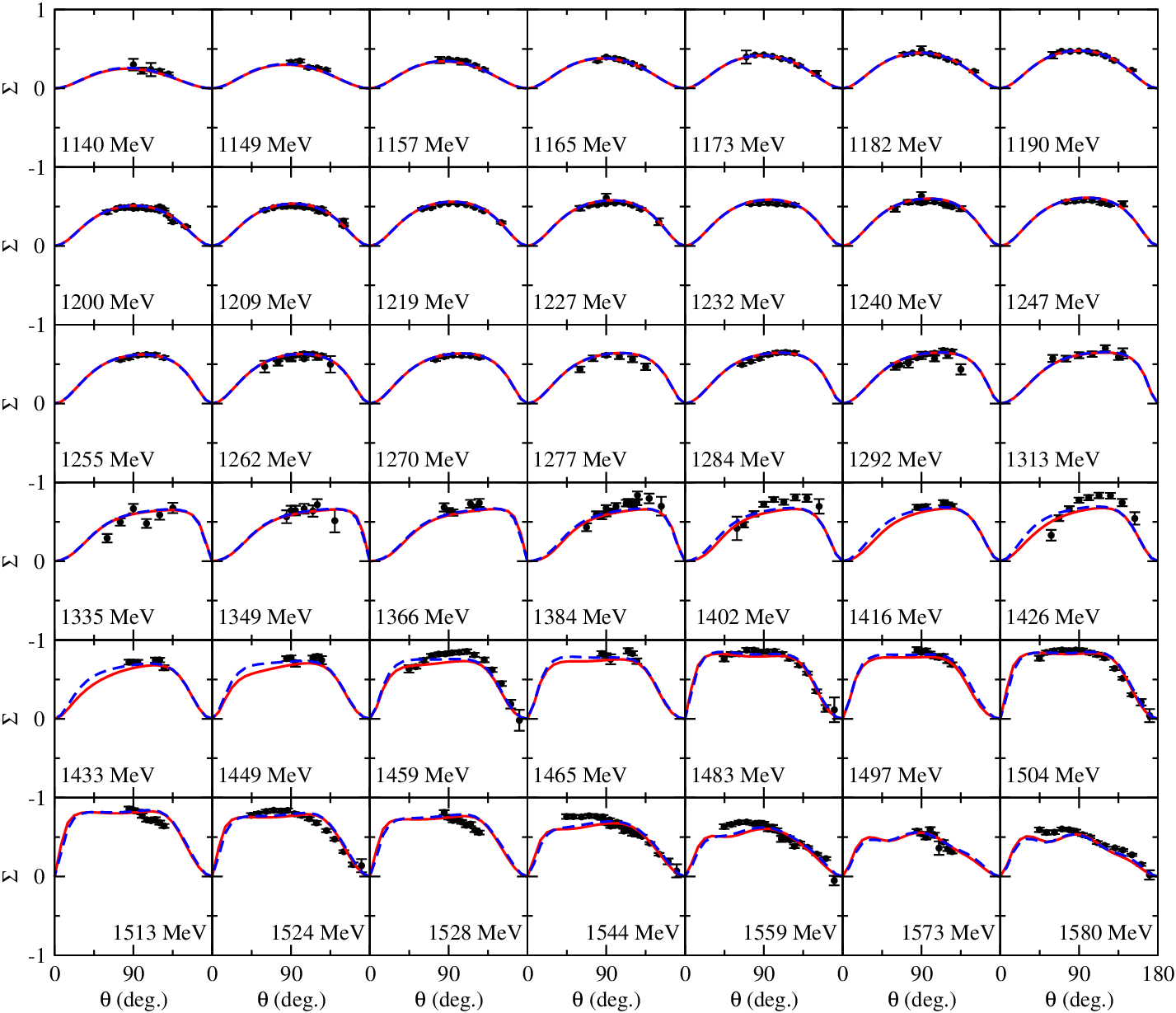}
\includegraphics[clip,width=0.80\textwidth]{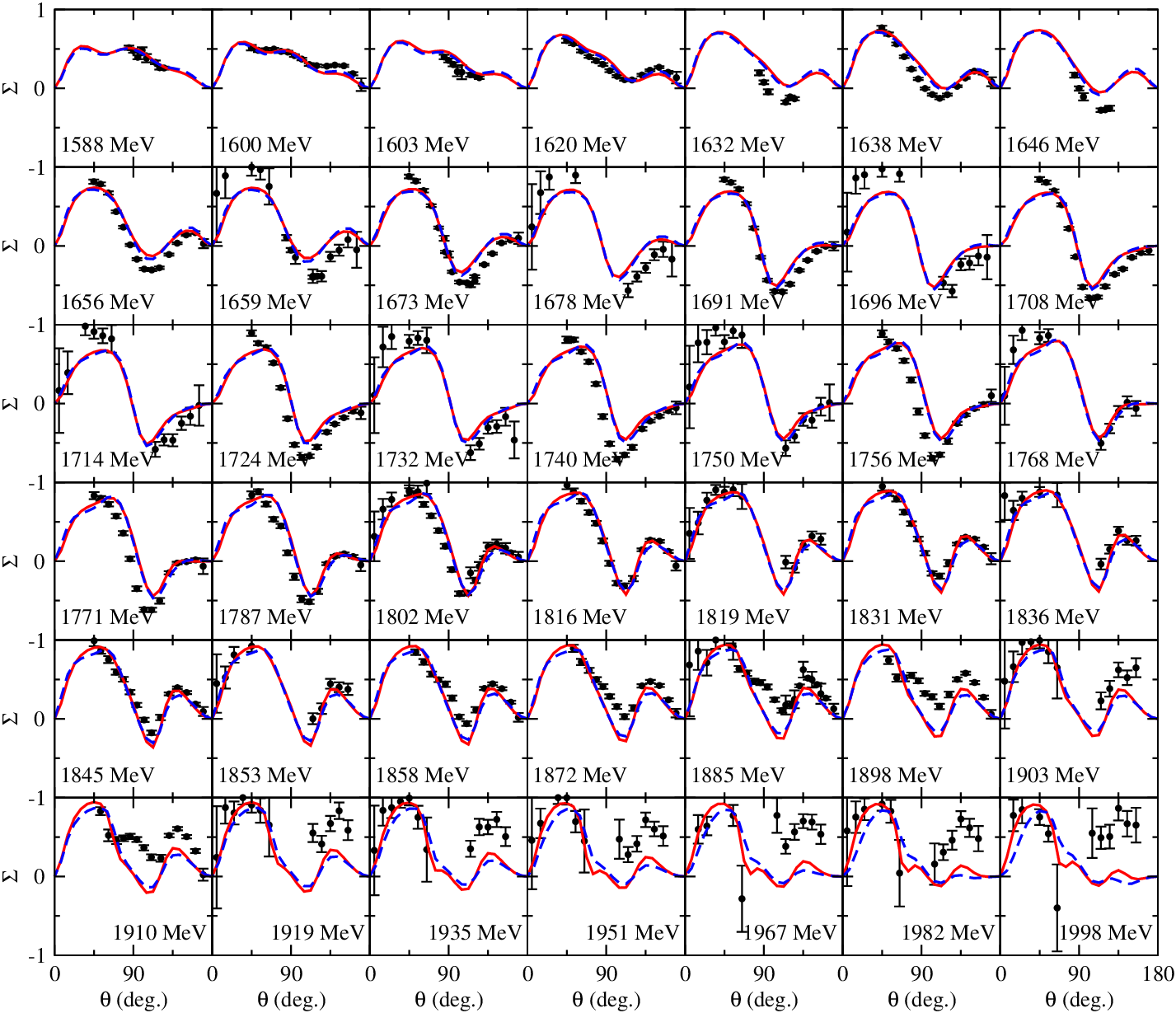}
\caption{\label{fig:gp-pi0p-s}
The photon asymmetries $\Sigma$ of $\gamma p\to \pi^0 p$.
The red solid curves are the current results while the blue dashed curves are
from our previous analysis.
}
\end{figure}

\clearpage
\subsection{$\gamma p \rightarrow \pi^+ n$}
\begin{figure}[h]
\includegraphics[clip,width=0.80\textwidth]{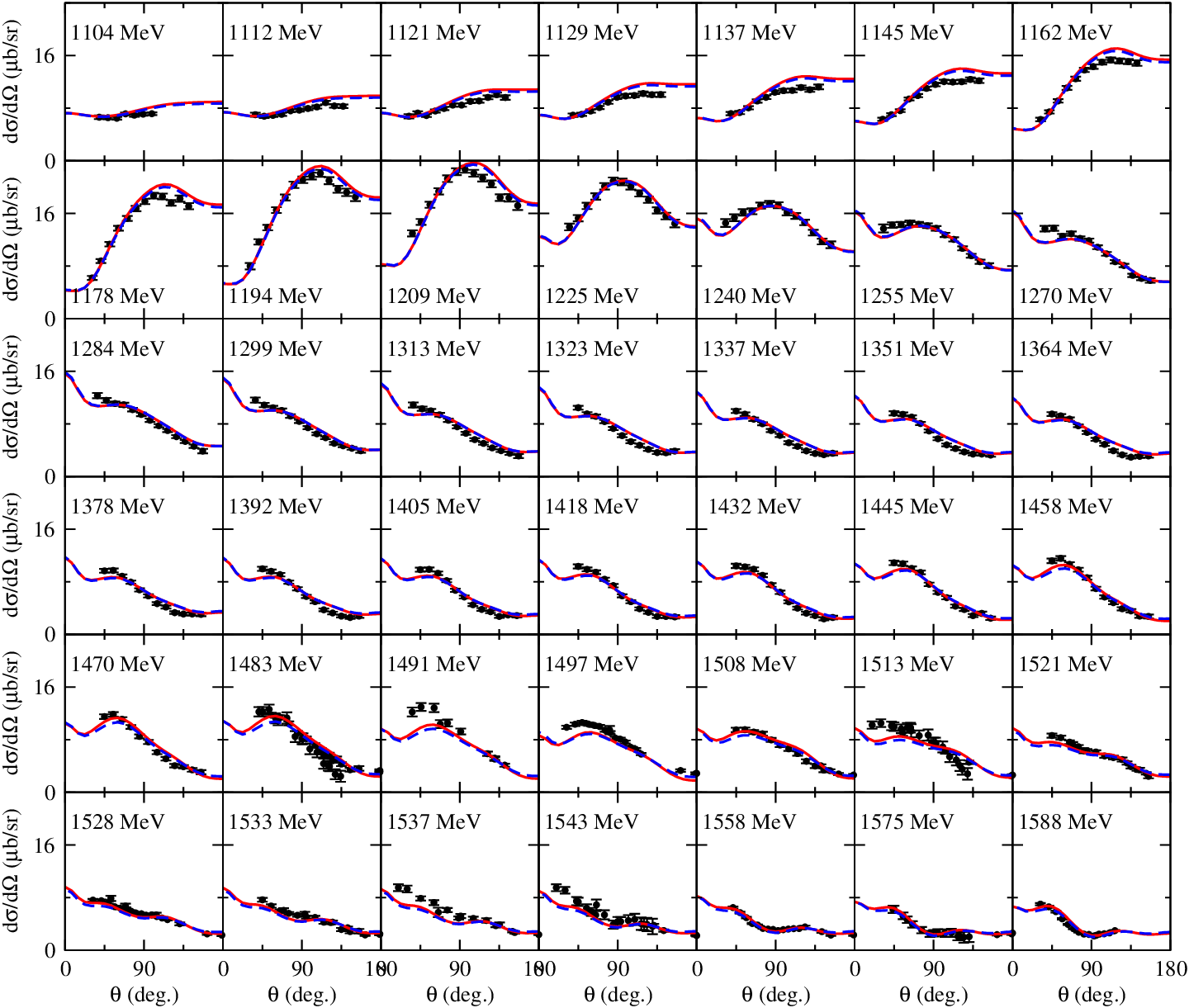}
\includegraphics[clip,width=0.80\textwidth]{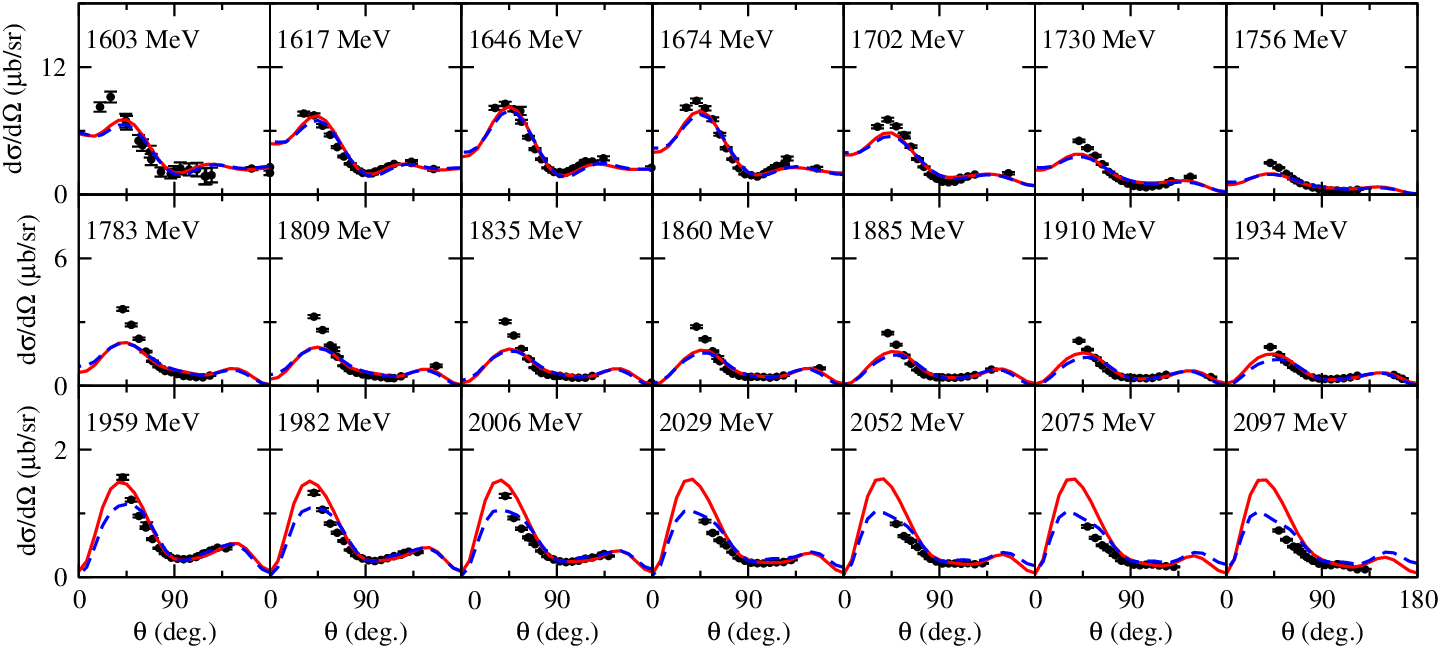}
\caption{\label{fig:gp-pipn-dcs}
Differential cross sections of $\gamma p\to \pi^+ n$.
The red solid curves are the current results while the blue dashed curves are
from our previous analysis.
}
\end{figure}

\begin{figure}[h]
\includegraphics[clip,width=0.80\textwidth]{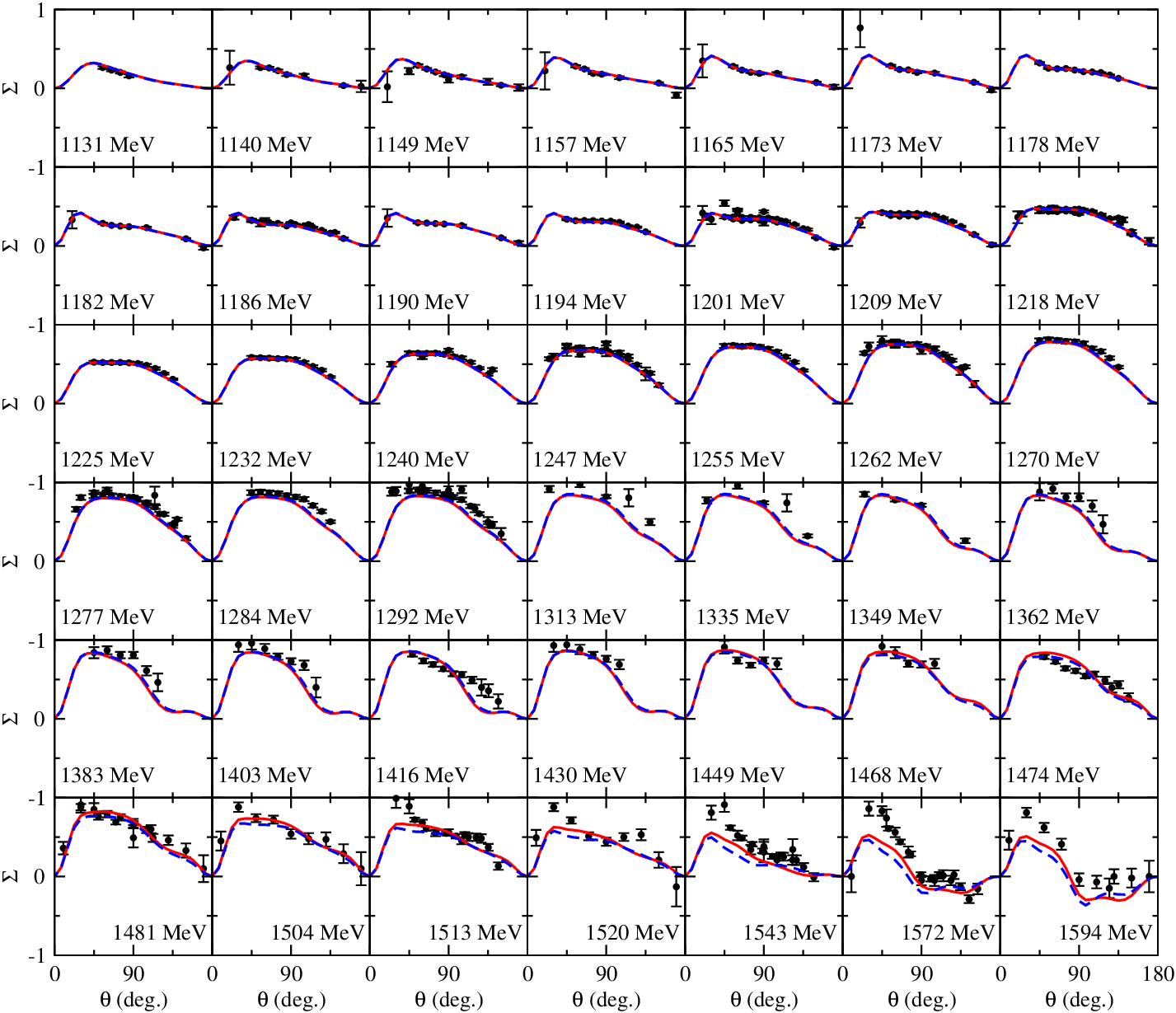}
\includegraphics[clip,width=0.80\textwidth]{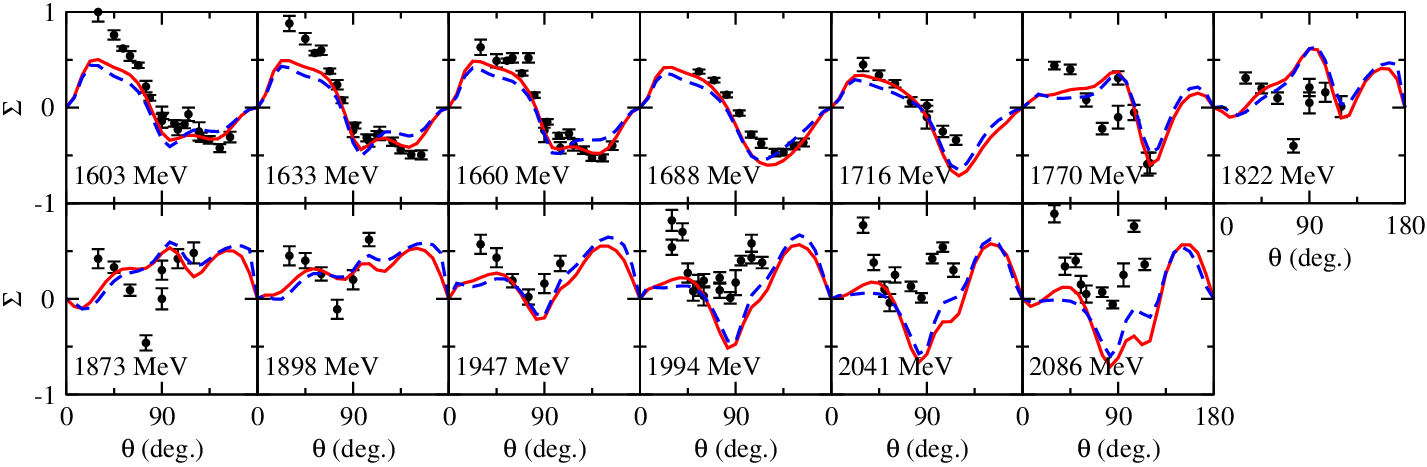}
\caption{\label{fig:gp-pipn-s}
The photon asymmetries $\Sigma$ of $\gamma p\to \pi^+ n$.
The red solid curves are the current results while the blue dashed curves are
from our previous analysis.
}
\end{figure}

\clearpage
\subsection{$\gamma p \rightarrow \eta p$}
\begin{figure}[h]
\includegraphics[clip,width=0.80\textwidth]{gpep-DC}
\includegraphics[clip,width=0.80\textwidth]{gpep-DC-2}
\caption{Differential cross sections of $\gamma p\to \eta p$.
The red solid curves are the current results while the blue dashed curves are
from our previous analysis.
}
\label{fig:gn-etan-dcs}
\end{figure}

\begin{figure}[h]
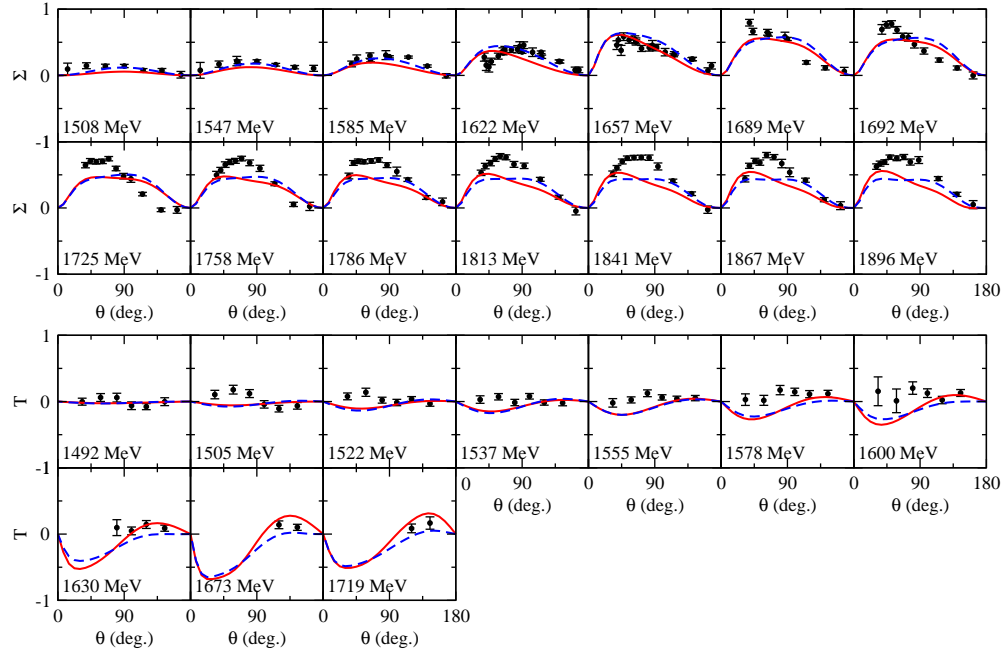

\includegraphics[clip,width=0.80\textwidth]{gpep-S}
\includegraphics[clip,width=0.80\textwidth]{gpep-T}
\caption{$\Sigma$  and $T$ of $\gamma p\to \eta p$.
The red solid curves are the current results while the blue dashed curves are
from our previous analysis.}
\label{fig:gn-etan-s-t}
\end{figure}
\clearpage

\subsection{$\gamma p \rightarrow K^+\Lambda$}
\begin{figure}[h]
\includegraphics[clip,width=0.80\textwidth]{gpkl-DC}
\caption{Differential cross sections
of $\gamma p\to K^+ \Lambda$.
The red solid curves are the current results while the blue dashed curves are
from our previous analysis.}
\label{fig:gn-pkl0-dcs}
\end{figure}

\begin{figure}[h]
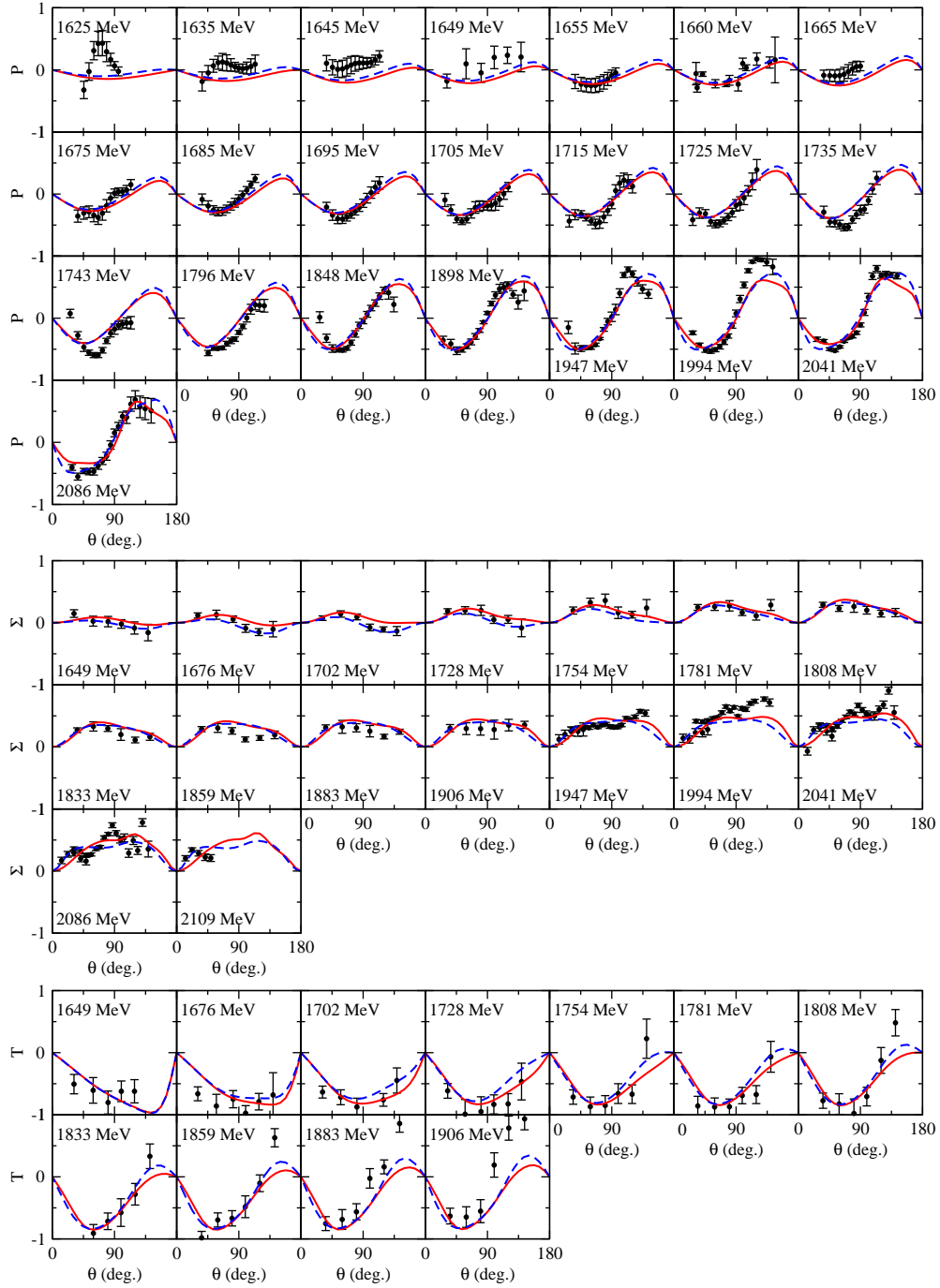

\includegraphics[clip,width=0.80\textwidth]{gpkl-P}
\includegraphics[clip,width=0.8\textwidth]{gpkl-S}
\includegraphics[clip,width=0.8\textwidth]{gpkl-T}
\caption{Observables $P$, $\Sigma$, $T$ of
of $\gamma p\to K^+ \Lambda$.
The red solid curves are the current results while the blue dashed curves are
from our previous analysis.}
\label{fig:gn-pkl0-s-t}
\end{figure}

\clearpage

\subsection{$\gamma p \rightarrow K\Sigma$}
\begin{figure}[h]
\includegraphics[clip,width=0.80\textwidth]{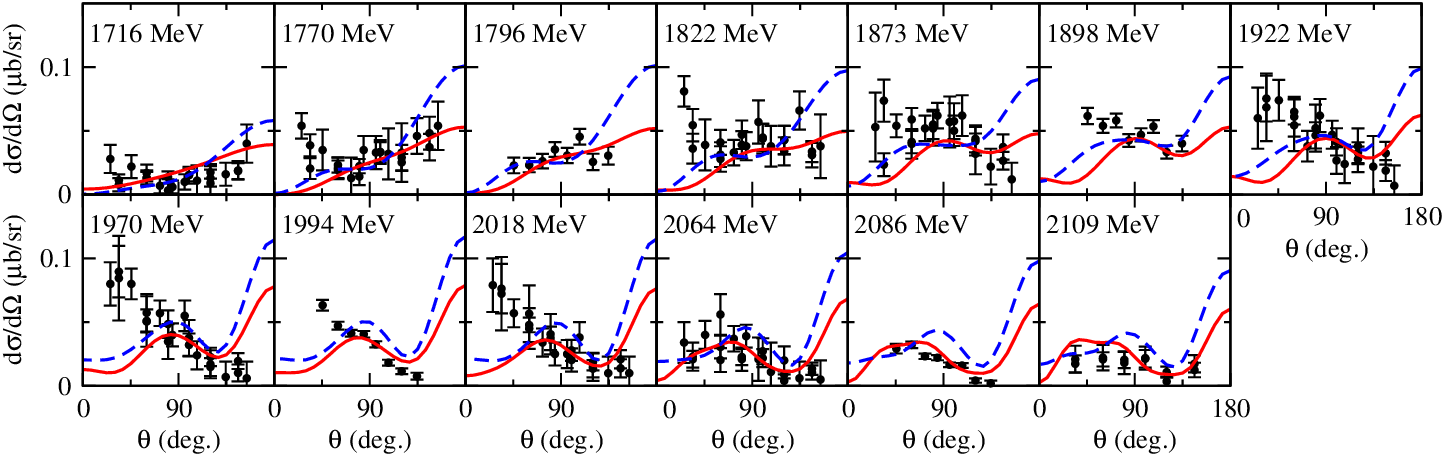}
\caption{Differential cross sections of
$\gamma p\to K^0\Sigma^+$.
The red solid curves are the current results while the blue dashed curves are
from our previous analysis.}
\label{fig:gpkps0-dcs}
\end{figure}
\clearpage

\begin{figure}[h]
\includegraphics[clip,width=0.80\textwidth]{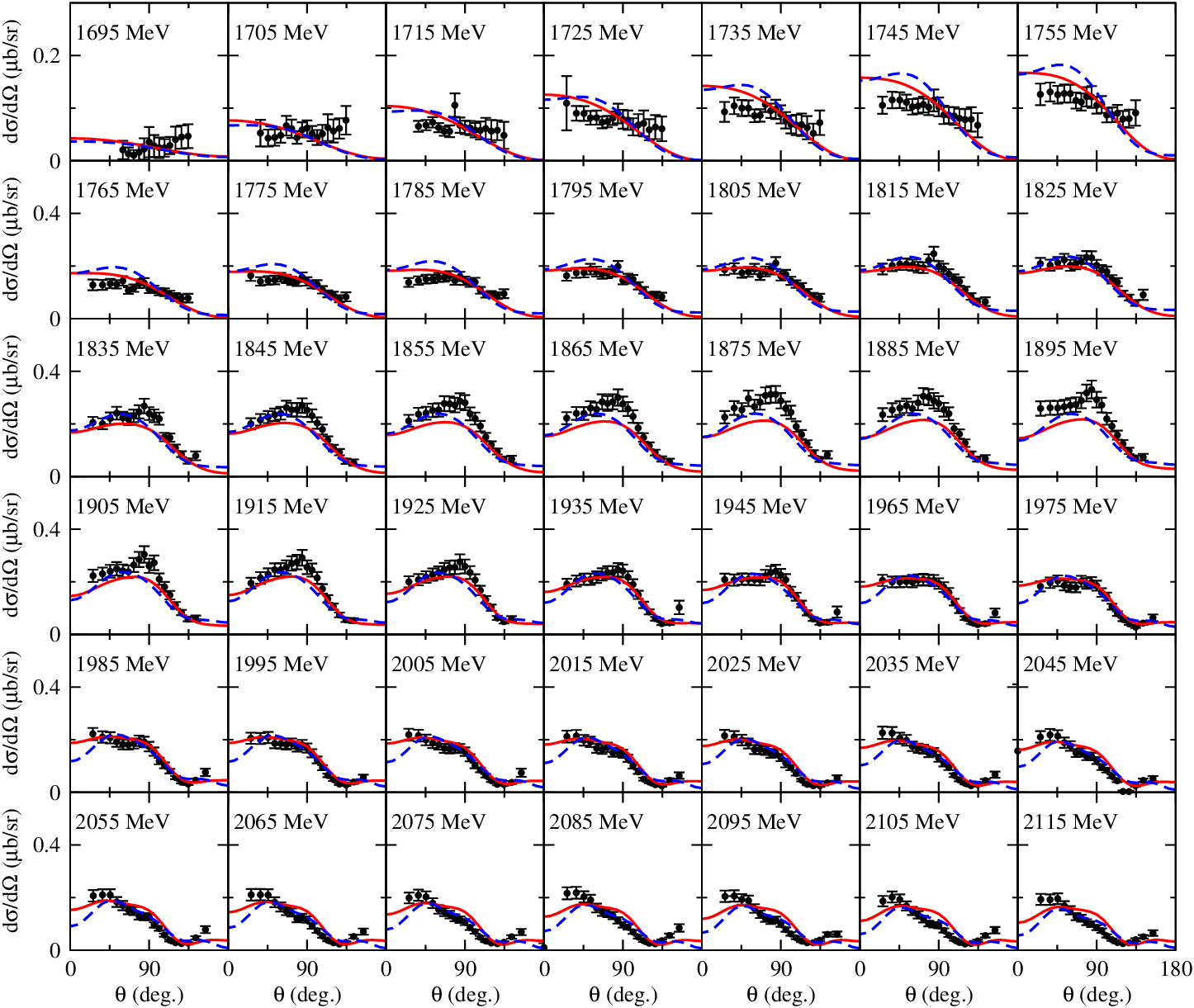}
\caption{Differential cross sections of $\gamma p\to K^+ \Sigma^0$.
The red solid curves are the current results while the blue dashed curves are
from our previous analysis.}
\label{fig:gpkps0-dcs-2}
\end{figure}

\clearpage
\section{Differential Cross sections of $\gamma^* p\rightarrow \pi N $ reactions}
\subsection{$\gamma^* p \rightarrow \pi^0 p$}

\begin{figure}[h]
\includegraphics[clip,width=1\textwidth]{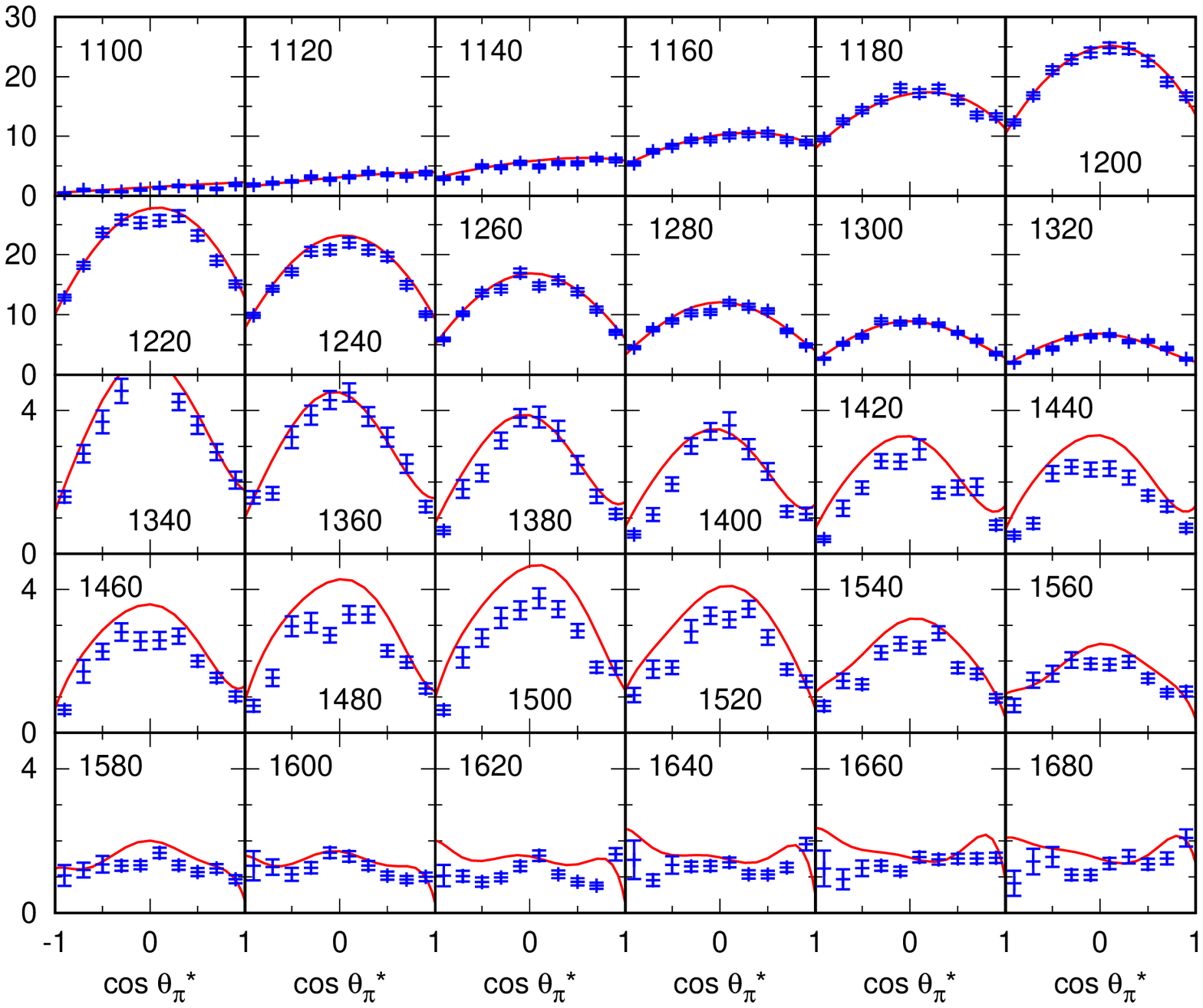}
\caption{Differential cross section ($\sigma_T+\epsilon\sigma_L$) for
 $\gamma^* p \rightarrow \pi^0 p$ at $Q^2=0.4$ (GeV/$c$)$^2$.
}
\label{fig:vg-pi0-dcs-0p4}
\end{figure}

\begin{figure}[h]
\includegraphics[clip,width=1.0\textwidth]{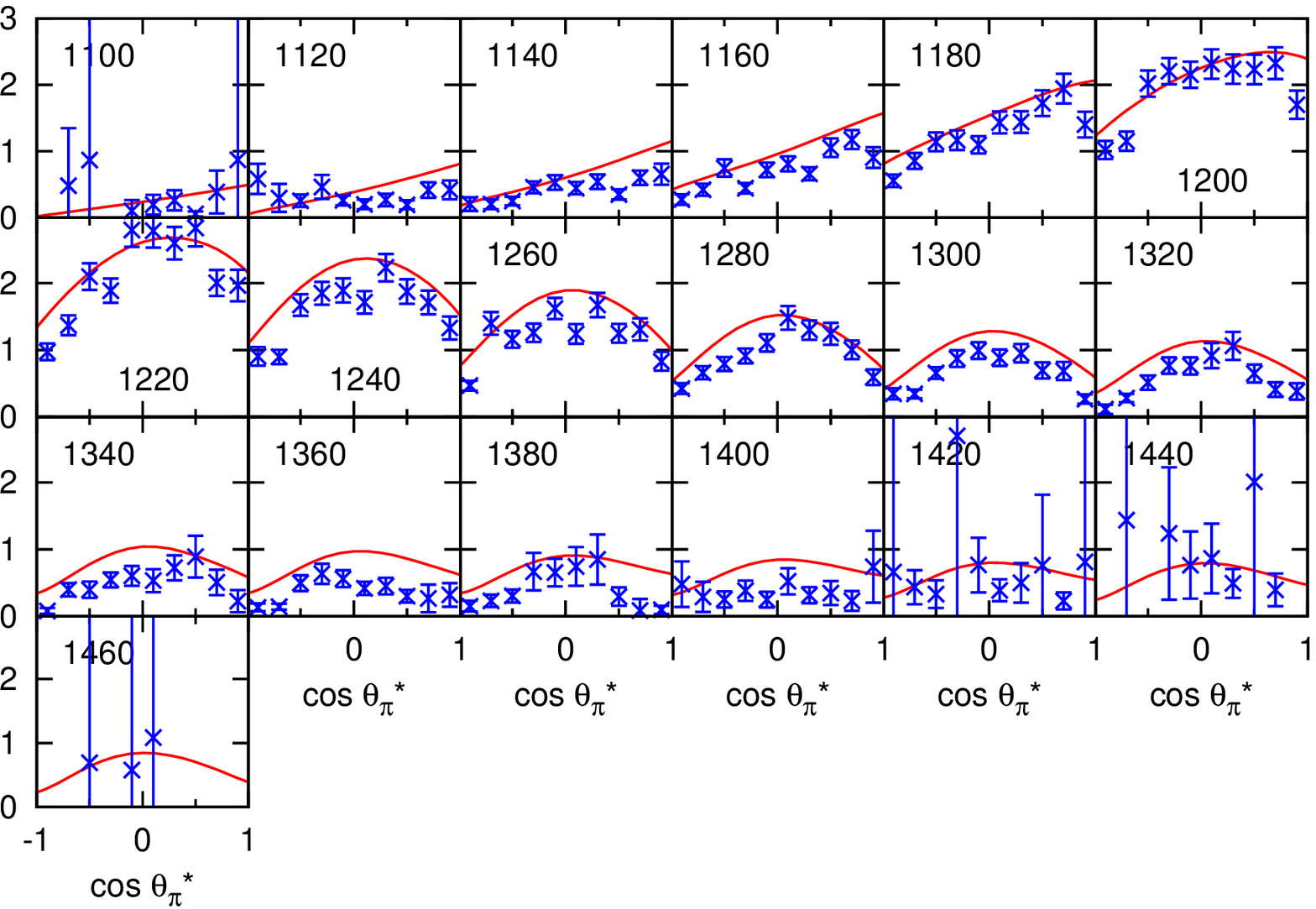}
\caption{Differential cross section ($\sigma_T+\epsilon\sigma_L$) for
 $\gamma^* p \rightarrow \pi^0 p$ at $Q^2=1.76$ (GeV/$c$)$^2$.
}
\label{fig:vg-pi0-dcs-1p76}
\end{figure}

\begin{figure}[h]
\includegraphics[clip,width=1.0\textwidth]{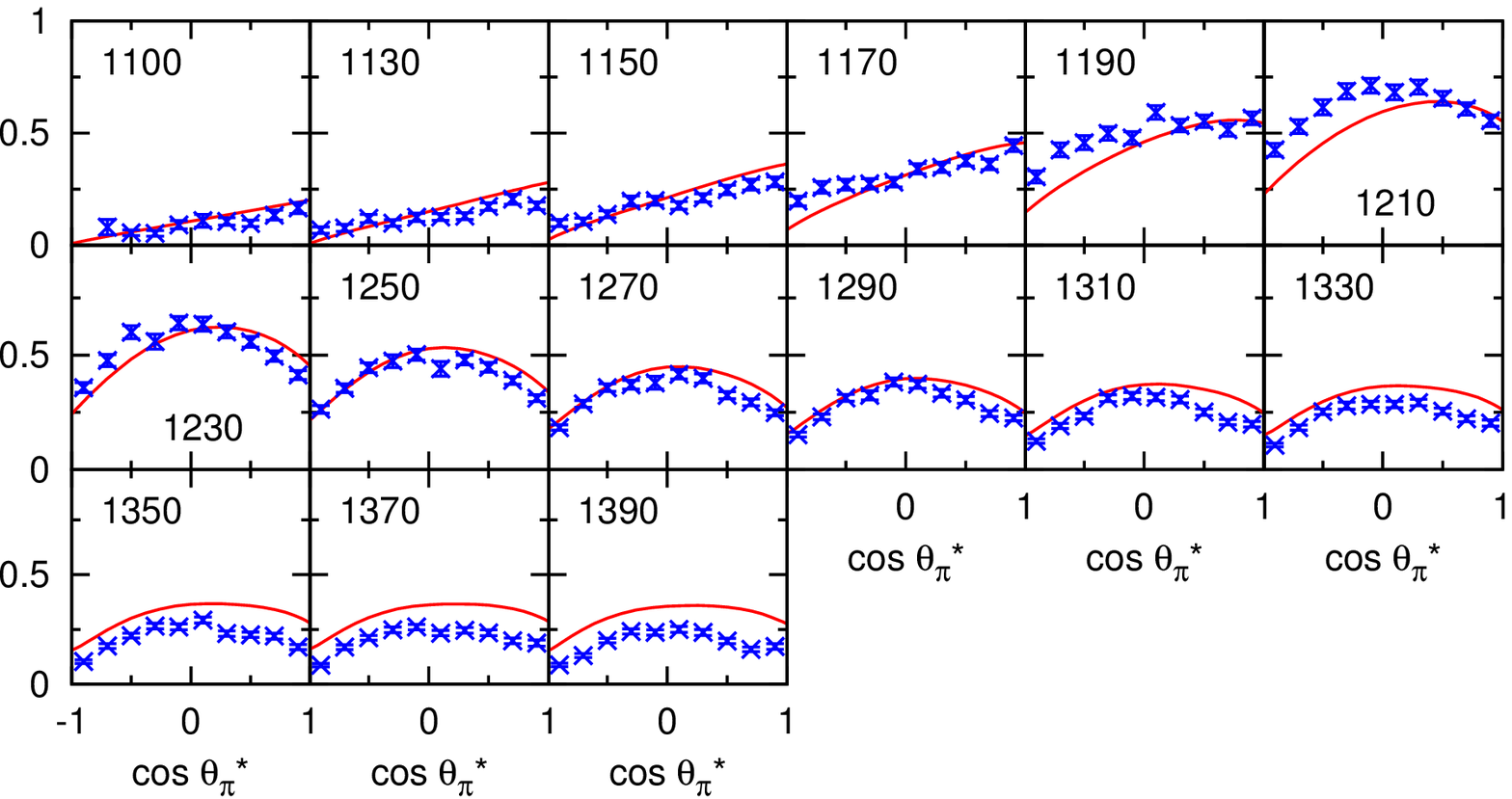}
\caption{Differential cross section ($\sigma_T+\epsilon\sigma_L$) for
 $\gamma^* p \rightarrow \pi^0 p$ at $Q^2=3.00$ (GeV/$c$)$^2$.
}
\label{fig:vg-pi0-dcs-3p00}
\end{figure}
\clearpage
\subsection{ $\gamma^* p \rightarrow \pi^+ n$}
\begin{figure}[h]
\includegraphics[clip,width=1.0\textwidth]{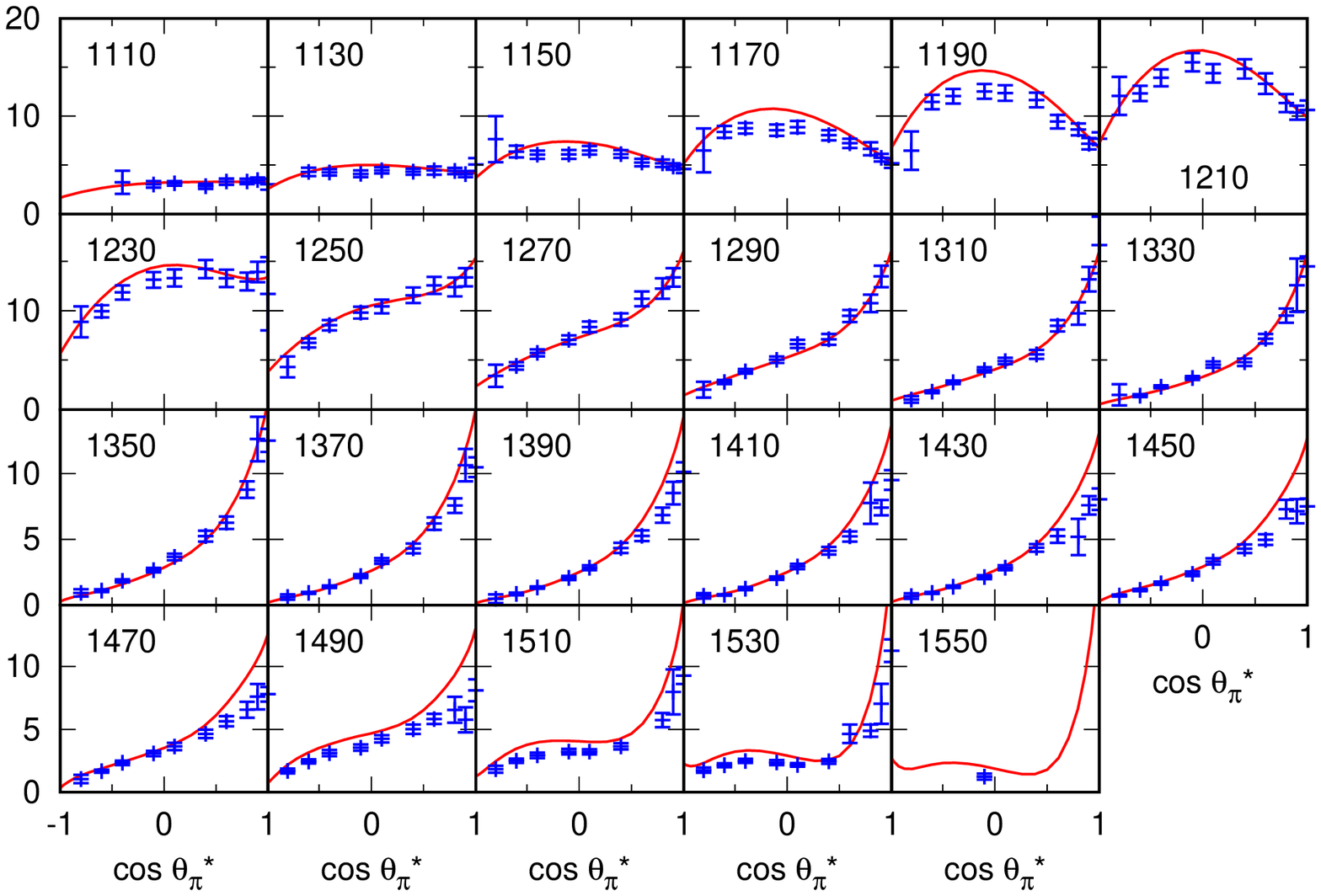}
\caption{Differential cross section ($\sigma_T+\epsilon\sigma_L$) for
 $\gamma^* p \rightarrow \pi^+ n$ at $Q^2=0.4$ (GeV/$c$)$^2$.
}
\label{fig:vg-pip-dcs-0p4}
\end{figure}
\begin{figure}[h]
\includegraphics[clip,width=1.0\textwidth]{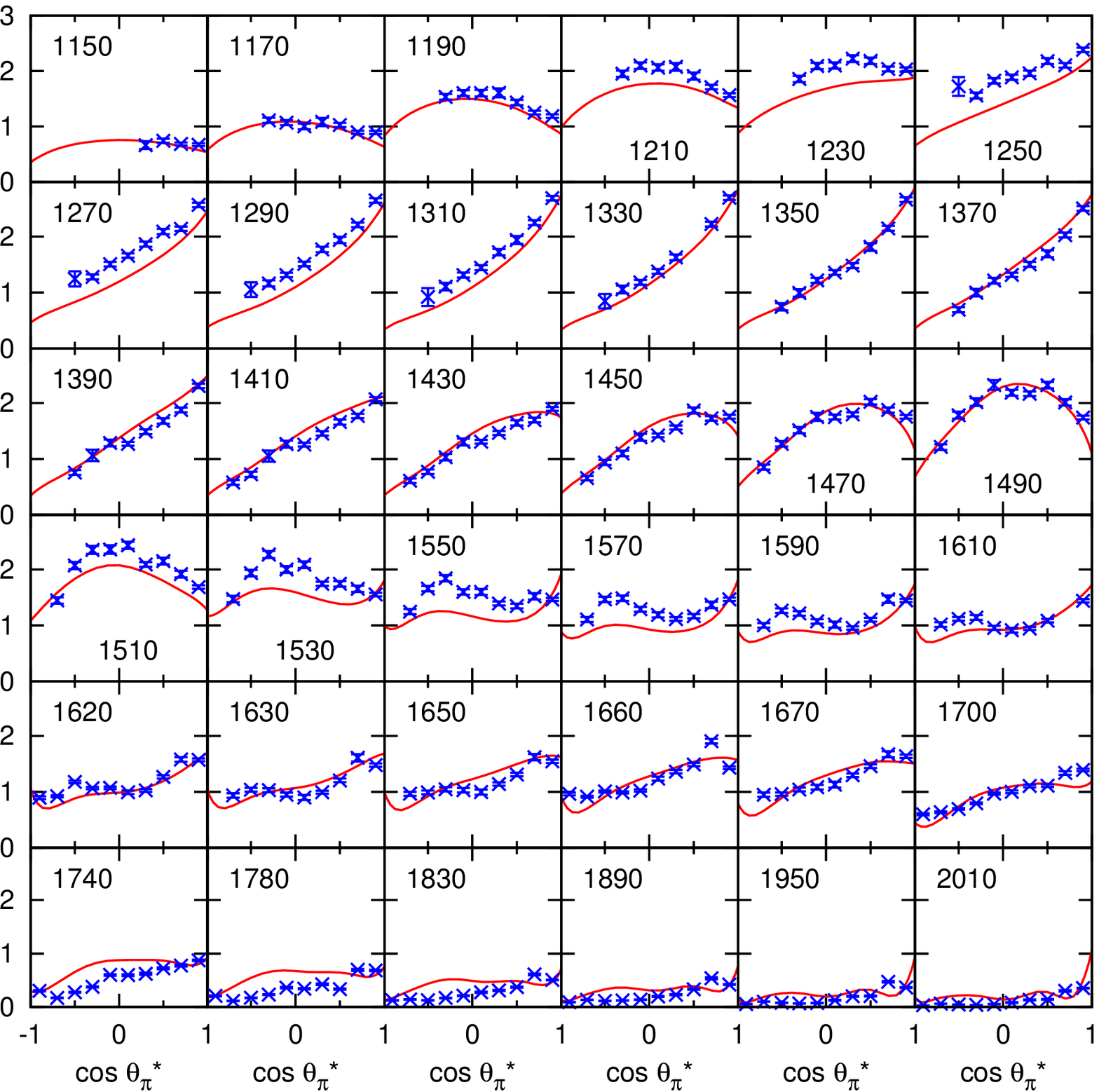}
\caption{Differential cross section ($\sigma_T+\epsilon\sigma_L$) for
 $\gamma^* p \rightarrow \pi^+ n$ at $Q^2=1.76$ (GeV/$c$)$^2$.
}
\label{fig:vg-pip-dcs-1p76}
\end{figure}

\begin{figure}[h]
\includegraphics[clip,width=1.0\textwidth]{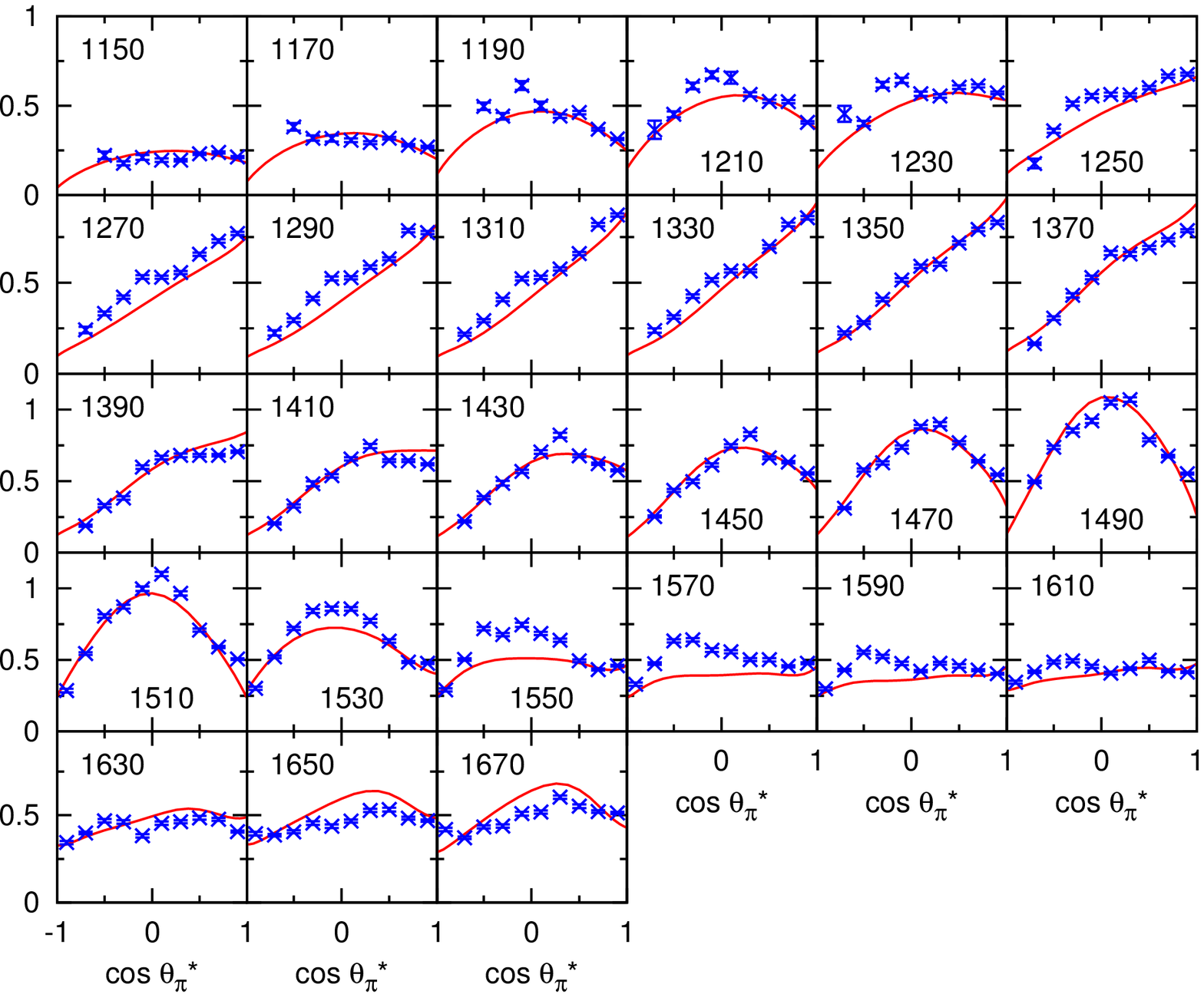}
\caption{Differential cross section ($\sigma_T+\epsilon\sigma_L$) for
 $\gamma^* p \rightarrow \pi^+ n$ at $Q^2=2.91$ (GeV/$c$)$^2$.
}
\label{fig:vg-pip-dcs-2p91}
\end{figure}

\clearpage
\section{Inclusive cross section of $p(e,e')$}
\begin{figure}[h]
\includegraphics[clip,width=0.4\textwidth]{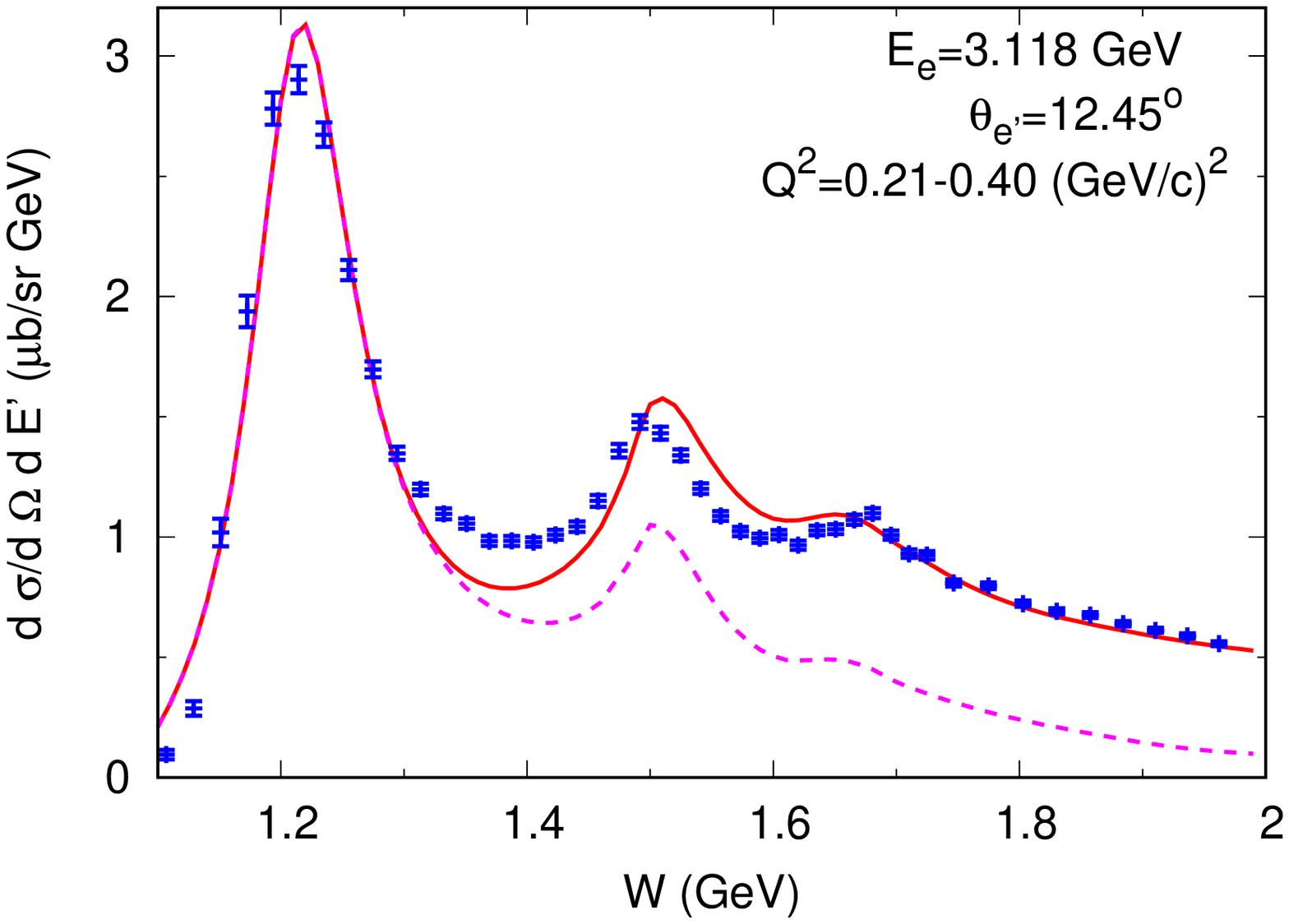}
\includegraphics[clip,width=0.4\textwidth]{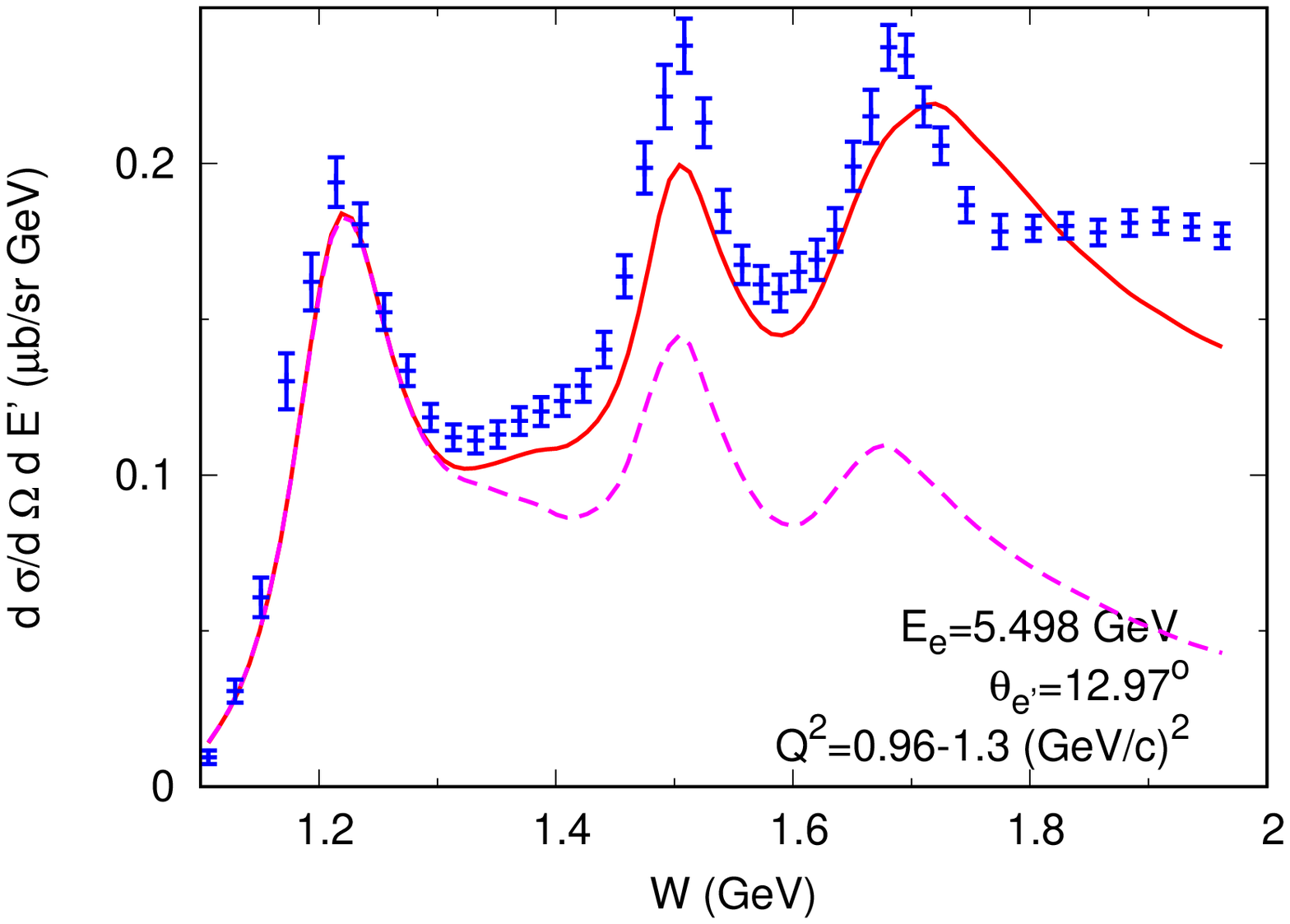}
\includegraphics[clip,width=0.4\textwidth]{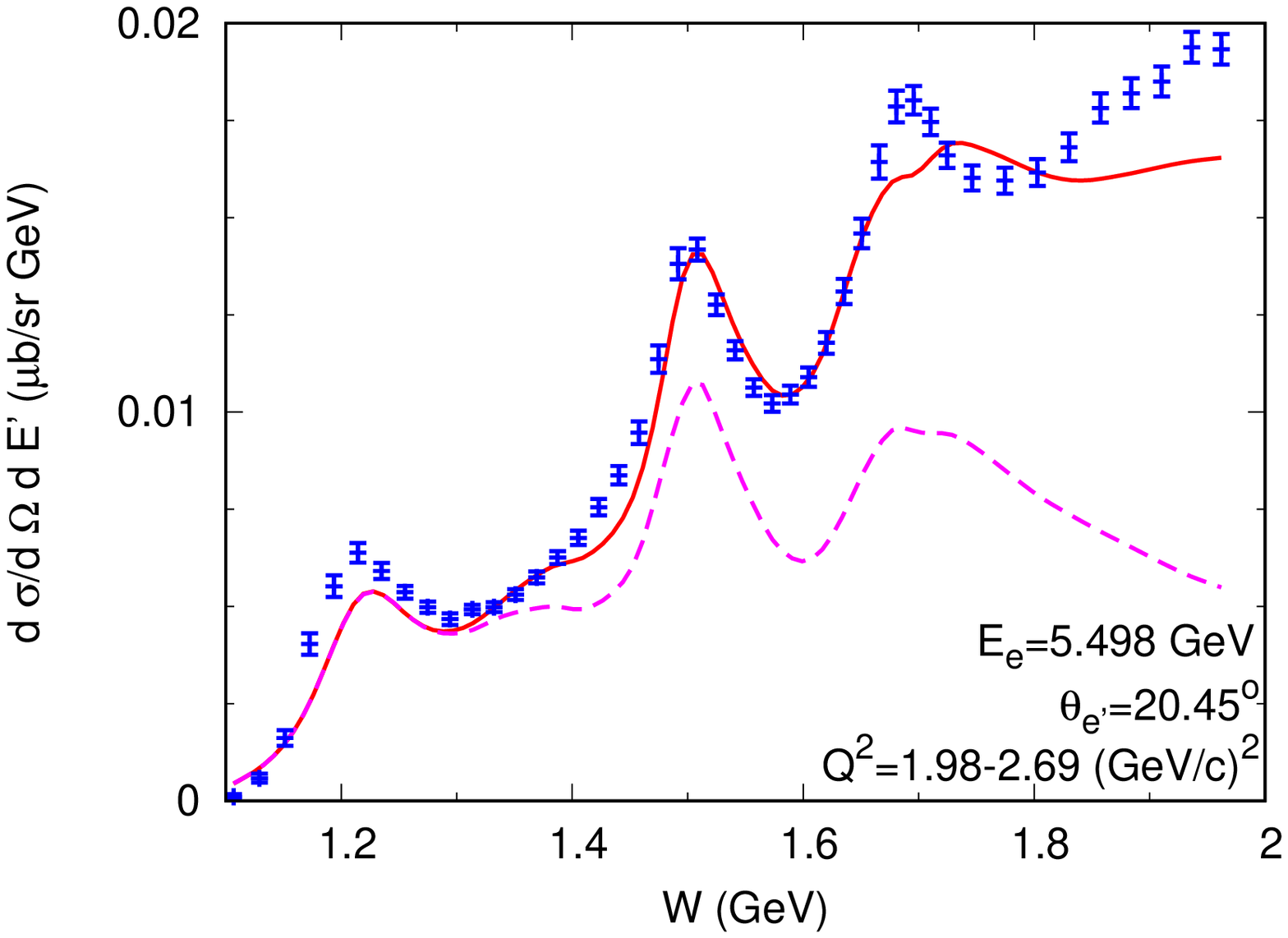}
\caption{Differential cross sections of $p(e,e')X$ at $Q^2\sim 0.3, 1.1, 2.3$ (GeV/$c$)$^2$.
Dashed curves are from the contributions of $p(e,e'\pi)N$.
}
\label{fig:ee-incl-0p3}
\end{figure}



\clearpage

\begin{thebibliography}{99}
\bibitem{sl96}
Meson-exchange model of $\pi N$ scattering and $\gamma N \rightarrow \pi N$ reaction

T. Sato, T.-S. H. Lee, Phys. Rev. C {\bf 34}, 2660 (1996)

\bibitem{sl01}
Dynamical study of $\Delta$ excitation in $N(e,e'\pi)$ reaction

T. Sato, T.-S. H. Lee, Phys. Rev. C {\bf 63}, 055201 (2001)

\bibitem{jlss07}
Extraction and Interpretation of$\gamma N \rightarrow \Delta$
form Factors within a Dynamical Model

B. Julia-Diaz, T.-S.H. Lee, T. Sato, L.C. Smith,
 Phys. Rev. C {\bf 75}, 015205 (2007)

\bibitem{msl07}
Dynamical coupled-channel model of meson production reaction in the nucleon resonance region

A. Matsuyama, T. Sato, T.-S. H. Lee, Phys. Rept {\bf 439}, 193 (2007)

\bibitem{jlms07}
Dynamical coupled-channel model of $\pi N$ scattering in the $W < 2$ GeV nucleon resonance region

B. Julia-Diaz, T.-S. H. Lee, A. Matsuyama, T. Sato, Phys. Rev. C {\bf 76}, 065201 (2007)

 
\bibitem{jlmss08}
Dynamical coupled-channel effects on pion photoproduction

B. Julia-Diaz, T.-S. H. Lee, A. Matsuyama, T. Sato, L.C. Smith,
	Phys. Rev. C {\bf 77}, 045201 (2008)

\bibitem{djlss08}
Dynamical coupled-channel study of $\pi^- p \rightarrow \eta n$ process

J. Durand, B. Julia-Diaz, T.-S. H. Lee, B. Saghai, T. Sato, Phys. Rev. C
	{\bf 78}, 025204 (2008)

\bibitem{ssl09}
Extraction of resonances from meson-nucleon reaction

N. Suzuki, T. Sato, T.-S. H. Lee, Phys. Rev. C {\bf 79}, 025205 (2009)

\bibitem{kjlms09}
Dynamical coupled-channels study of $\pi N \rightarrow \pi\pi N$ reactions

	H. Kamano, B. Julia-Diaz, T.-S. H. Lee, A. Matsuyama, T. Sato,
	Phys. Rev. C {\bf 79}, 025206 (2009)


\bibitem{jklmts09}
Dynamical coupled-channels analysis of $^1H(e,e'\pi)N$ reactions

B. Julia-Diaz, H. Kamano, T.-S. H. Lee, A. Matsuyama, T. Sato, N. Suzuki
Phys. Rev. C {\bf 80}, 025207 (2009)


\bibitem{kjlms09-2}
Double and single pion photoproduction within a dynamical coupled-channels model

H. Kamano, B. Julia-Diaz, T.-S.H. Lee, A. Matsuyama, T. Sato,
Phys. Rev. C {\bf 80}, 065203 (2009)


\bibitem{sjklms10}
Disentangling the dynamical origin of $P_{11}$ nucleon resonances

N. Suzuki, B. Julia-Diaz, H. Kamano, T.-S. H. Lee, A. Matsuyama, T. Sato,
 Phys. Rev. Lett. {\bf 104}, 042302 (2010)


\bibitem{knls10}
Extraction of $P_{11}$ resonances from $\pi N$ data

H. Kamano, S.X. Nakamura, T.-S.H. Lee, T. Sato,
Phys. Rev. C {\bf 81}, 065207 (2010)

\bibitem{ssl10}
Extraction of electromagnetic transition form factors for nucleon resonances within a dynamical coupled-channels model

 N. Suzuki, T. Sato, T.-S.H. Lee,
 Phys. Rev. C {\bf 82}, 045206 (2010)

\bibitem{shkl11}
Determining pseudoscalar meson photo-production amplitudes from complete experiments

M. Sandorfi, S. Hoblit, H. Kamano, T.-S. H. Lee,
J. Phys. G {\bf 38}, 053001 (2011)

\bibitem{knls13}
Nucleon resonances within a dynamical coupled-channel model of $\pi N$ and $\gamma N$ reactions

H. Kamano, S.X. Nakamura, T.-S.H. Lee, T. Sato, Phys. Rev. C {\bf 88}, 035209 (2013)

\bibitem{knls16}
Isospin decomposition of $\gamma^*N \rightarrow N^*$  transitions within a dynamical coupled-channels model

H. Kamano, S.X. Nakamura, T.S.H. Lee, T. Sato,
 Phys. Rev. C {\bf 94}, 015201 (2016)

\bibitem{nks15}
Dynamical coupled-channels model for neutrino-induced meson productions in resonance region

S.X. Nakamura, H. Kamano, T. Sato,
Phys. Rev. D {\bf 92}, 074024 (2015)


\end{thebibliography}
\end{document}